\documentclass[%
 reprint,
superscriptaddress,
 amsmath,amssymb,
 aps,
prb,
]{revtex4-2}

\usepackage{booktabs}
\usepackage{multirow}
\usepackage{color}
\usepackage{graphicx}
\usepackage{dcolumn}
\usepackage{bm}
\usepackage{hyperref}
\usepackage[mathlines]{lineno}

\begin{document}

\title{Intrinsic and tunable quantum anomalous Hall effect and magnetic topological phases in \textit{XY}Bi${}_2$Te${}_5$}

\author{Xin-Yi Tang}
\affiliation{%
 State Key Laboratory of Low Dimensional Quantum Physics, Department of Physics, Tsinghua University, Beijing 100084, China
}%

\author{Zhe Li}%
 \email{lizhe21@iphy.ac.cn}
\affiliation{%
 Beijing National Research Center for Condensed Matter Physics, and Institute of Physics, Chinese Academy of Sciences, Beijing 100190, China
}%

\author{Feng Xue}
\affiliation{%
 State Key Laboratory of Low Dimensional Quantum Physics, Department of Physics, Tsinghua University, Beijing 100084, China
}%
\affiliation{%
 Beijing Academy of Quantum Information Sciences, Beijing 100193, China
}%

\author{Pengfei Ji}
\affiliation{%
 State Key Laboratory of Low Dimensional Quantum Physics, Department of Physics, Tsinghua University, Beijing 100084, China
}%

\author{Zetao Zhang}
\affiliation{%
 State Key Laboratory of Low Dimensional Quantum Physics, Department of Physics, Tsinghua University, Beijing 100084, China
}%

\author{Xiao Feng}
\affiliation{%
 State Key Laboratory of Low Dimensional Quantum Physics, Department of Physics, Tsinghua University, Beijing 100084, China
}%
\affiliation{%
Beijing Academy of Quantum Information Sciences, Beijing 100193, China
}%
\affiliation{%
 Frontier Science Center for Quantum Information, Beijing 100084, China
}%

\author{Yong Xu}
\affiliation{%
 State Key Laboratory of Low Dimensional Quantum Physics, Department of Physics, Tsinghua University, Beijing 100084, China
}%
\affiliation{%
 Frontier Science Center for Quantum Information, Beijing 100084, China
}%
\affiliation{%
Tencent Quantum Laboratory, Tencent, Shenzhen, Guangdong 518057, China
}%
\affiliation{%
RIKEN Center for Emergent Matter Science (CEMS), Wako, Saitama 351-0198, Japan
}%

\author{Quansheng Wu}
\affiliation{%
 Beijing National Research Center for Condensed Matter Physics, and Institute of Physics, Chinese Academy of Sciences, Beijing 100190, China
}%
\affiliation{%
 University of Chinese Academy of Sciences, Beijing 100049, China
}%

\author{Ke He}
\affiliation{%
 State Key Laboratory of Low Dimensional Quantum Physics, Department of Physics, Tsinghua University, Beijing 100084, China
}%
\affiliation{%
 Beijing Academy of Quantum Information Sciences, Beijing 100193, China
}%
\affiliation{%
 Frontier Science Center for Quantum Information, Beijing 100084, China
}%
\affiliation{%
Hefei National Laboratory, Hefei 230088, China
}%

\date{\today}

\begin{abstract}
By first-principles calculations, we study the magnetic and topological properties of \textit{XY}Bi${}_2$Te${}_5$-family (\textit{X}, \textit{Y} = Mn, Ni, V, Eu) compounds. The strongly coupled double magnetic atom-layers can significantly enhance the magnetic ordering temperature while keeping the topologically non-trivial properties. Particularly, NiVBi${}_2$Te${}_5$ is found to be a magnetic Weyl semimetal in bulk and a Chern insulator in thin film with both the Curie temperature ($\sim$150 K) and full gap well above 77 K. Ni${}_2$Bi${}_2$Te${}_5$, MnNiBi${}_2$Te${}_5$, NiVBi${}_2$Te${}_5$ and NiEuBi${}_2$Te${}_5$ exhibits intrinsic dynamic axion state. Among them, MnNiBi${}_2$Te${}_5$ has a Néel temperature over 200 K and Ni${}_2$Bi${}_2$Te${}_5$ even demonstrates antiferromagnetic order above room temperature. These results indicate an approach to realize high temperature quantum anomalous Hall effect and other topological quantum effects for practical applications.
\end{abstract}

\maketitle
\section{introduction}
In the past two decades, topological states of materials have grown into a large branch of modern condensed matter physics \cite{hasan-RevModPhys.82.3045,qi-RevModPhys.83.1057,wsm-rev-RevModPhys.90.015001,Tokura2019,axion-review,QAHE-RevModPhys.95.011002}. Emergent topological phases, especially the ones associated with magnetism, like the quantum anomalous Hall effect (QAHE) \cite{haldane-PhysRevLett.61.2015,yurui-doi:10.1126/science.1187485,qahe-doi:10.1126/science.1234414,apl-doi:10.1063/1.4935075,Chang2015}, axion insulators (AxIs) \cite{tft-PhysRevB.78.195424,daf-Li2010,wj-axi-PhysRevB.92.081107,nagaosa-axi-PhysRevB.92.085113,tokura-axi-Mogi2017,chang-axi-PhysRevLett.120.056801} and magnetic Weyl semimetals (WSMs) \cite{wan-wsm-PhysRevB.83.205101,Co3Sn2S2-doi:10.1126/science.aav2873,Co3Sn2S2-Liu2018}, have attracted lots of attention. The QAHE has been experimentally achieved in Cr-doped topological insulator thin films initially in 2013 \cite{qahe-doi:10.1126/science.1234414}. However, the reliance on extremely low temperature hindered its applications. Many candidate materials that may realize high temperature QAH insulators, magnetic WSMs or AxIs have been proposed, such as Co${}_3$Sn${}_2$S${}_2$ \cite{Co3Sn2S2-doi:10.1126/science.aav2873,Co3Sn2S2-Liu2018}, EuIn${}_2$As${}_2$ \cite{euin2as2-PhysRevLett.122.256402}, LiFeSe \cite{lifese-PhysRevLett.125.086401}, PdBr${}_3$-family materials \cite{pdbr3-PhysRevApplied.12.024063}, NiAsO${}_3$ and PdSbO${}_3$ \cite{qiao-PhysRevLett.129.036801}. Although some of them have been prepared and well-studied, quantized transport properties—the crucial mile stone—have not been achieved. Thin films of the intrinsic magnetic topological insulator MnBi${}_2$Te${}_4$ \cite{mbt-exp-Yan,mbt-exp-Otrokov2019} are a high temperature QAH system that has shown quantized transport properties in experiment. The magnetically induced surface state gap (briefly, magnetic gap) of MnBi${}_2$Te${}_4$ thin films can reach $\sim$50 meV \cite{mbt-sciadv-doi:10.1126/sciadv.aaw5685,ljh-maggap-PhysRevB.100.121103}, more than enough to support QAHE above 77 K. Zero-field quantization of the anomalous Hall resistance have been observed in odd septuple-layer (SL) MnBi${}_2$Te${}_4$ thin flakes at 1.4 K \cite{mbt-exp-zyb-doi:10.1126/science.aax8156}, higher than that in magnetically doped (Bi,Sb)${}_2$Te${}_3$. Amazingly, under high magnetic field, nearly quantized Hall resistance has been observed in some MnBi${}_2$Te${}_4$ thin flake samples at a temperature as high as $\sim$40 K \cite{nsr-10.1093/nsr/nwaa089}. Other magnetic topological phases such as magnetic WSM and AxI can be realized in MnBi${}_2$Te${}_4$ with certain magnetic configurations and thicknesses \cite{mbt-sciadv-doi:10.1126/sciadv.aaw5685,ljh-maggap-PhysRevB.100.121103,mbt-otrokov-PhysRevLett.122.107202,mbt-wj-PhysRevLett.122.206401,mbt-exp-Liu2020,nsr-10.1093/nsr/nwaa089}. Obviously, MnBi${}_2$Te${}_4$ provides a solid base for the exploration of high-temperature QAHE and other related topological quantum effects.

The temperature to achieve the QAHE is determined by both the magnetic ordering temperature and the magnetic gap size. Since the magnetic gap size of MnBi${}_2$Te${}_4$ thin films is as large as $\sim$50 meV, the bottleneck is its magnetic ordering temperature which is only $\sim$25 K \cite{mbt-exp-Otrokov2019}. Such a low temperature originates not only from the weak magnetic coupling strength but also from magnetic fluctuations, largely due to the two-dimensional (2D) nature of its magnetism \cite{mermin-PhysRevLett.17.1133}. A straightforward way to solve the two problems is to make the strongly coupled magnetic atomic layer a little thicker while not thick enough to destroy the 2D topological properties. Actually, the Curie or Néel temperature of magnetic thin films is significantly elevated as the thickness increases from monolayer to bilayer \cite{fewlayers-PhysRevB.49.3962}. 

In fact, there exists such a material: Mn${}_2$Bi${}_2$Te${}_5$, which can be considered as MnBi${}_2$Te${}_4$ with an additional MnTe bilayer inserted in each septuple layer (SL). The material was predicted to be a possible host of QAHE, topological magnetoelectric effect (TME) or AFM topological insulator (TI) phase depending on its layer magnetization \cite{225cpl-Zhang_2020,225prb-PhysRevB.102.121107,otrokov-prb-PhysRevB.105.195105}. Mn${}_2$Bi${}_2$Te${}_5$ single crystals have recently been successfully prepared \cite{225exp-PhysRevB.104.054421}. However, as discussed below, the inherent magnetic couplings of Mn${}_2$Bi${}_2$Te${}_5$ are not strong and thus its magnetic ordering temperature is just 20 K. We can replace one of Mn atom layer in Mn${}_2$Bi${}_2$Te${}_5$ with other magnetic atoms (partially discussed in Ref. \cite{nanoscale}), or even both, namely \textit{XY}Bi${}_2$Te${}_5$ where \textit{X} and \textit{Y} represent magnetic atoms, to implement the above idea. 

In this paper, via systematic first-principles calculations, we demonstrate our findings on high-temperature topological phases in some compounds of \textit{XY}Bi${}_2$Te${}_5$-family materials, where \textit{X}, \textit{Y} = Mn, Ni, V, Eu. We predict that NiVBi${}_2$Te${}_5$ can be a FM WSM in the bulk phase with a Curie temperature of $ \sim $150 K and a Chern insulator in thin films that can possibily show high-temperature QAHE. Furthermore, the bulk phases of Ni${}_2$Bi${}_2$Te${}_5$ and MnNiBi${}_2$Te${}_5$ are predicted to be high-temperature dynamic AxIs, in which the first one even shows magnetic ordering temperature above room temperature.

\section{methods}
The density functional theory (DFT) calculations were carried out via the Vienna \textit{Ab initio} Simulation Package (VASP) \cite{vasp-PhysRevB.54.11169}. The projector-augmented wave (PAW) method and the plane-wave basis with an energy cutoff of 350 eV were utilized. The exchange-correlation energy is described by Perdew-Burke-Ernzerhof (PBE) functional under the generalized gradient approximation (GGA) \cite{gga-PhysRevLett.77.3865}. To treat the localized $ d $- and $ f $-orbitals, the GGA+$ U $  approach was adopted. $ U $ parameters were selected to be 4 eV, 4 eV, 3 eV and 5 eV for Mn-$ 3d $, Ni-$ 3d $, V-$ 3d $ and Eu-$ 4f $ orbitals, respectively. In addition, the modified Becke-Johnson (mBJ) functional was used in bulk band computations \cite{mbj-doi:10.1063/1.2213970}. In order to correctly describe the inherent van der Waal (vdW) interactions, the DFT-D3 method \cite{dft-3d-doi:10.1063/1.3382344} was considered. Moreover, VASPKIT \cite{vaspkit} and PHONOPY \cite{phonopy} codes were used for data post-processing. Plus, structural visualization was achieved with the help of VESTA \cite{VESTA}.

The $ \Gamma $-centered Monkhorst-Pack $k$-point meshes of 9$ \times $9$ \times $3 and 13$ \times $13$ \times $1 were adopted for bulk and thin-film structures respectively. Denser grids of 11$ \times $11$ \times $3 and 23$ \times $23$ \times $5 were used to calculate magnetocrystalline anisotropy energies (MAEs). For energy and band calculations, the geometry optimizations were performed until the Hellmann-Feynman force on each atom is smaller than 0.01 eV/Å, while the energy convergence criterion was chosen to be 1.0$ \times $10$ ^{-6} $ eV. As for phonon spectra, structural relaxations were done with threshold of 1.0$ \times $10$ ^{-4} $ eV/Å and 1.0$ \times $10$ ^{-7} $ eV for higher accuracy and dispersion relations were calculated by density functional perturbation theory (DFPT).

The interlayer and interatomic-layer exchange energies ($ E_{ex} $) are defined as the energy subtraction between antiferromagnetic (AFM) and ferromagnetic (FM) configurations, i.e. $E_{ex}=E_{\text{AFM}}-E_{\text{FM}}$. Notice that when we calculated the interlayer (interatomic-layer) $ E_{ex} $, the interatomic-layer (interlayer) couplings were kept unchanged. The MAEs are represented by the energy subtraction between out-of-plane and in-plane configurations, i.e. $E_{\text{MAE}}=E_{\text{in-plane}}-E_{\text{out-of-plane}}$. The in-plane magnetic moments are set along $ x $-axis, since there is neglecting energy difference between $ x $-axis and $ y $-axis polarizations in all materials within the scope of this paper. Collinear spin-polarized calculations without spin-orbital couplings (SOCs) and non-collinear ones with SOCs were respectively conducted for the calculations of exchange energies and MAEs.

We also obtained tight-binding Hamiltonians of bulk materials based on the maximally localized Wannier functions (MLWFs) by the Wannier90 package \cite{wannier90-MOSTOFI20142309,mlwf1-PhysRevB.56.12847,mlwf2-PhysRevB.65.035109}. Thin-film Hamiltonians were extracted from corresponding bulk ones, originated from mBJ-functional calculations. Next, all the topological properties, including Fermi surface visualizations, and Weyl points (WPs) characterizations of bulk materials, as well as edge states and Chern-number calculations of thin films, were accomplished using the tight-binding Hamiltonian method as implemented in the WannierTools package \cite{wanniertools-WU2018405} based on MLWFs.

\section{results}
\subsection{Lattice structures}
\begin{figure}
    \centering
    \includegraphics[width=1\linewidth]{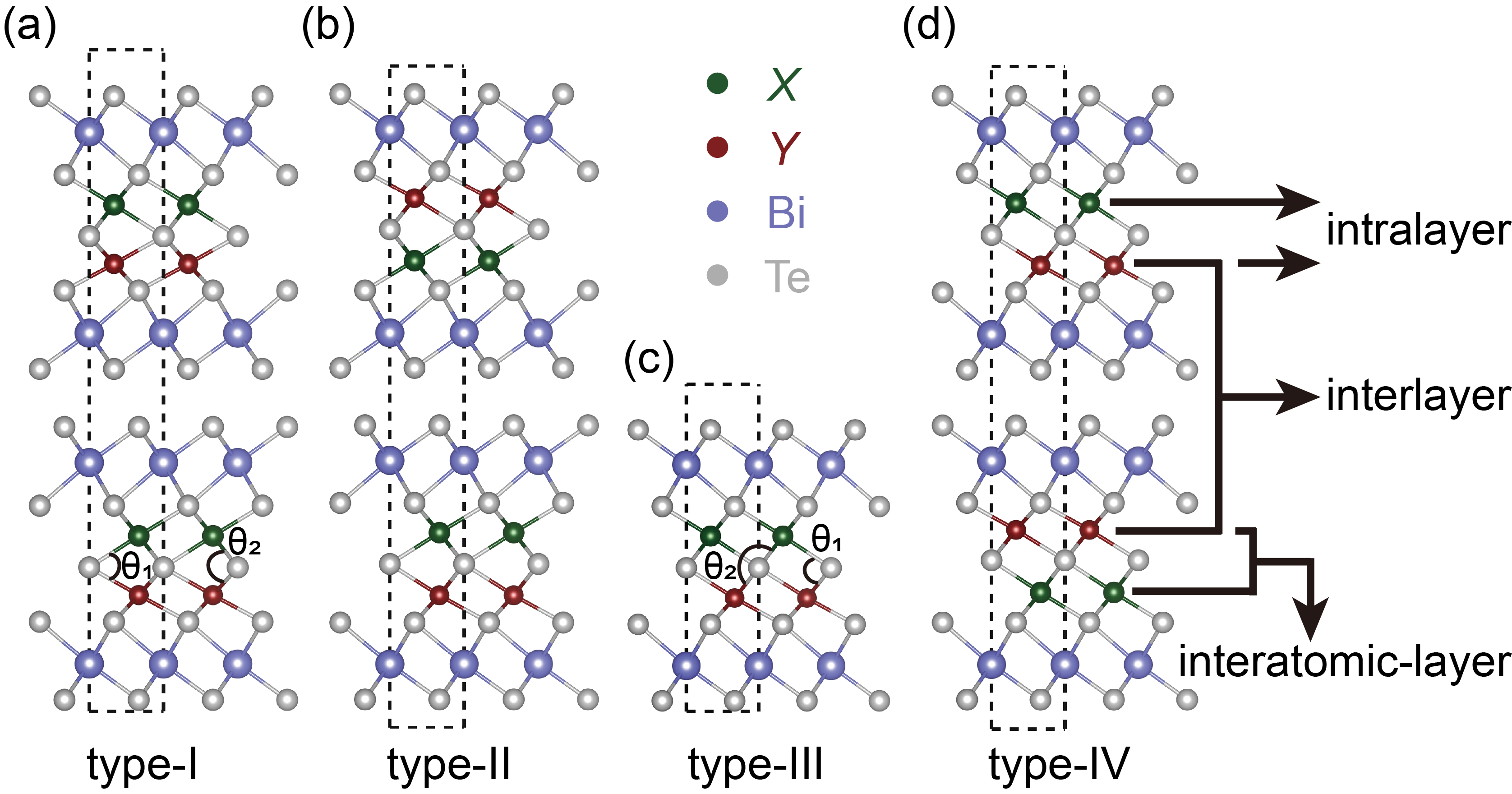}
    \caption{Four typical structures of bulk \textit{XY}Bi${}_2$Te${}_5$. Dashed lines stands for unit cells. If \textit{X}=\textit{Y}, type-I and type-II (also type-III and type-IV) become the same. $\theta_1$ and $\theta_2$ in panels (a) and (c) are $90^{\circ}$-like and $180^{\circ}$-like angles, which are determining factors of magnetic coupling between \textit{X} and \textit{Y} atomic layers inside one NL. Panel (d) illustrates the three kinds of magnetic couplings discussed in Sec. \ref{mag} by using type-IV lattice structure as an example.}
    \label{fig1:Lattice}
\end{figure}

The unit cell of \textit{XY}Bi${}_2$Te${}_5$ consists of Te-Bi-Te-\textit{X}-Te-\textit{Y}-Te-Bi-Te nonuple layers (NLs), with a vdW vacuum gap between neighboring NLs. There have been published theoretical works \cite{225cpl-Zhang_2020,225prb-PhysRevB.102.121107} focusing on Mn${}_2$Bi${}_2$Te${}_5$, one special case of \textit{XY}Bi${}_2$Te${}_5$ where \textit{X}=\textit{Y}=Mn. They assumed an \textit{ABC}-stacking structure of the internal Mn-Te bilayer structures, which is similar with the well-known MnBi${}_2$Te${}_4$. There are also different viewpoints on what the most stable structure should be like. As the number of Mn-Te bilayers increases, these atomic layers may tend to form an \textit{ABAC}-stacking structure \cite{otrokov-nanolett-doi:10.1021/acs.nanolett.8b03057,otrokov-nc-Hirahara2020} which resembles the bulk phase of MnTe, and this seems to be the case in Mn${}_2$Bi${}_2$Te${}_5$-family materials \cite{otrokov-prb-PhysRevB.105.195105}.

Beyond MnTe, bulk phases of NiTe and VTe also appear as \textit{ABAC}-stacking, while EuTe remains \textit{ABC}-stacking \cite{MaterialsProject}. This suggests that \textit{ABAC}-stacking and \textit{ABC}-stacking lattice structures (NiAs-type and NaCl-type respectively \cite{otrokov-prb-PhysRevB.105.195105}) may coexist in \textit{XY}Bi${}_2$Te${}_5$ when different magnetic elements are included.

Therefore, we took both types of lattice structures into consideration. Another important question is the ordering of magnetic atomic layers when \textit{X}$ \neq $\textit{Y}. To answer it, we studied two simple bulk models with the sequence of magnetic atomic layers being like "$ \cdots $-\textit{XY}-\textit{XY}-\textit{XY}-\textit{XY}-$ \cdots $" (\textit{XY}-\textit{XY}BT) or "$ \cdots $-\textit{XY}-\textit{YX}-\textit{XY}-\textit{YX}-$ \cdots $" (\textit{XY}-\textit{YX}BT). Thus a total of four lattice structures, named as type-I, type-II, type-III and type-IV, were investigated, as depicted in Fig. \ref{fig1:Lattice}. These four typical structures are NiAs-type \textit{XY}-\textit{XY}BT (type-I), NiAs-type \textit{XY}-\textit{YX}BT (type-II), NaCl-type \textit{XY}-\textit{XY}BT (type-III) and NaCl-type \textit{XY}-\textit{YX}BT (type-IV). Their corresponding space group is shown in Table \ref{tab1:spacegroup}. Monolayer \textit{XY}Bi${}_2$Te${}_5$, however, can just reflect the stacking order inside one single NL, so only two types of structures should be considered (type-I and type-III). 

\begin{table}
    \caption{\label{tab1:spacegroup} Space group of different lattice structures. Magnetic moments have been ignored. When \textit{X}=\textit{Y}, type-I and type-II (also type-III and type-IV) become the same, so we omit the space group of type-II (and type-IV) in this table.}
    \begin{ruledtabular}
    \begin{tabular}{ccccc}
        & type-I       & type-II      & type-III      & type-IV     \\
        \colrule
        \textit{X}=\textit{Y}& No. 194& & No. 164& \\
        \textit{X}$ \neq $\textit{Y}& No. 186      & No. 164      & No. 156       & No. 164    
        \end{tabular}
    \end{ruledtabular}
    \end{table}

\begin{figure*}
    \centering
    \includegraphics[width=1.0\linewidth]{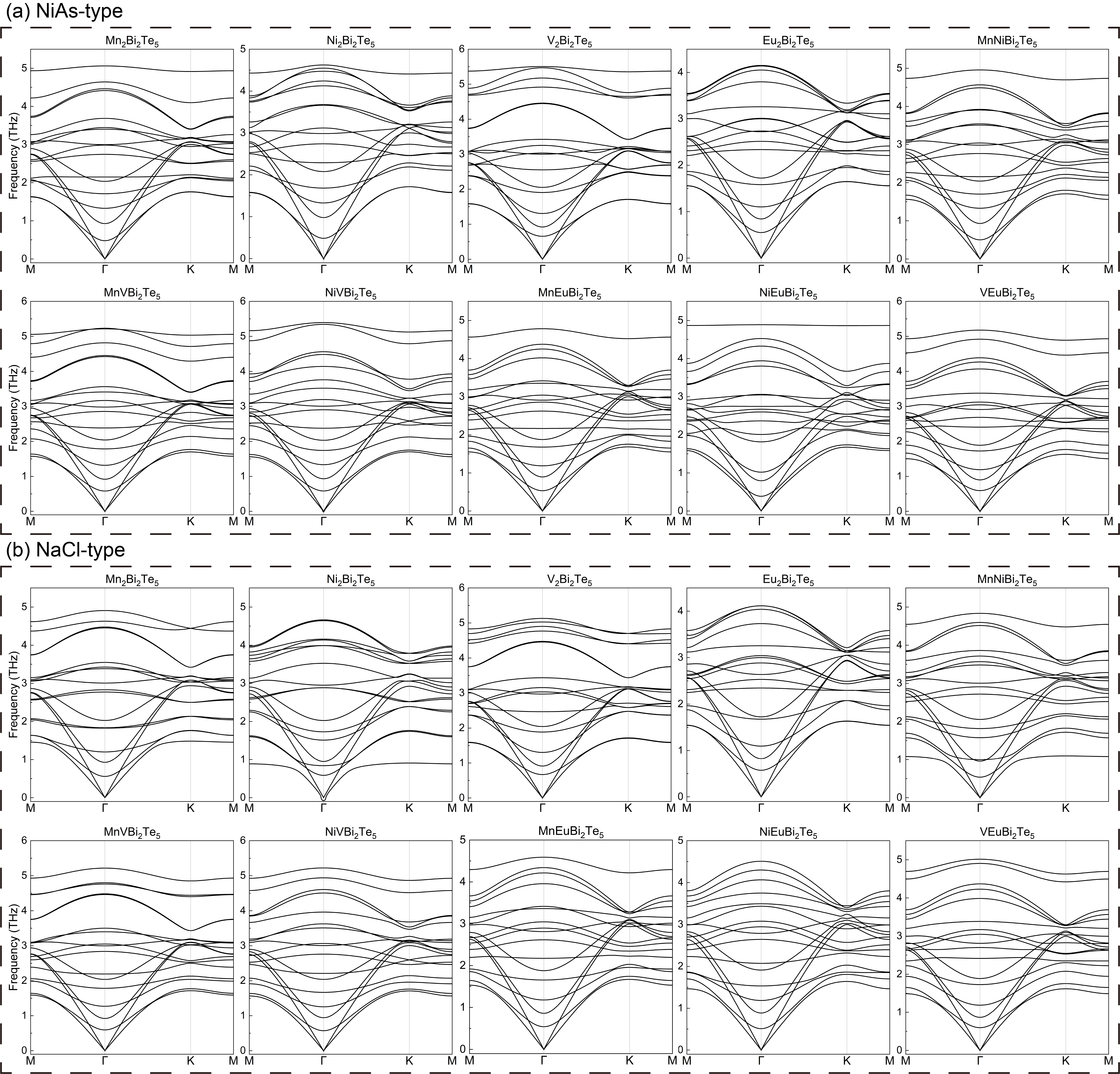}
    \caption{Phonon dispersions of (a) type-I (NiAs-type) and (b) type-III (NaCl-type) monolayer \textit{XY}Bi${}_2$Te${}_5$-family materials.}
    \label{fig2:Phonon}
\end{figure*}

The relaxed lattice constants of both monolayer and bulk \textit{XY}Bi${}_2$Te${}_5$ (\textit{X}, \textit{Y}=Mn, Ni, V, Eu) compounds are demonstrated in Appendix \ref{latticeconstants}. To confirm the structural stability, we performed calculations on phonon dispersions. Due to the layered-stacking nature of \textit{XY}Bi${}_2$Te${}_5$, we only carried out phonon-dispersion computations on monolayers. Monolayers with type-I and type-III lattice structures can be seen as basic building blocks of all the structures discussed here. Therefore, we analyzed the twenty monolayer structures as a whole. Their phonon spectra are shown in Fig. \ref{fig2:Phonon}. For the majority of them, no virtual frequency exists, indicating that either NiAs-type or NaCl-type \textit{XY}Bi${}_2$Te${}_5$ should be stable. The only exception appears in NaCl-type Ni${}_2$Bi${}_2$Te${}_5$ since there is slight virtual frequency of the acoustic phonons around the $ \Gamma $ point [see Fig. \ref{fig2:Phonon}(b)]. This is not a rare case in first-principles calculations and might be a result of insufficient accuracy \cite{mbt-sciadv-doi:10.1126/sciadv.aaw5685,Qiao-PhysRevB.103.245403,Yan-PhysRevLett.127.046401}. For this reason, the stability of NaCl-type Ni${}_2$Bi${}_2$Te${}_5$ may be possibly acceptable (perhaps metastable) and we still include it in the following sections also for the completeness of this work.

We also examined the influence of \textit{X}-\textit{Y} mixing (for $ X\neq Y $ only) on structural stability. Here \textit{X}-\textit{Y} mixing refers to situations where the two distinguishable kinds of magnetic atoms are mixed instead of locating at separate atomic layers [see Fig. \ref{fig1:Lattice}]. Our systematic investigation suggests that \textit{X}-\textit{Y} mixing will result in significant virtual frequencies on phonon spectra, which is direct evidence for structural instability. Therefore, we think it is reasonable to study \textit{XY}Bi${}_2$Te${}_5$ structures without \textit{X}-\textit{Y} mixing problem. Computational details can be found in Appendix \ref{mix}.

\subsection{Magnetic properties \label{mag}}
We then come to discuss the magnetic properties of \textit{XY}Bi${}_2$Te${}_5$. Both of the ground state magnetic configurations and the coupling strength are determined by superexchange rules \cite{book}. In \textit{XY}Bi${}_2$Te${}_5$, three types of magnetic couplings determine the stable magnetic configurations, including couplings between \textit{X} and \textit{Y} atoms in neighboring NLs (interlayer couplings), between \textit{X} or \textit{Y} atoms in the same atomic layer plane (intralayer couplings), and between \textit{X} and \textit{Y} atoms in the same NL (interatomic-layer couplings). All of these couplings are illustrated in Fig. \ref{fig1:Lattice}(d), where type-IV lattice structure is used as an example. The interlayer and intralayer couplings resemble those in MnBi${}_2$Te${}_4$-family materials, originated from Goodenough-Kanamori 180${}^{\circ}$ and 90${}^{\circ}$ rules separately \cite{mbt-sciadv-doi:10.1126/sciadv.aaw5685,mbt-otrokov-PhysRevLett.122.107202,mbt-wj-PhysRevLett.122.206401,lizhe-PhysRevB.102.081107}, so we mainly need to understand the extra interatomic-layer couplings. 

\begin{table}
\caption{\label{tab2:exchangeenergy}Summary of interlayer/interatomic-layer $ E_{ex} $ (in unit of meV) and MAE (meV/NL) of bulk \textit{XY}Bi${}_2$Te${}_5$ materials. Positive (negative) values stand for FM (AFM) coupling in $E_{ex}$, as well as out-of-plane (in-plane) magnetization direction in MAE.}
\begin{ruledtabular}
    \begin{tabular}{ccrrrr}
        \multicolumn{2}{c}{\textit{X}-\textit{Y}} &
          \multicolumn{1}{c}{type-I} &
          \multicolumn{1}{c}{type-II\footnote{In Type-II and Type-IV structures, two kinds of interlayer couplings exist because of the \textit{XY}-\textit{YX} order. Hence, the interlayer $ E_{ex} $ here can be seen as the average values of them.}} &
          \multicolumn{1}{c}{type-III} &
          \multicolumn{1}{c}{type-IV\footnotemark[1]} \\ \hline
        \multirow{3}{*}{Mn-Mn} & Interlayer  & -1.48   &         & -1.74   &         \\
                               & Interatomic & -58.45  &         & 1.36    &         \\
                               & MAE         & 0.34    &         & 0.18    &         \\ \hline
        \multirow{3}{*}{Ni-Ni} & Interlayer  & -9.77   &         & -10.82  &         \\
                               & Interatomic & -283.38 &         & -266.89 &         \\
                               & MAE         & 2.59    &         & 5.22    &         \\ \hline
        \multirow{3}{*}{V-V}   & Interlayer  & -0.56   &         & -0.68   &         \\
                               & Interatomic & -156.41 &         & -96.10  &         \\
                               & MAE         & -0.46    &         & -0.17   &         \\ \hline
        \multirow{3}{*}{Eu-Eu} & Interlayer  & -0.14   &         & -0.11   &         \\
                               & Interatomic & -0.74   &         & -3.17   &         \\
                               & MAE         & -0.08   &         & -0.06   &         \\ \hline
        \multirow{3}{*}{Mn-Ni} & Interlayer  & -3.62   & -8.06   & -4.14   & -8.23   \\
                               & Interatomic & -132.65 & -135.32 & -104.69 & -106.89 \\
                               & MAE         & 1.90    & 1.33    & 2.18    & 2.53    \\ \hline
        \multirow{3}{*}{Mn-V}  & Interlayer  & 1.66    & -1.44   & 1.33    & -0.85   \\
                               & Interatomic & -33.07  & -32.42  & 34.40   & 34.90   \\
                               & MAE         & -0.14   & -0.13   & -0.03   & -0.07   \\ \hline
        \multirow{3}{*}{Ni-V}  & Interlayer  & 1.41    & -11.55  & 1.59    & -8.89   \\
                               & Interatomic & 56.69   & 58.10   & 84.46   & 85.25   \\
                               & MAE         & 0.94    & 1.08    & 0.85    & 1.00    \\ \hline
        \multirow{3}{*}{Mn-Eu} & Interlayer  & 0.25    & -0.57   & 0.38    & -0.56   \\
                               & Interatomic & 15.25   & 15.38   & 11.68   & 11.38   \\
                               & MAE         & 0.05    & 0.06    & 0.10    & 0.11    \\ \hline
        \multirow{3}{*}{Ni-Eu} & Interlayer  & 0.27    & -9.19   & 0.80    & -12.48  \\
                               & Interatomic & 28.24   & 26.81   & 24.52   & 23.76   \\
                               & MAE         & 0.82    & 0.96    & 1.02    & 1.21    \\ \hline
        \multirow{3}{*}{V-Eu}  & Interlayer  & -0.32   & -0.44   & -0.30   & -0.52   \\
                               & Interatomic & -1.01   & -2.91   & -13.43  & -13.71  \\
                               & MAE         & -0.13   & -0.11   & -0.09   & -0.06  
        \end{tabular}
\end{ruledtabular}
\end{table}

\begin{figure*}
    \centering
    \includegraphics[width=0.8\linewidth]{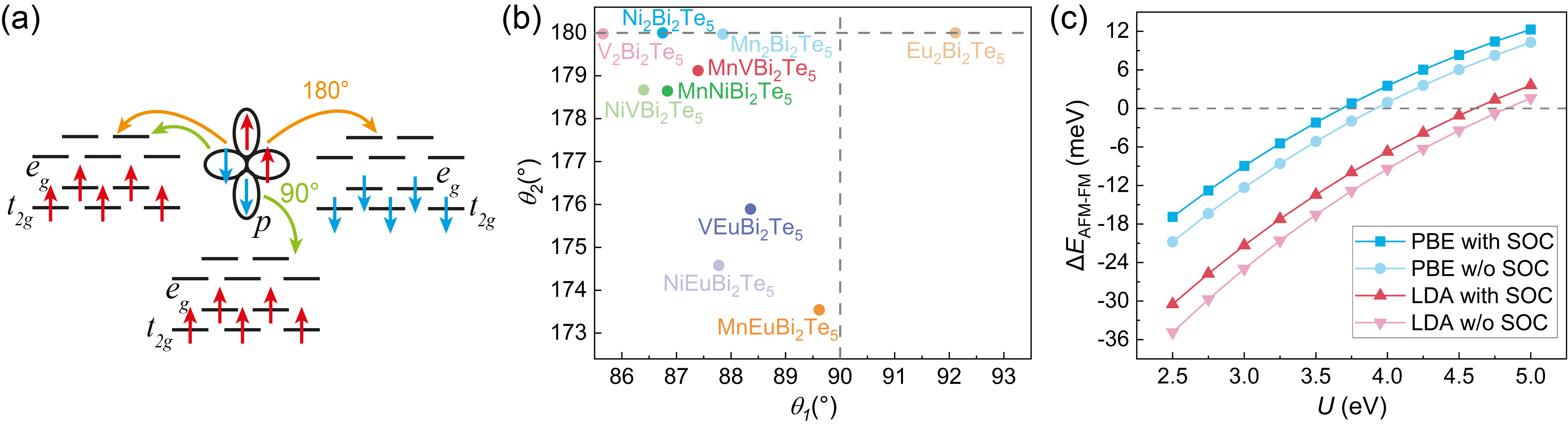}
    \caption{Interatomic-layer couplings of type-III \textit{XY}Bi${}_2$Te${}_5$. (a) Schematic illustration of the Goodenough-Kanamori 180${}^{\circ}$ and 90${}^{\circ}$ superexchange mechanisms. Here we adopt $d$ orbitals of Mn as an example. (b) Bond angle values of type-III \textit{XY}Bi${}_2$Te${}_5$. (c) Energy difference between FM and AFM interatomic-layer couplings of type-III Mn${}_2$Bi${}_2$Te${}_5$ obtained with different $ U $ and exchange-correlation functionals.}
    \label{fig3:Exchange}
\end{figure*}

Table \ref{tab2:exchangeenergy} shows values of interlayer/interatomic-layer $ E_{ex} $, and also MAE of \textit{XY}Bi${}_2$Te${}_5$-family materials. Similar to the MnBi${}_2$Te${}_4$-family materials, the sign and relative strength of the magnetic couplings can also be understood with the Goodenough-Kanamori rules \cite{lizhe-PhysRevB.102.081107}. Here, we focus on the interatomic-layer couplings which are expected to be much stronger than the interlayer ones. Firstly, we analyze the type-III structure, whose inherent \textit{ABC}-stacking structure resembles that of well-known MnBi${}_2$Te${}_4$. Its interatomic-layer couplings are contributed by two kinds of hopping channels of \textit{X}-Te-\textit{Y} bonds, one with the bond angle near $90^{\circ}$ [$ \theta_1 $  in Fig. \ref{fig1:Lattice}(c)], and the other with the bond angle near $180^{\circ}$ [$ \theta_2 $  in Fig. \ref{fig1:Lattice}(c)]. The signs of the couplings contributed by the two channels (referred as $ \theta_1 $ and $ \theta_2 $ channels below, respectively) are opposite [see Fig. \ref{fig3:Exchange}(a)]. $ \theta_1 $ and $ \theta_2 $ values of type-III \textit{XY}Bi${}_2$Te${}_5$ are summarized in Fig. \ref{fig3:Exchange}(b). For Ni${}_2$Bi${}_2$Te${}_5$ and V${}_2$Bi${}_2$Te${}_5$, $ \theta_2 $ is very close to $180^{\circ}$, meanwhile $ \theta_1 $ deviates several degrees from $90^{\circ}$. Therefore, the coupling is dominated by the $ \theta_2 $ channel which is AFM. The interatomic-layer coupling of NiVBi${}_2$Te${}_5$ is also dominated by the $ \theta_2 $ channel, which however gives a FM ground state since the number of $ 3d $  electron is above five in Ni and below five in V \cite{lizhe-PhysRevB.102.081107}. In Mn${}_2$Bi${}_2$Te${}_5$, the interatomic-layer coupling is even smaller than the interlayer one. It is because the couplings via the $ \theta_1 $ and $ \theta_2 $ channels are largely compensated with each other. Note that by varying the on-site Coulomb repulsion $ U $ or the exchange-correlation functionals [see Fig. \ref{fig3:Exchange}(c)], the interatomic-layer coupling of Mn${}_2$Bi${}_2$Te${}_5$ can be changed from FM to A-type AFM. This does not contradict previous work on Mn${}_2$Bi${}_2$Te${}_5$ \cite{225cpl-Zhang_2020,225prb-PhysRevB.102.121107,otrokov-prb-PhysRevB.105.195105}. Nonetheless, whether the interatomic-layer coupling in Mn${}_2$Bi${}_2$Te${}_5$ is FM or AFM has not been checked by experiment \cite{225exp-PhysRevB.104.054421}.

The values of $ \theta_1 $ in \textit{X}EuBi${}_2$Te${}_5$ also approach 90${}^{\circ}$, but the reverse (caused by 90${}^{\circ}$-rule) of interatomic-layer coupling ground state which is mainly determined by 180${}^{\circ}$-rule does not appear. It can be understood in the following way. According to typical superexchange mechanism, the coupling strength of the 180${}^{\circ}$ rule and the 90${}^{\circ}$ rule can be expressed as \cite{book}
\begin{equation}
    J_{180}=\frac{4 t_{p d}^4}{\left(U_d+\Delta_{p d}\right)^2}\left(\frac{1}{U_d}+\frac{1}{U_d+\Delta_{p d}}\right), \label{180}
\end{equation}
\begin{equation}
    J_{90}=-\frac{4 t_{p d}^4}{\left(U_d+\Delta_{p d}\right)^2} \frac{2 J_{x y}}{4\left(U_d+\Delta_{p d}\right)^2-J_{x y}^2}, \label{90}
\end{equation}
\begin{equation}
    J_{90} / J_{180} \approx-\frac{-U_{d} J_{x y}}{2\left(U_d+\Delta_{p d}\right)\left(2 U_d+\Delta_{p d}\right)}. \label{90-180}
\end{equation}

Here $ t_{pd} $ represents hopping strength between $ p $ and $ d $ orbitals. $ U_d $ stands for Hubbard $ U $ term of magnetic atoms, while $ \Delta_{pd} $ can be obtained by subtracting the energy of occupied $ p $ orbitals from that of occupied $ d $ orbitals. $ J_{xy} $ is the Coulomb exchange term between two orthogonal $ p $ orbitals. Eq. (\ref{90-180}) is extracted by considering $ J_{xy}\ll U_d+\Delta_{pd} $. The $ 5d $ orbitals of Eu, compared with $ 3d $ orbitals of Mn, V and Ni, locate quite far away from the Fermi level \cite{lizhe-PhysRevB.102.081107}. So, $ \Delta_{pd} $ of materials containing Eu is obviously larger than that of others, leading to weaker contributions from 90${}^{\circ}$ rule. Therefore, there is no anomaly in the interatomic-layer couplings of \textit{X}EuBi${}_2$Te${}_5$.

The above analysis on type-III structure is also suitable for type-IV, since they bare resembling \textit{X}-Te-\textit{Y} bond structures and also interatomic-layer $ E_{ex} $. However, due to the $ ABAC $-stacking order in type-I and type-II structures, the $ \theta_1 $ and $ \theta_2 $ stay far away from either 180${}^{\circ}$ or 90${}^{\circ}$ [see Fig. \ref{fig1:Lattice}(a)] and it becomes hard to qualitatively settle the interatomic-layer couplings via Goodenough-Kanamori 180${}^{\circ}$ and 90${}^{\circ}$ rules. Also, the shortest distance between magnetic atoms in the two separate layers becomes smaller in NiAs-type structure ($\lesssim$4 Å) than that in NaCl-type structure and even the distance between the nearest magnetic atoms within one atomic layer ($\textgreater$4 Å), which means the direct exchange interaction should not be ignored, making the interatomic-layer couplings more complicated. Based on DFT calculations, we found that, besides Mn${}_2$Bi${}_2$Te${}_5$ mentioned in Ref. \cite{otrokov-prb-PhysRevB.105.195105}, the interatomic-layer coupling of MnVBi${}_2$Te${}_5$ also changes (both from FM to AFM) when the lattice structures evolve from NaCl-Type to NiAs-type, while others remain unchanged.

\begin{figure*}
    \centering
    \includegraphics[width=1.0\linewidth]{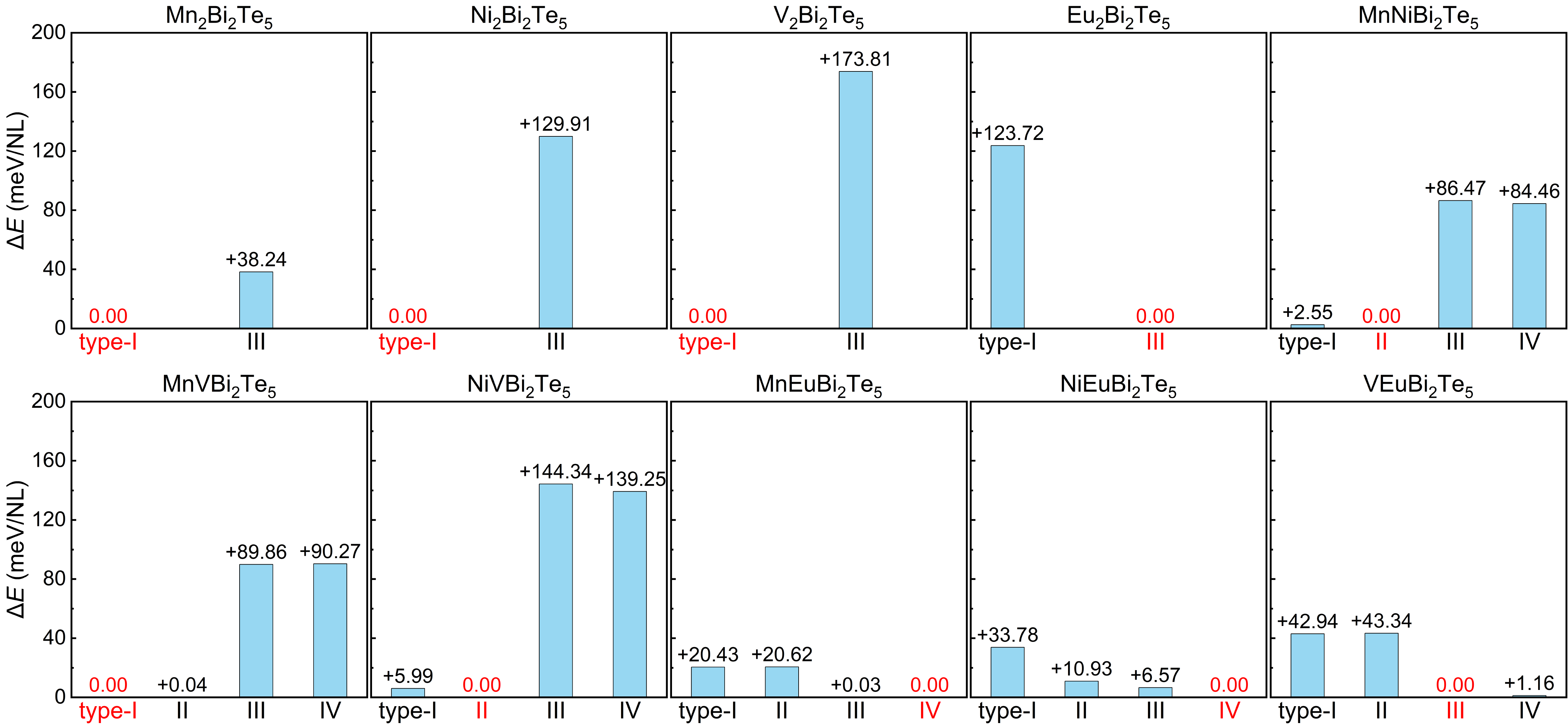}
    \caption{Energy comparisons of magnetic ground states based on four types of bulk phase \textit{XY}Bi${}_2$Te${}_5$ lattice structures (two types indeed when \textit{X}=\textit{Y}). The lowest energies are equally set to zero and marked in red in every compound of  \textit{XY}Bi${}_2$Te${}_5$. Type-II MnVBi${}_2$Te${}_5$ and type-III MnEuBi${}_2$Te${}_5$ may be quasi-ground states since they show slight difference compared to type-I MnVBi${}_2$Te${}_5$ and type-IV MnEuBi${}_2$Te${}_5$ respectively, which is beyond the precision of computations. }
    \label{fig4:GroundState}
\end{figure*}

After figuring out the magnetic couplings and especially the ground state configurations of all \textit{XY}Bi${}_2$Te${}_5$, we tried to make it clear which of the four lattice structures proposed in this paper, should be energetically the most favorable. As can be seen in Fig. \ref{fig4:GroundState}, we made comparisons of ground state energies within every kind of \textit{XY}Bi${}_2$Te${}_5$. A NiAs-type lattice structure is preferred when Eu is excluded from the component magnetic elements, while a NaCl-type lattice structure is preferred once Eu is included. This may be a result of the lackness of \textit{ABAC}-stacking structure in bulk phase EuTe, as mentioned before. With the increasing number of magnetic-cation and Te-anion bilayers, the local chemical environment around magnetic atoms will approach that in their corresponding bulk phases, leading to similar stacking orders eventually.

Another finding is about the sequence of magnetic atomic layers. It is influenced by the strength of interlayer couplings. For example, the energetically favorite structures in MnNiBi${}_2$Te${}_5$ (type-II), NiVBi${}_2$Te${}_5$ (type-II) and NiEuBi${}_2$Te${}_5$ (type-IV) are all \textit{XY}-\textit{YX}BT. This is because the interlayer coupling between Ni atomic layers is much stronger than others [see Table \ref{tab2:exchangeenergy}], which is beneficial for lowering the total energy. This rule is not apparent or applicable to cases where the strength of interlayer couplings between \textit{X}-\textit{X}, \textit{Y}-\textit{Y} and \textit{X}-\textit{Y} are comparable, since their relative values may be sensitive to even slight change in lattice structures. MnVBi${}_2$Te${}_5$ and MnEuBi${}_2$Te${}_5$ are two typical examples. Negligible energy difference is found between their type-I(III) and type-II(IV) structures.

\begin{figure}
    \centering
    \includegraphics[width=1\linewidth]{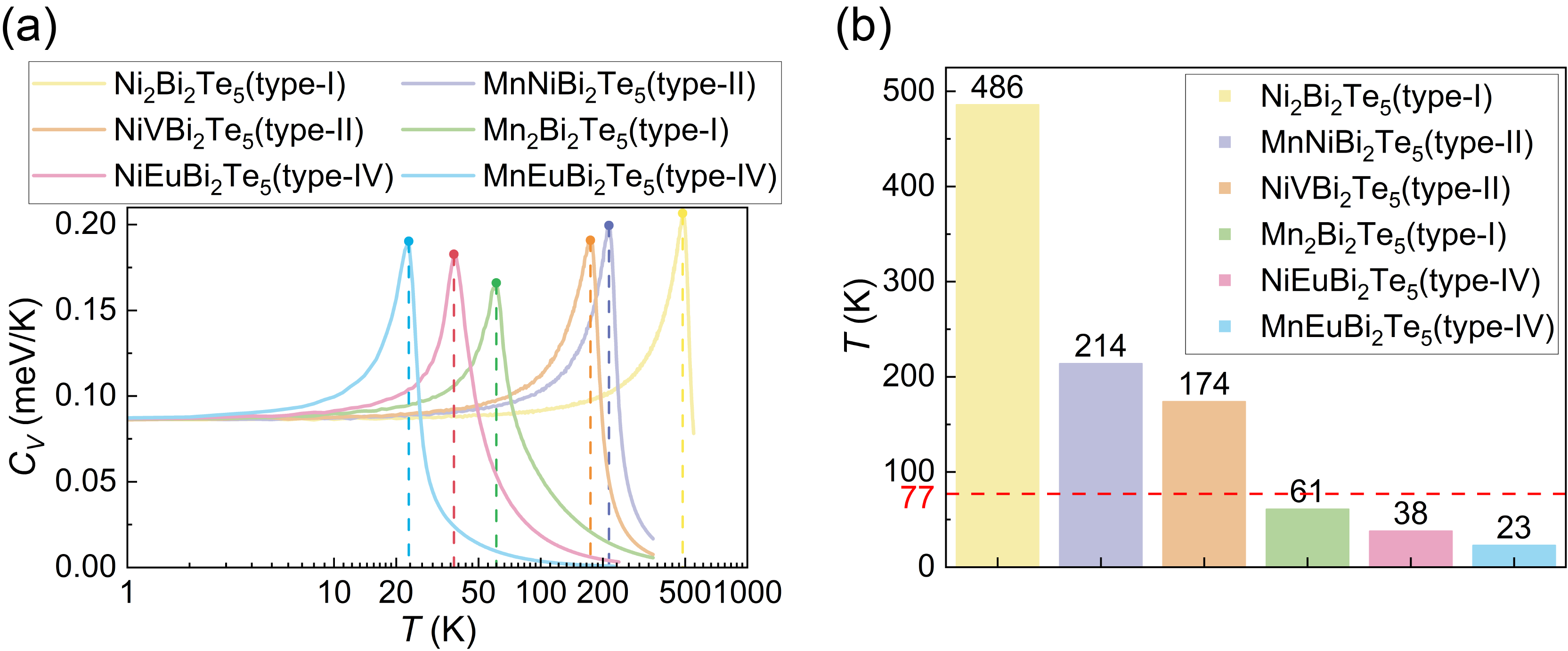}
    \caption{Néel/Curie temperatures of six kinds of \textit{XY}Bi${}_2$Te${}_5$ materials with out-of-plane magnetic ground states. (a) $C_V$-$T$ curves of those \textit{XY}Bi${}_2$Te${}_5$ combinations obtained by Monte Carlo simulations. (b) Their corresponding Néel/Curie temperatures are extracted and shown in a bar chart for clarity. The boiling temperature of liquid nitrogen (77 K) is marked with a red dashed line.}
    \label{fig5:Tc}
\end{figure}

We estimated the Néel/Curie temperatures of selected bulk \textit{XY}Bi${}_2$Te${}_5$ materials with Monte Carlo (MC) simulations, mainly including the six \textit{XY}Bi${}_2$Te${}_5$ with out-of-plane magnetic ground states [see Table \ref{tab2:exchangeenergy}]. A Heisenberg model was employed to describe their magnetic behaviors, which involves the aforementioned three kinds of magnetic couplings up to next-nearest neighbors and single-ion anisotropy energy.

\begin{equation}
    \begin{aligned}
        &H=  \sum_{i \in X, Y} A_i(S_i^z)^2+\sum_{i, j \in X} J_{i j}^X \boldsymbol{S}_{i} \cdot \boldsymbol{S}_{j}+\sum_{i, j \in Y} J_{i j}^Y \boldsymbol{S}_{i} \cdot \boldsymbol{S}_{j} \\
        &+  \sum_{i, j \in \text{interlayer}} J_{i j}^I \boldsymbol{S}_{i} \cdot \boldsymbol{S}_{j}+\sum_{i, j \in \text{interatomic-layer}} J_{i j}^S \boldsymbol{S}_{i} \cdot \boldsymbol{S}_{j}.
    \end{aligned}
    \label{Heisenberg}
\end{equation}

$ A_i $, $ J_{i j}^X $ and $ J_{i j}^Y $ are single ion anisotropy energy and the
intralayer couplings in \textit{X} and \textit{Y} atomic layers respectively. $ J_{i j}^I $  and $ J_{i j}^S $ refer to interlayer and interatomic-layer interaction parameters. Note that in type-II and type-IV structures, $ J_{i j}^I $ can be further decomposed into $ J_{i j}^{IX} $ and $ J_{i j}^{IY} $ in order to describe the two kinds of interlayer couplings. 

During simulations, we firstly adopted 2$ \times $10$ ^{5} $ MC steps for equilibrating the system at each temperature, and next 1$ \times $10$ ^{6} $ steps for computing physical quantity such as heat capacity. A supercell of 12$ \times $12$ \times $3 unit cells as well as periodic boundary conditions were used. Larger supercell were also tested, and we find in general a supercell of 12$ \times $12$ \times $3 is enough.

As shown in Fig. \ref{fig5:Tc}, type-I Ni${}_2$Bi${}_2$Te${}_5$, type-II NiVBi${}_2$Te${}_5$ and type-II MnNiBi${}_2$Te${}_5$ have Néel/Curie temperatures above 77 K, in which type-I Ni${}_2$Bi${}_2$Te${}_5$ even has a Néel temperature above room temperature. The result coincides with their strong interatomic-layer couplings (all above 50 meV per unit cell) shown in Tab. \ref{tab2:exchangeenergy}.

Our calculation for type-I Mn${}_2$Bi${}_2$Te${}_5$ shows a moderate value of its Néel temperature (61 K), which is much higher than what was obtained in experiment \cite{225exp-PhysRevB.104.054421} and another theoretical work \cite{otrokov-prb-PhysRevB.105.195105}. However, it's worth mentioning that our calculations for type-III Mn${}_2$Bi${}_2$Te${}_5$ (27 K) is close to the experimental result (20 K) \cite{225exp-PhysRevB.104.054421} (see Appendix \ref{Tc}). This might imply that the Mn${}_2$Bi${}_2$Te${}_5$ samples synthesized in laboratories possessed a stable but not energetically favorite structure (type-III, not type-I), which is also likely to happen in other \textit{XY}Bi${}_2$Te${}_5$ combinations. We think more experimental information, like images of atomically resolved high-angle annular dark field STEM, is necessary before the end of this structural puzzle.

\begin{figure}
    \centering
    \includegraphics[width=1.0\linewidth]{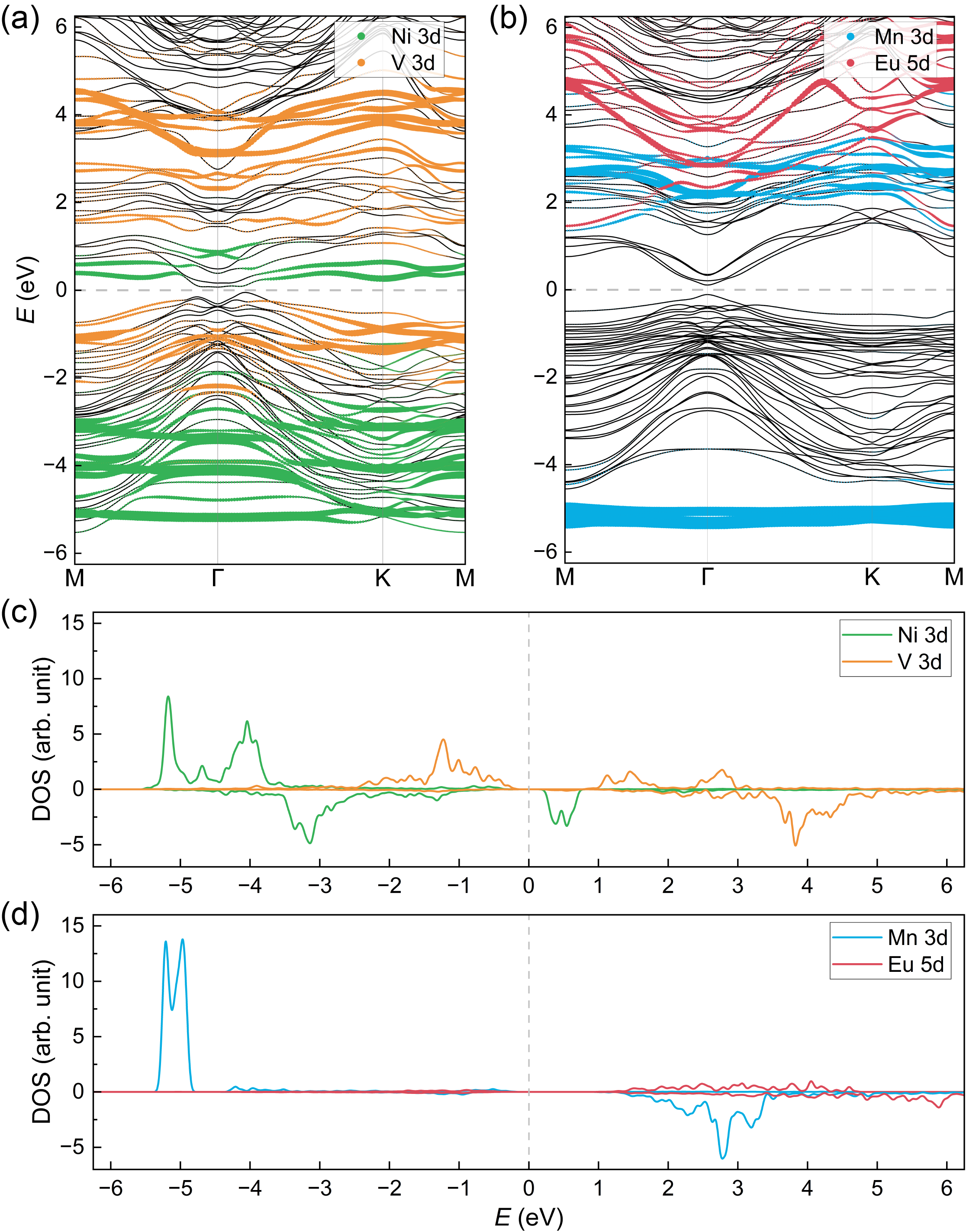}
    \caption{Band structures along high symmetry lines (with SOC) and partial DOS distributions (without SOC) using monolayer type-I NiVBi${}_2$Te${}_5$ (a, c) and type-III MnEuBi${}_2$Te${}_5$ (b, d) as examples. We have highlighted the $ d $-orbital components related to magnetic couplings.}
    \label{fig6:pDOS}
\end{figure}

One can see from Fig. \ref{fig5:Tc} and Tab. \ref{tab2:exchangeenergy} that the compounds including Ni atoms have higher Néel/Curie temperature. The projected band structures and partial density of states (DOS) of Ni, V, Mn and Eu are plotted in Fig. \ref{fig6:pDOS} respectively. Appreciably, the distributions of Ni $3d$ orbitals locate near the Fermi level, which benefit the virtual hopping processes and greatly enhance the magnetic coupling strength \cite{book}. With unoccupied $e_g$ orbitals situating even nearer the Fermi level, Ni behaves much better than V. In contrast, the Mn $3d$ and Eu $5d$ orbitals locate quite far from the Fermi level, generating much weaker coupling strength.

\subsection{Band topologies}
Now we are ready to talk about both the intrinsic and tunable magnetic topological properties of \textit{XY}Bi${}_2$Te${}_5$-family materials. Up to now, experimental results \cite{225exp-PhysRevB.104.054421} are not sufficient to decide which type of lattice structures can be achieved in laboratory. Besides, due to the quite weak vdW interactions, it may be possible to artificially tune the interlayer stacking orders of \textit{X} and \textit{Y} atomic layers by methods like exfoliation. Therefore, although it is theoretically reasonable to exclusively discuss the structures with the lowest energies, we extend our discussions beyond this range and consider other dynamically stable lattice structures if necessary. 

Firstly, we discuss about a kind of symmetry-protected topological phase. It was reported that in a system whose low-energy physics can be described by a Bi${}_2$Te${}_3$-like effective Hamiltonian \cite{Bi2Te3Zhang2009,modelPhysRevB.82.045122}, a $\mathcal{P}$-breaking, $\mathcal{T}$-breaking and $\mathcal{PT}$-conserving perturbation brings chances on achieving dynamic axion states \cite{daf-Li2010}. This has been theoretically shown in type-III Mn${}_2$Bi${}_2$Te${}_5$ with A-type AFM magnetic configuration \cite{225cpl-Zhang_2020,225prb-PhysRevB.102.121107}.

\begin{figure}
    \centering
    \includegraphics[width=1.0\linewidth]{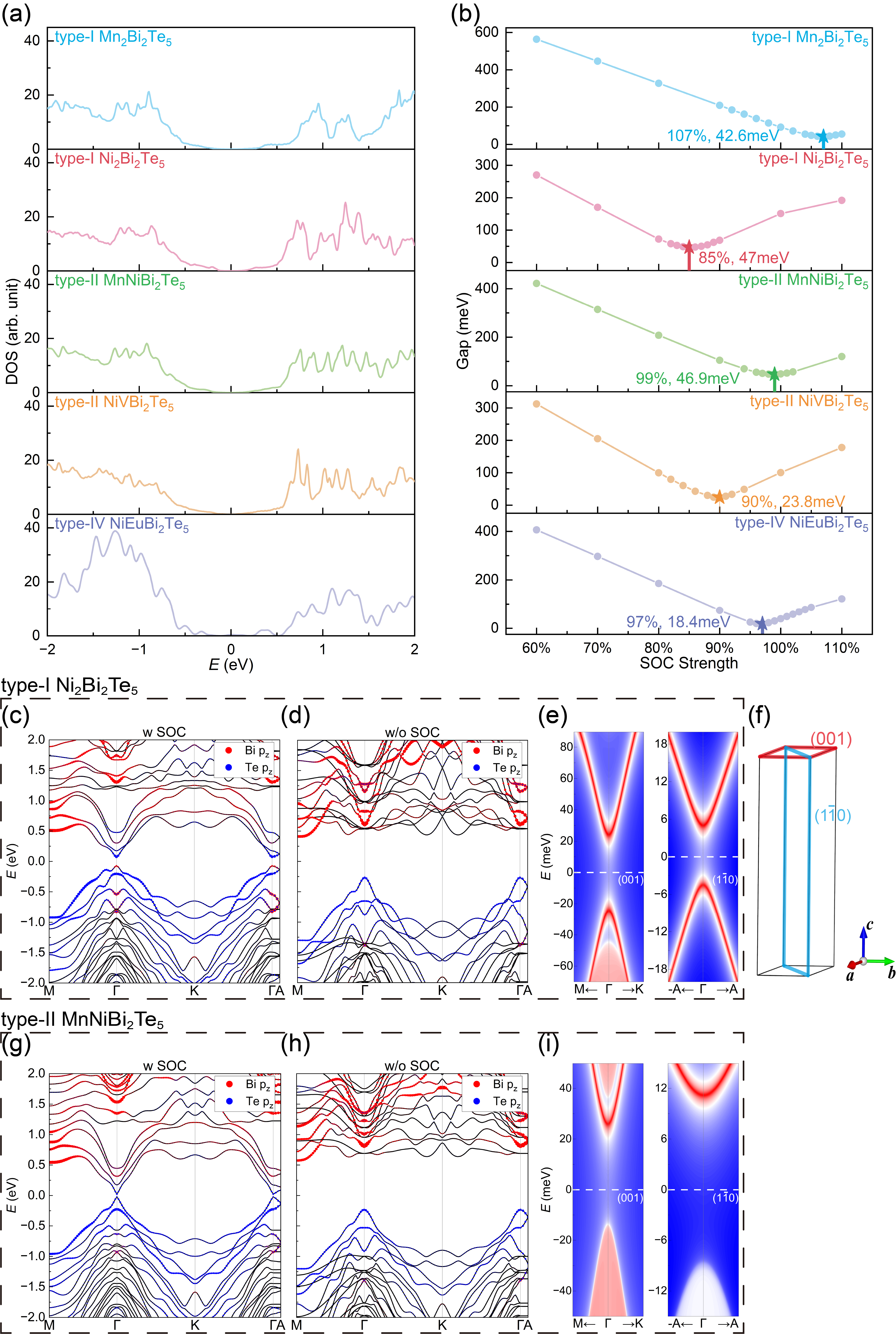}
    \caption{Intrinsic dynamic axion states in ground state \textit{XY}Bi${}_2$Te${}_5$. (a) DOS of the five candidates. SOC is considered. (b) The evolution of band gap under increasing SOC strength. (c, d) Band structures of type-I Ni${}_2$Bi${}_2$Te${}_5$ with SOC (w SOC) and without SOC (w/o SOC), respectively. (e) Calculated surface states of type-I Ni${}_2$Bi${}_2$Te${}_5$ on its (001) and (1$\bar{1}$0) planes. (f) Schematic illustration of these two surfaces. Panels (g) to (i) are similar with panels (c) to (e), but demonstrate situations in type-II MnNiBi${}_2$Te${}_5$.}
    \label{fig7:Axion}
\end{figure}

Beyond this, we also found intrinsic dynamic axion states in type-I Ni${}_2$Bi${}_2$Te${}_5$, type-II MnNiBi${}_2$Te${}_5$, type-II NiVBi${}_2$Te${}_5$ and type-IV NiEuBi${}_2$Te${}_5$, but not in type-I Mn${}_2$Bi${}_2$Te${}_5$ by mBJ+$ U $ calculations. They are all stable and energetically favored lattice structures. Besides, their ground state magnetic couplings [see Table \ref{tab2:exchangeenergy} and Fig. \ref{fig4:GroundState}], together with the lattice structures, satisfied the aforementioned symmetry constraints. Although they belong to different space groups from Bi${}_2$Te${}_3$ (No. 166), the threefold rotation along the $ c $ direction, twofold rotation along the $ a $ direction and also inversion symmetry (when magnetic moments are ignored) remain, which means a perturbed Bi${}_2$Te${}_3$-like effective Hamiltonian is still applicable.

As reported previously, a band inversion process without gap-closing point and gapped surface states on all surfaces are supposed to appear in materials preserving $\mathcal{PT}$ symmetry but breaking individual $\mathcal{P}$ and $\mathcal{T}$ symmetries, serving as the signatures on the emergence of dynamic axion fields \cite{pt-PhysRevB.101.081109}. Figure \ref{fig7:Axion}(a) clearly illustrates the insulating behavior of type-I Mn${}_2$Bi${}_2$Te${}_5$, type-I Ni${}_2$Bi${}_2$Te${}_5$, type-II MnNiBi${}_2$Te${}_5$, type-II NiVBi${}_2$Te${}_5$ and type-IV NiEuBi${}_2$Te${}_5$. Ground state magnetic configurations and SOC are considered. Then we artificially tuned the SOC strength and studied their band gap evolution [see Fig. \ref{fig7:Axion}(b)]. By increasing the SOC strength from 60\% to 100\%, in which 100\% represents the realistic value, the band gaps in four of them firstly decreases, reaching their minimum and then reincreases. The turning point of type-II MnNiBi${}_2$Te${}_5$ (99\%) and type-IV NiEuBi${}_2$Te${}_5$ (97\%) is very close to the realistic value, thus implying large inherent dynamic axion field. However, type-I Mn${}_2$Bi${}_2$Te${}_5$ can only demonstrate this topological phase transition under a bit larger SOC strength (107\%), so we claim it to be trivial.

Figures \ref{fig7:Axion}(c) and \ref{fig7:Axion}(d) show the orbital-projected band structures of type-I Ni${}_2$Bi${}_2$Te${}_5$ with and without SOC, respectively. Band inversion between Bi $6p_z$ and Te $5p_z$ indicates its topologically non-trivial character. Surface states on the (001) and (1$\bar{1}$0) planes were calculated and the minimum gap of surface states reaches approximately 6 meV ($\sim$74 K) [see Figs. \ref{fig7:Axion}(e) and \ref{fig7:Axion}(f)]. Figures \ref{fig7:Axion}(g), \ref{fig7:Axion}(h) and \ref{fig7:Axion}(i) shows the dynamic AxI features in type-II MnNiBi${}_2$Te${}_5$. Specially, the minimum gap of surface states in type-II MnNiBi${}_2$Te${}_5$ is larger than 17 meV ($\sim$210 K), which is comparable with its high enough Néel temperature ($\sim$214 K). The band inversion characters and surface state visualizations of the rest two candidates can be found in Appendix \ref{axion}.

\begin{figure}
    \centering
    \includegraphics[width=1.0\linewidth]{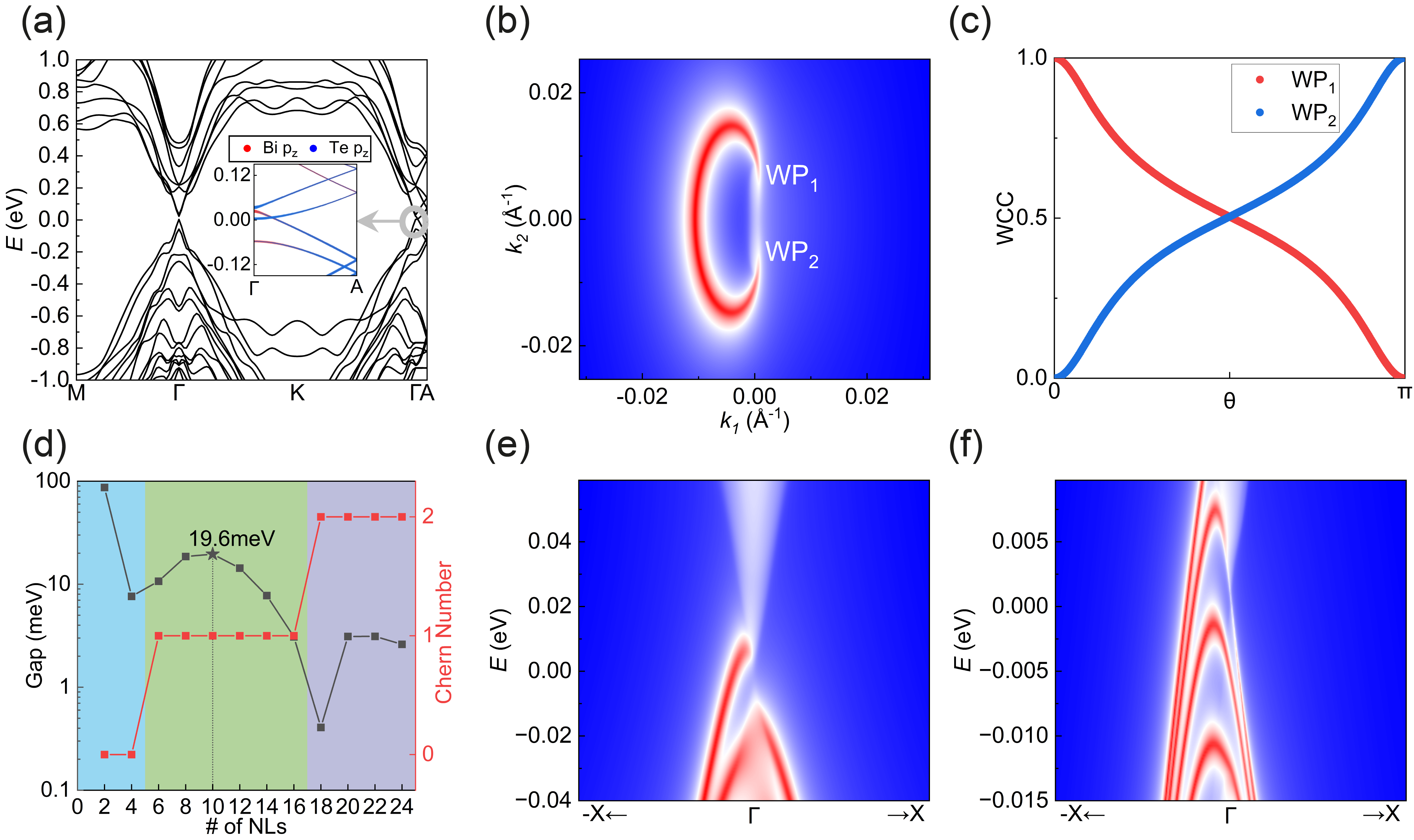}
    \caption{Topological features of type-I NiVBi${}_2$Te${}_5$ in bulk and thin films. (a) Band structure of WSM type-I NiVBi${}_2$Te${}_5$. Details of projected band near one of the WPs are shown in the inset. (b) Surface states on the (100) plane, showing a Fermi arc connecting the two WPs. (c) The motions of the sum of WCCs on spheres surrounding each WP in the momentum space. (d) The evolution of band gap and Chern number with thickness of type-I NiVBi${}_2$Te${}_5$ slab. Chiral edge states of (e) 6-NL ($\mathcal{C}=1$) and (f) 20-NL ($\mathcal{C}=2$) slabs.}
    \label{fig8:Weyl}
\end{figure}

In addition, we identified intrinsic magnetic topological phases beyond the aforementioned dynamic AxIs. Figure \ref{fig8:Weyl} shows the topological properties of type-I NiVBi${}_2$Te${}_5$, a \textit{XY}Bi${}_2$Te${}_5$ compound with FM ground state and a high Curie temperature of $\sim$152 K [see Appendix \ref{Tc}]. The band structure is demonstrated in Fig. \ref{fig8:Weyl}(a), and one pair of Weyl points (WPs) emerge along the -$A-\Gamma-A$ line, with one of them shown in the inset. Our surface state calculations confirm that one pair of WPs exist due to $\mathcal{T}$-symmetry breaking. Figure \ref{fig8:Weyl}(b) clearly shows that a Fermi arc connects two WPs that are symmetrically located along the -$A-\Gamma-A$ on the (100) surface plane. Two WPs demonstrate opposite chirality by checking the motions of the sum of Wannier charge centers (WCCs) [see Fig. \ref{fig8:Weyl}(c)], indicating that type-I NiVBi${}_2$Te${}_5$ exhibits a intrinsic WSM state in bulk under zero external magnetic field. The strong magnetic coupling makes NiVBi${}_2$Te${}_5$ a charming magnetic WSM with high enough Curie temperature (up to 152 K). Beyond type-I NiVBi${}_2$Te${}_5$, intrinsic high-temperature WSM phase can also appear in type-III NiVBi${}_2$Te${}_5$. See Appendix \ref{weyl} for more discussions.

The intrinsic bulk WSM phase in type-I (also type-III) NiVBi${}_2$Te${}_5$ indicates that its thin films can host Chern insulator phases with Chern number growing with increasing thickness \cite{HgCr2Se4-PhysRevLett.107.186806}. Figure. \ref{fig8:Weyl}(d) shows the evolution of the gap size and the Chern number with thickness, which were calculated using maximally localized Wannier function (MLWF) based tight-binding models. A type-I NiVBi${}_2$Te${}_5$ film is a normal insulator below 6-NL, Chern insulator with $\mathcal{C}=1$ between 6-NL and 16-NL, and high Chern insulator ($\mathcal{C}\geqslant 2$) above 16-NL. The relatively short distance between the two WPs leads to slow growth of Chern number, compared to that of FM state MnBi${}_2$Te${}_4$ \cite{nsr-10.1093/nsr/nwaa089}. Naturally, the slab band gaps experience a close and reopen process every time the Chern number changes. In the Chern insulator region, which supports QAHE, the full gap reaches its maximum value of 19.6 meV in 10-NL slabs, corresponding to 227 K. Figures \ref{fig8:Weyl}(e) and \ref{fig8:Weyl}(f) demonstrate the calculated chiral edge states of 6-NL and 20-NL respectively, validating the Chern insulator phases with Chern number $\mathcal{C}=1$ and $\mathcal{C}=2$ respectively.

In order to further understand the origin of WSM in bulk phase NiVBi${}_2$Te${}_5$, and also the tunability of different topological phases, we investigated the topological properties of NiVBi${}_2$Te${}_5$ under hydrostatic pressures from -1.0 GPa to +1.0 GPa, where ambient pressure is labeled as 0.0 GPa and experimentally unavailable negative pressure conditions were adopted to complete the study of phase transition. We studied the electronic structures of bulk NiVBi${}_2$Te${}_5$ with out-of-plane FM magnetization. This magnetic configuration is exactly the magnetic ground state of type-I and type-III NiVBi${}_2$Te${}_5$. It can also be achieved in type-II and type-IV NiVBi${}_2$Te${}_5$ if their AFM interlayer couplings are flipped under strong enough external magnetic field.

\begin{figure}
    \centering
    \includegraphics[width=1.0\linewidth]{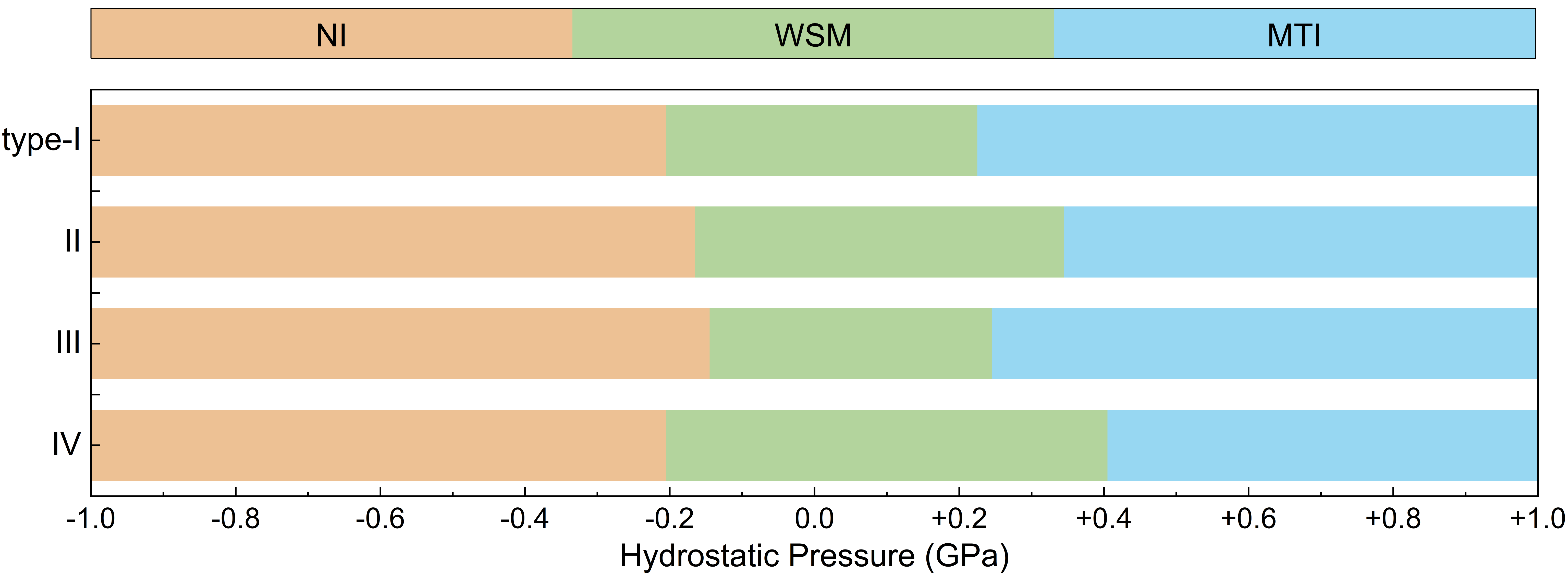}
    \caption{Phase diagrams of out-of-plane FM NiVBi${}_2$Te${}_5$ obtained by tuning external pressures. The orange, green and blue zones stand for NI, WSM and MTI phase, respectively.}
    \label{fig9:Pressure}
\end{figure}

Three topologically distinguishable types of pressure-induced topological phases are identified in Fig. \ref{fig9:Pressure}, including normal insulator (NI), WSM and magnetic topological insulator (MTI). Different phases of bulk NiVBi${}_2$Te${}_5$ corresponds to different 2D phases in thin films. As the thin films grow thicker, the Chern number of NIs fixes at zero and that of WSMs grows as analyzed in Fig. \ref{fig8:Weyl}(d), while that of MTI jumps from zero to one at certain thickness and remains unchanged afterwards. In the pressure range discussed here, all types of NiVBi${}_2$Te${}_5$ behaves like NI at low pressure, evolves into WSM near the ambient pressure region and steps into MTI eventually at high pressure.

Theoretically there have been well-established phase diagrams of NI-WSM(QAH)-MTI transition based on magnetically doped NI-TI superlattices \cite{burkov-prl-PhysRevLett.107.127205,burkov-pt-PhysRevB.85.165110}. In these models, intralayer coupling strength $ \Delta_S $ between top and bottom surface states (SSs), interlayer coupling strength $ \Delta_D $ between bottom and top SSs of neighboring layers, and Zeeman splitting denoted as $ m $  decide the final topological character. Increasing (Decreasing) the external pressure can shrink (enlarge) the ratio between $ \Delta_S $ and $ \Delta_D $, leading to phase transitions. Besides, symmetries play a crucial rule, and the basic model in Ref. \cite{burkov-prl-PhysRevLett.107.127205} requires the presence of inversion symmetry $ \mathcal{P} $. Once $ \mathcal{P} $ is broken, two extra parameters including the electrostatic potential difference $ V $ between bottom and top SSs, and Dresselhaus-like SOC interaction $ \lambda $, should be considered to handle the symmetry change \cite{burkov-pt-PhysRevB.85.165110}.

For NiVBi${}_2$Te${}_5$, $ \mathcal{P} $ is conserved in type-II and type-IV structures, while broken in type-I and type-III structures. However, Ni and V atoms share very similar size and thus SOC strength, leading to ignorable $ \lambda $. Based on Ref. \cite{burkov-pt-PhysRevB.85.165110}, the phase diagram remains almost unchanged as long as we replace $ m $ with an effective Zeeman splitting parameter $ \sqrt{m^2-V^2} $. Therefore, it’s reasonable that all these four compounds can experience a phase transition from NI to WSM and MTI under different external pressure.

\section{conclusions}
In conclusion, we systematically study the structural, magnetic and topological properties of \textit{XY}Bi${}_2$Te${}_5$-family materials. Interatomic-layer exchange couplings play a crucial role in keeping magnetic order of some compounds above 77 K (several compounds even above 150 K), while interlayer couplings and hybridization between top and bottom surfaces determine the band topology. We find type-I(III) NiVBi${}_2$Te${}_5$ to be an emergent material with high Chern number and possible high-temperature ($\sim$152 K) QAHE state and type-I Ni${}_2$Bi${}_2$Te${}_5$ as another candidate demonstrating above-room-temperature AFM order. Ni${}_2$Bi${}_2$Te${}_5$, as well as some other kinds of \textit{XY}Bi${}_2$Te${}_5$ like type-II MnNiBi${}_2$Te${}_5$, type-II NiVBi${}_2$Te${}_5$ and type-IV NiEuBi${}_2$Te${}_5$, also demonstrates nontrivial dynamic axion states. The surface state gaps of, for example, type-II MnNiBi${}_2$Te${}_5$ ($ \sim $17 meV) can be large enough for high-temperature explorations. Under external pressure or magnetic field, possible high-temperature WSM and QAH phases can also be tunable.

\begin{acknowledgments}
    We thank Boxuan Li for helpful discussions and Jiaheng Li for technical support. This work was supported by the National Natural Science Foundation of China (92065206).

    X.-Y.T. and Z.L. contributed equally to this work.
\end{acknowledgments}
\newpage

\appendix

\begin{figure*}
    \centering
    \includegraphics[width=1.0\linewidth]{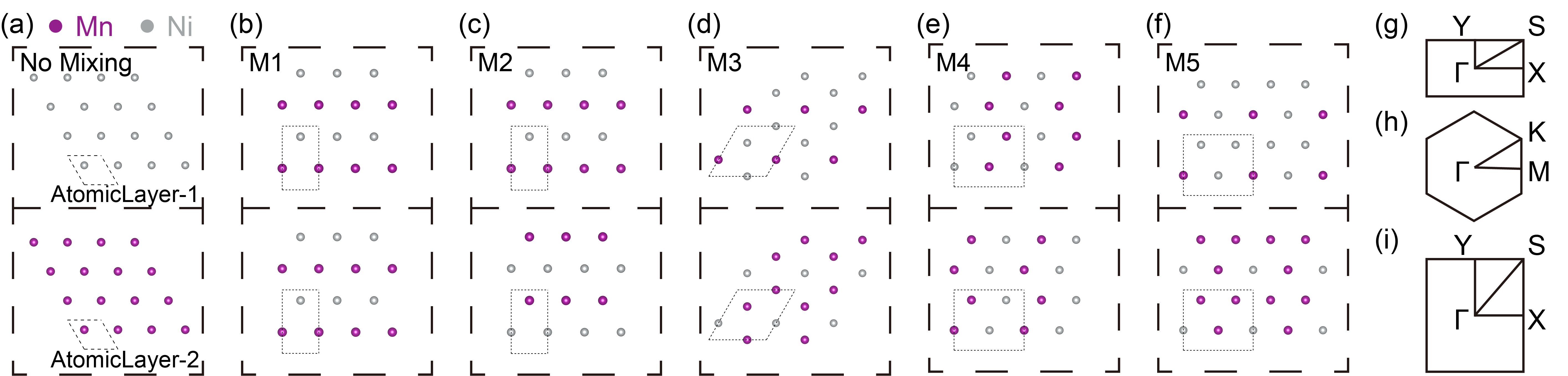}
    \caption{Illustration of \textit{X}-\textit{Y} mixing using monolayer type-I MnNiBi${}_2$Te${}_5$ as an example. (a) Pristine lattice structure with only Ni (AtomicLayer-1) and Mn (AtomicLayer-2) atomic layers plotted. (b-f) Five considered lattice structures with \textit{X}-\textit{Y} mixing are denoted as M1-M5. Only the two magnetic atomic layers are plotted for clarity. We marked the unit cells inside every layer as thin dashed lines. (g-i) Schematic first Brillouin zones of M1-M2, M3 and M4-M5.} 
    \label{fig10:MixingType}
\end{figure*}

\begin{figure*}
	\centering
	\includegraphics[width=0.95\linewidth]{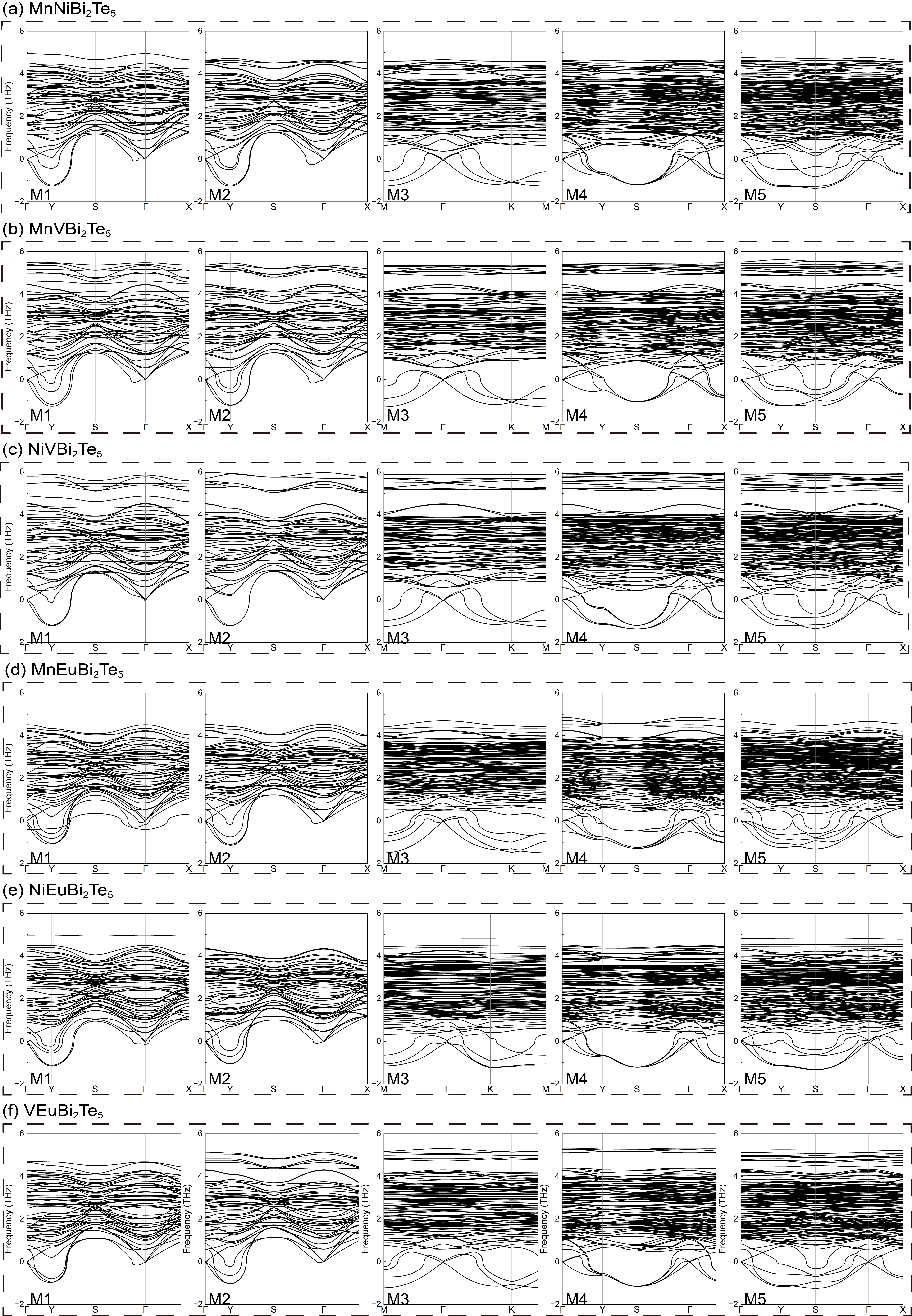}
	\caption{Phonon dispersions of lattice structures with \textit{X}-\textit{Y} mixing.}
	\label{Fig11:Mixing}
\end{figure*}

\begin{figure}
    \centering
    \includegraphics[width=\linewidth]{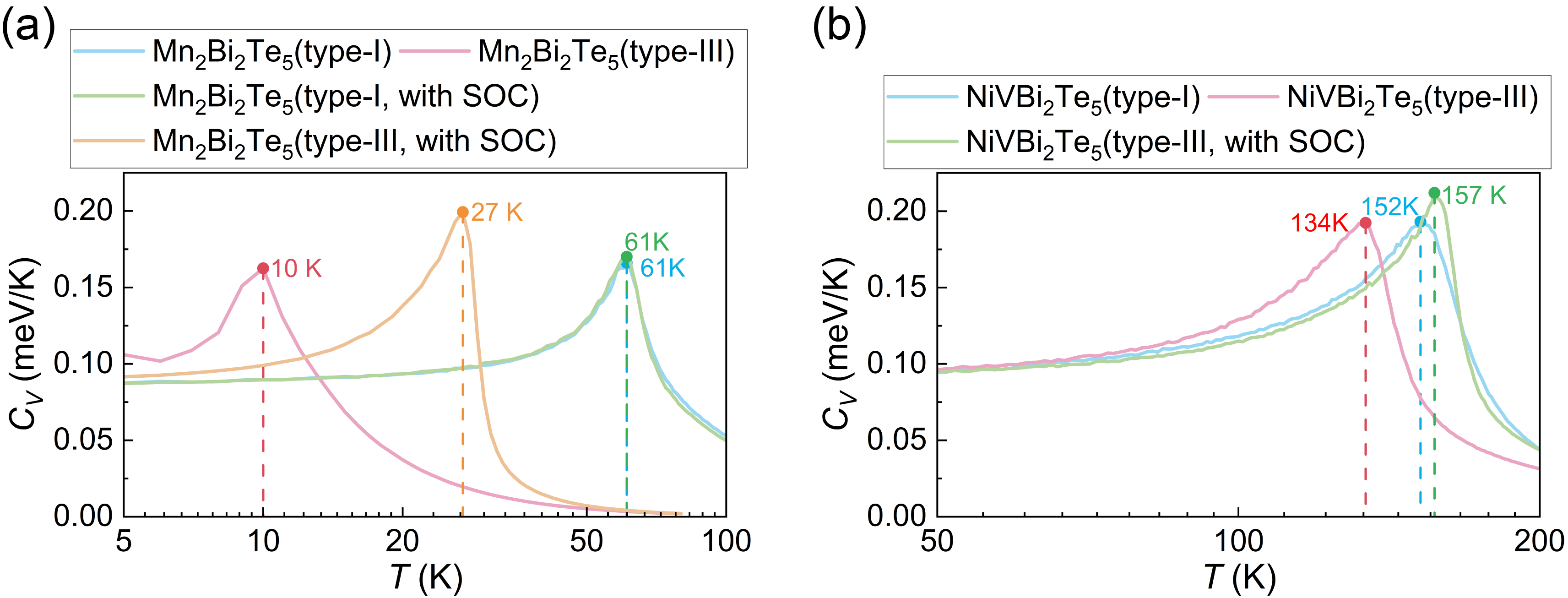}
    \caption{$C_V$-$T$ curves of (a) type-I and type-III Mn${}_2$Bi${}_2$Te${}_5$ and (b) type-I and type-III NiVBi${}_2$Te${}_5$.}
    \label{Fig12:Tc}
\end{figure}

\section{LATTICE CONSTANTS OF MONOLAYER AND BULK \textit{XY}Bi${}_2$Te${}_5$ \label{latticeconstants}}
The lattice constants are almost the same under different magnetic configurations. Thus we only demonstrate the data calculated at magnetic ground states. Detailed discussions on intrinsic magnetism of \textit{XY}Bi${}_2$Te${}_5$ can be found in Sec. \ref{mag}. For Mn${}_2$Bi${}_2$Te${}_5$, all presented data in this paper are based on theoretical lattice constants. The difference between theoretical and reported experimental lattice constants \cite{225exp-PhysRevB.104.054421} of bulk Mn${}_2$Bi${}_2$Te${}_5$ is less than 1\%.

\begin{table}[h]
    \label{tabA}
    \caption{Lattice constants of all ten kinds of \textit{XY}Bi${}_2$Te${}_5$ combinations. Here $ a $ and $ c $ refer to in-plane and out-of-plane lattice constants individually. Both are in unit of Å.}
    \begin{ruledtabular}
        \begin{tabular}{cccrrrr}
            \multicolumn{3}{c}{\textit{X}-\textit{Y}} & \multicolumn{1}{c}{type-I} & \multicolumn{1}{c}{type-II} & \multicolumn{1}{c}{type-III} & \multicolumn{1}{c}{type-IV} \\ \hline
            Mn-Mn                & monolayer & $a$                     & 4.338  &        & 4.333  &        \\
            \multicolumn{1}{l}{} & bulk      & \multicolumn{1}{l}{$a$} & 4.310  &        & 4.307  &        \\
                                 &           & $c$                     & 33.951 &        & 17.045 &        \\ \hline
            Ni-Ni                & monolayer & $a$                     & 4.222  &        & 4.189  &        \\
            \multicolumn{1}{l}{} & bulk      & \multicolumn{1}{l}{$a$} & 4.289  &        & 4.243  &        \\
                                 &           & $c$                     & 32.393 &        & 16.423 &        \\ \hline
            V-V                  & monolayer & $a$                     & 4.337  &        & 4.325  &        \\
            \multicolumn{1}{l}{} & bulk      & \multicolumn{1}{l}{$a$} & 4.315  &        & 4.307  &        \\
                                 &           & $c$                     & 33.377 &        & 16.819 &        \\ \hline
            Eu-Eu                & monolayer & $a$                     & 4.537  &        & 4.549  &        \\
            \multicolumn{1}{l}{} & bulk      & \multicolumn{1}{l}{$a$} & 4.519  &        & 4.532  &        \\
                                 &           & $c$                     & 35.864 &        & 17.734 &        \\ \hline
            Mn-Ni                & monolayer & $a$                     & 4.276  &        & 4.257  &        \\
            \multicolumn{1}{l}{} & bulk      & \multicolumn{1}{l}{$a$} & 4.285  & 4.285  & 4.261  & 4.259  \\
                                 &           & $c$                     & 33.196 & 33.256 & 16.718 & 33.452 \\ \hline
            Mn-V                 & monolayer & $a$                     & 4.336  &        & 4.323  &        \\
            \multicolumn{1}{l}{} & bulk      & \multicolumn{1}{l}{$a$} & 4.311  & 4.311  & 4.301  & 4.302  \\
                                 &           & $c$                     & 33.771 & 33.786 & 16.994 & 33.994 \\ \hline
            Ni-V                 & monolayer & $a$                     & 4.283  &        & 4.258  &        \\
            \multicolumn{1}{l}{} & bulk      & \multicolumn{1}{l}{$a$} & 4.263  & 4.262  & 4.265  & 4.266  \\
                                 &           & $c$                     & 32.925 & 33.113 & 16.697 & 33.393 \\ \hline
            Mn-Eu                & monolayer & $a$                     & 4.405  &        & 4.438  &        \\
            \multicolumn{1}{l}{} & bulk      & \multicolumn{1}{l}{$a$} & 4.400  & 4.405  & 4.419  & 4.420  \\
                                 &           & $c$                     & 35.049 & 35.071 & 17.404 & 34.830 \\ \hline
            Ni-Eu                & monolayer & $a$                     & 4.363  &        & 4.362  &        \\
            \multicolumn{1}{l}{} & bulk      & \multicolumn{1}{l}{$a$} & 4.333  & 4.347  & 4.355  & 4.356  \\
                                 &           & $c$                     & 34.646 & 34.602 & 17.175 & 34.340 \\ \hline
            V-Eu                 & monolayer & $a$                     & 4.499  &        & 4.427  &        \\
            \multicolumn{1}{l}{} & bulk      & \multicolumn{1}{l}{$a$} & 4.398  & 4.398  & 4.409  & 4.355  \\
                                 &           & $c$                     & 34.938 & 34.907 & 17.326 & 34.472
            \end{tabular}
\end{ruledtabular}
\end{table}

\section{DISCUSSIONS ON \textit{XY}Bi${}_2$Te${}_5$ WITH \textit{X}-\textit{Y} MIXING\label{mix}}
We considered the situations where \textit{X} and \textit{Y} atoms (\textit{X}$ \neq $\textit{Y}) are mixed instead of locating at separate atomic layers. We showed that for some combinations of \textit{X} and \textit{Y}, \textit{X}-\textit{Y} mixing is neither energetically favored nor structurally stable. For the others, despite \textit{X}-\textit{Y} mixing leads to energy gain, the structural instability remains.

Taking monolayer \textit{XY}Bi${}_2$Te${}_5$ as an example, we constructed five representative structures with \textit{X}-\textit{Y} mixing, denoted as M1-M5 [see Figs. \ref{fig10:MixingType}(b-f)]. Each magnetic atomic layer in M1(M2) shows stripy structures. The only difference between M1 and M2 is the interatomic-layer stacking order. M3 simulates the mixing problem using a $ \sqrt{3}\times\sqrt{3} $ supercell. M4 possesses a zigzag structure while M5 is derived from M4 but with a different mixing ratio of 3:1 in each single layer. Note that the mixed structures are all based on monolayer ground-state lattice structures, which means we selected type-I MnNiBi${}_2$Te${}_5$, MnVBi${}_2$Te${}_5$ and NiVBi${}_2$Te${}_5$ but type-III MnEuBi${}_2$Te${}_5$, NiEuBi${}_2$Te${}_5$ and VEuBi${}_2$Te${}_5$ instead.

\begin{table}[h]
    \caption{\label{tab3:mixingenergy}Energy comparisons between mixed (M1-M5) and pristine structures of monolayer \textit{XY}Bi${}_2$Te${}_5$ (\textit{X}$ \neq $\textit{Y}). The ground state energies of pristine structures (without mixing) are set as zero points and all values are in unit of meV. Negative (positive) values refer to occasions where mixing leads to energy gain (cost). All these values have been averaged to cells as large as original primitive cells (including one \textit{X} and one \textit{Y}) for better comparison.}
    \begin{ruledtabular}
        \begin{tabular}{crrrrr}
            \textit{X}-\textit{Y} &
              \multicolumn{1}{c}{\text{M1}} &
              \multicolumn{1}{c}{\text{M2}} &
              \multicolumn{1}{c}{\text{M3}} &
              \multicolumn{1}{c}{\text{M4}} &
              \multicolumn{1}{c}{\text{M5}} \\ \hline
            Mn-Ni & 55.08  & 60.07  & 64.47  & 60.68  & 48.71  \\
            Mn-V  & -5.01  & -18.57 & -9.46  & -11.19 & -5.98  \\
            Ni-V  & -71.15 & -73.78 & -59.87 & -68.99 & -52.66 \\
            Mn-Eu & 134.80 & 4.89 & 64.08  & 74.12  & 37.05  \\
            Ni-Eu & 114.92 & 82.43  & 109.42 & 55.33  & 102.38 \\
            V-Eu  & 230.44 & 27.26  & 111.52 & 147.97 & 78.15 
            \end{tabular}
    \end{ruledtabular}
    \end{table}

We calculated the magnetic ground state energies of different mixed structures. Similar to calculations in Sec. \ref{mag}, we also used collinear calculations and ignored SOC. The data are summarized in Table \ref{tab3:mixingenergy}. As we can see, the ground state energy may be further reduced in MnVBi${}_2$Te${}_5$ and NiVBi${}_2$Te${}_5$ once Mn(Ni) and V atoms are mixed, while in other \textit{XY}Bi${}_2$Te${}_5$ considered here, separate \textit{X} and \textit{Y} atomic layers are energetically favored.

Then we studied the phonon dispersions of mixed structures [see Fig. \ref{Fig11:Mixing}]. We found that no matter how the \textit{X} and \textit{Y} atoms are mixed, significant virtual frequency across the first Brillouin zone (FBZ) emerges, indicating highly unstable structures. Specifically, the appearance of virtual frequency is not sensitive to the interatomic-layer stacking order, but highly related to the in-plane direction along which mixing happens [see M1 and M2 in Fig. \ref{Fig11:Mixing}]. The latter conclusion can be made when the phonon dispersions along $ \Gamma-X $ and $ \Gamma-Y $ are compared with each other in cases of M1 and M2. Therefore, we think \textit{X}-\textit{Y} mixing should not be a severe problem theoretically and \textit{XY}Bi${}_2$Te${}_5$ materials are supposed to inherently establish separate magnetic atomic layers.

\section{SUPPLEMENTARY $C_V$-$T$ CURVES of Mn${}_2$Bi${}_2$Te${}_5$ and NiVBi${}_2$Te${}_5$\label{Tc}}
Extra $C_V$-$T$ curves of Mn${}_2$Bi${}_2$Te${}_5$ and NiVBi${}_2$Te${}_5$ are demonstrated in Fig. \ref{Fig12:Tc}. The Curie temperatures of type-I and type-III NiVBi${}_2$Te${}_5$ (152 K and 134 K respectively) are also well above 77 K, firmly supporting the description of high-temperature WSM in the main text. We also tested the influence of SOC on the results using type-III NiVBi${}_2$Te${}_5$, type-I and type-III Mn${}_2$Bi${}_2$Te${}_5$ as examples. Typically, the introduction of SOC leads to a slight increase in magnetic critical temperature. As for type-III Mn${}_2$Bi${}_2$Te${}_5$, since the estimated Néel temperature is not high and close to experimental results, we decided to consider an extra next-next-nearest intralayer coupling term in our simulation and adopted experimental lattice in the calculation of exchange constants. As can be seen in Fig \ref{Fig12:Tc}(a), its weak interatomic-layer $ E_{ex} $ leads to a Néel temperature of 10 K when SOC is ignored, while 27 K when SOC is considered, which is much closer to the experimental result than type-I Mn${}_2$Bi${}_2$Te${}_5$. 

\section{BAND INVERSION FEATURES AND SURFACE STATES OF DYNAMIC AXION INSULATORS\label{axion}}
\begin{figure}[h]
	\centering
	\includegraphics[width=1.0\linewidth]{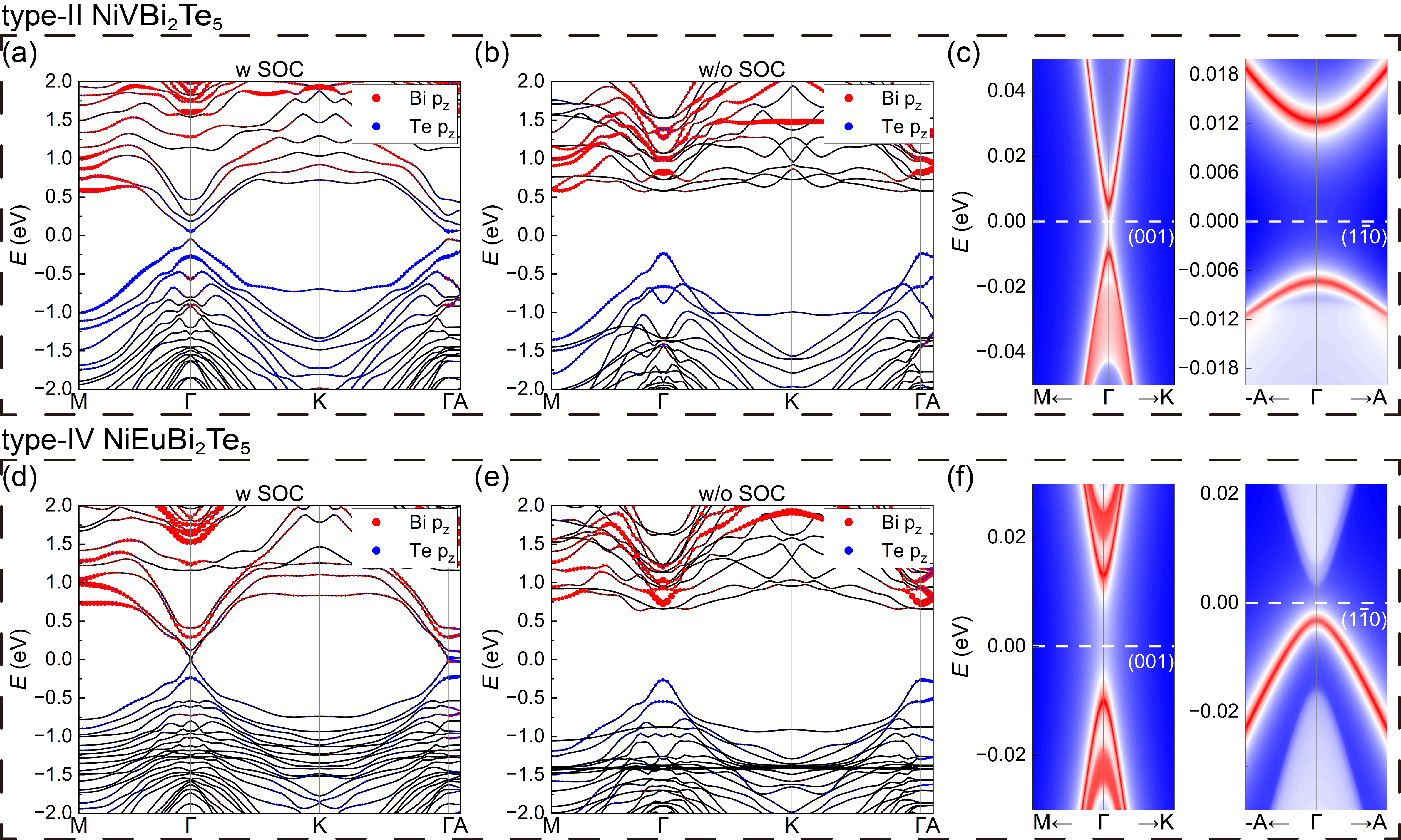}
	\caption{Band structures and surface states of (a-c) type-II NiVBi${}_2$Te${}_5$ and (d-f) type-IV NiEuBi${}_2$Te${}_5$. (a,d) Band structures with SOC. (b,e) Band structures without SOC. (c,f) Surface states on (001) and (1$\bar{1}$0) planes.}
	\label{Fig13:RestAxion}
\end{figure}

The bulk band structures and also surface states of type-II NiVBi${}_2$Te${}_5$ and type-IV NiEuBi${}_2$Te${}_5$ are shown in Fig. \ref{Fig13:RestAxion}. Clearly, the band inversion features and gapped surface states can be found out. Combining these characters with information provided in Fig. \ref{fig7:Axion}, we can conclude that these two kinds of \textit{XY}Bi${}_2$Te${}_5$ are also dynamic AxIs.

\section{WEYL SEMIMETAL AND CHERN INSULATOR PHASES OF TYPE-III NiVBi${}_2$Te${}_5$\label{weyl}}
\begin{figure}
	\centering
	\includegraphics[width=1.0\linewidth]{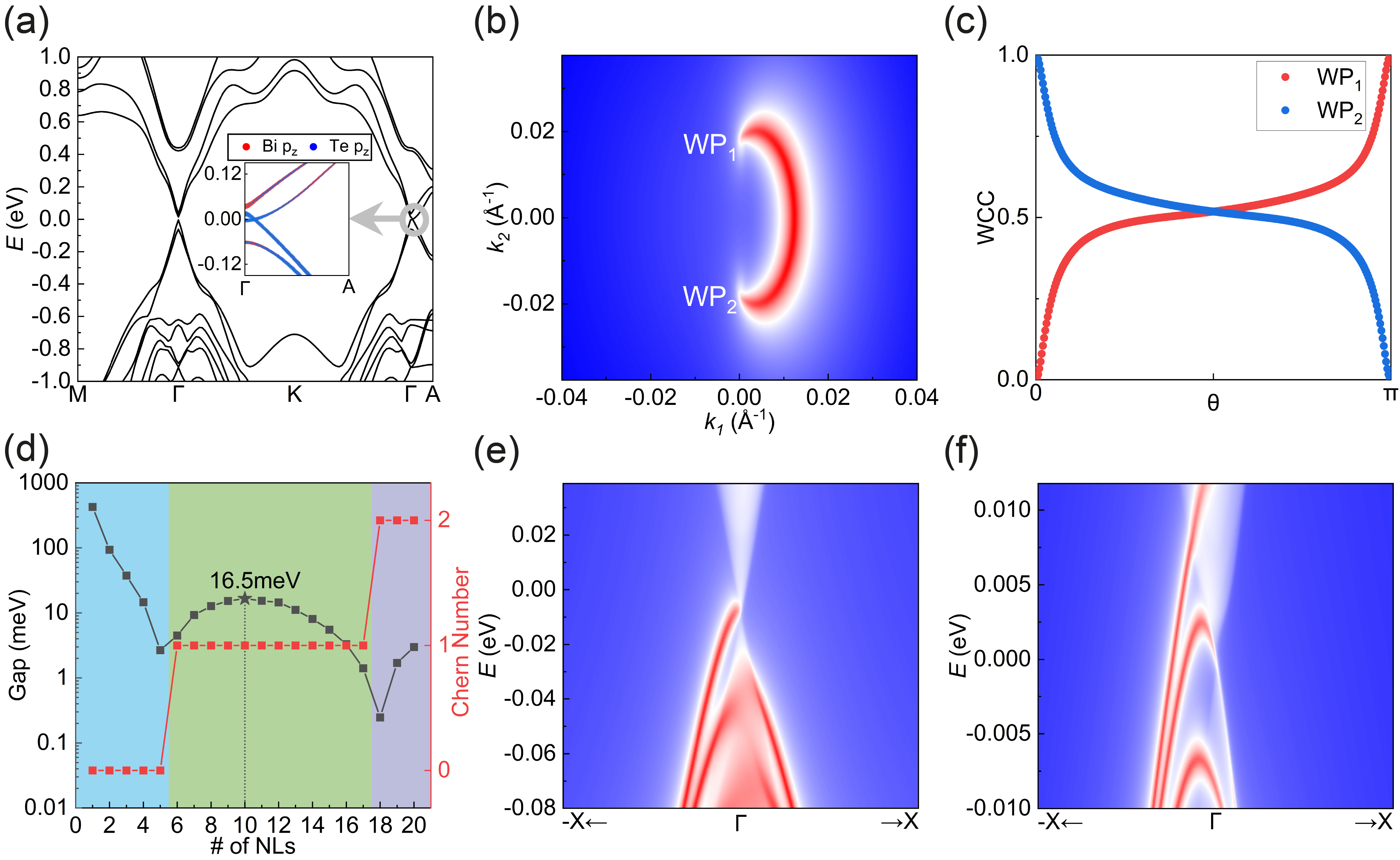}
	\caption{Topological features of type-III NiVBi${}_2$Te${}_5$ in bulk and thin films. (a) Band structure of WSM type-III NiVBi${}_2$Te${}_5$. Details of projected band near one of the WPs are shown in the inset. (b) Surface states on the (100) plane, showing a Fermi arc connecting the two WPs. (c) The motions of the sum of WCCs on spheres surrounding each WP in the momentum space. (d) The evolution of band gap and Chern number with thickness of type-III NiVBi${}_2$Te${}_5$ slab. Chiral edge states of (e) 6-NL ($\mathcal{C}=1$) and (f) 20-NL ($\mathcal{C}=2$) slabs.}
	\label{Fig14:RestWeyl}
\end{figure}

Similar to type-I NiVBi${}_2$Te${}_5$, type-III NiVBi${}_2$Te${}_5$ also has a FM ground state and behaves as intrinsic magnetic WSM. Figure \ref{Fig14:RestWeyl} summarizes the overall topological properties of type-III NiVBi${}_2$Te${}_5$ both in its 3D bulk phase and 2D thin films.

\newpage
\nocite{*}
\bibliography{main_text}

\providecommand{\noopsort}[1]{}\providecommand{\singleletter}[1]{#1}%
\begin{thebibliography}{64}%
\makeatletter
\providecommand \@ifxundefined [1]{%
 \@ifx{#1\undefined}
}%
\providecommand \@ifnum [1]{%
 \ifnum #1\expandafter \@firstoftwo
 \else \expandafter \@secondoftwo
 \fi
}%
\providecommand \@ifx [1]{%
 \ifx #1\expandafter \@firstoftwo
 \else \expandafter \@secondoftwo
 \fi
}%
\providecommand \natexlab [1]{#1}%
\providecommand \enquote  [1]{``#1''}%
\providecommand \bibnamefont  [1]{#1}%
\providecommand \bibfnamefont [1]{#1}%
\providecommand \citenamefont [1]{#1}%
\providecommand \href@noop [0]{\@secondoftwo}%
\providecommand \href [0]{\begingroup \@sanitize@url \@href}%
\providecommand \@href[1]{\@@startlink{#1}\@@href}%
\providecommand \@@href[1]{\endgroup#1\@@endlink}%
\providecommand \@sanitize@url [0]{\catcode `\\12\catcode `\$12\catcode
  `\&12\catcode `\#12\catcode `\^12\catcode `\_12\catcode `\%12\relax}%
\providecommand \@@startlink[1]{}%
\providecommand \@@endlink[0]{}%
\providecommand \url  [0]{\begingroup\@sanitize@url \@url }%
\providecommand \@url [1]{\endgroup\@href {#1}{\urlprefix }}%
\providecommand \urlprefix  [0]{URL }%
\providecommand \Eprint [0]{\href }%
\providecommand \doibase [0]{https://doi.org/}%
\providecommand \selectlanguage [0]{\@gobble}%
\providecommand \bibinfo  [0]{\@secondoftwo}%
\providecommand \bibfield  [0]{\@secondoftwo}%
\providecommand \translation [1]{[#1]}%
\providecommand \BibitemOpen [0]{}%
\providecommand \bibitemStop [0]{}%
\providecommand \bibitemNoStop [0]{.\EOS\space}%
\providecommand \EOS [0]{\spacefactor3000\relax}%
\providecommand \BibitemShut  [1]{\csname bibitem#1\endcsname}%
\let\auto@bib@innerbib\@empty
\bibitem [{\citenamefont {Hasan}\ and\ \citenamefont
  {Kane}(2010)}]{hasan-RevModPhys.82.3045}%
  \BibitemOpen
  \bibfield  {author} {\bibinfo {author} {\bibfnamefont {M.~Z.}\ \bibnamefont
  {Hasan}}\ and\ \bibinfo {author} {\bibfnamefont {C.~L.}\ \bibnamefont
  {Kane}},\ }\bibfield  {title} {\bibinfo {title} {Colloquium: Topological
  insulators},\ }\href {https://doi.org/10.1103/RevModPhys.82.3045} {\bibfield
  {journal} {\bibinfo  {journal} {Rev. Mod. Phys.}\ }\textbf {\bibinfo {volume}
  {82}},\ \bibinfo {pages} {3045} (\bibinfo {year} {2010})}\BibitemShut
  {NoStop}%
\bibitem [{\citenamefont {Qi}\ and\ \citenamefont
  {Zhang}(2011)}]{qi-RevModPhys.83.1057}%
  \BibitemOpen
  \bibfield  {author} {\bibinfo {author} {\bibfnamefont {X.-L.}\ \bibnamefont
  {Qi}}\ and\ \bibinfo {author} {\bibfnamefont {S.-C.}\ \bibnamefont {Zhang}},\
  }\bibfield  {title} {\bibinfo {title} {Topological insulators and
  superconductors},\ }\href {https://doi.org/10.1103/RevModPhys.83.1057}
  {\bibfield  {journal} {\bibinfo  {journal} {Rev. Mod. Phys.}\ }\textbf
  {\bibinfo {volume} {83}},\ \bibinfo {pages} {1057} (\bibinfo {year}
  {2011})}\BibitemShut {NoStop}%
\bibitem [{\citenamefont {Armitage}\ \emph {et~al.}(2018)\citenamefont
  {Armitage}, \citenamefont {Mele},\ and\ \citenamefont
  {Vishwanath}}]{wsm-rev-RevModPhys.90.015001}%
  \BibitemOpen
  \bibfield  {author} {\bibinfo {author} {\bibfnamefont {N.~P.}\ \bibnamefont
  {Armitage}}, \bibinfo {author} {\bibfnamefont {E.~J.}\ \bibnamefont {Mele}},\
  and\ \bibinfo {author} {\bibfnamefont {A.}~\bibnamefont {Vishwanath}},\
  }\bibfield  {title} {\bibinfo {title} {Weyl and {D}irac semimetals in
  three-dimensional solids},\ }\href
  {https://doi.org/10.1103/RevModPhys.90.015001} {\bibfield  {journal}
  {\bibinfo  {journal} {Rev. Mod. Phys.}\ }\textbf {\bibinfo {volume} {90}},\
  \bibinfo {pages} {015001} (\bibinfo {year} {2018})}\BibitemShut {NoStop}%
\bibitem [{\citenamefont {Tokura}\ \emph {et~al.}(2019)\citenamefont {Tokura},
  \citenamefont {Yasuda},\ and\ \citenamefont {Tsukazaki}}]{Tokura2019}%
  \BibitemOpen
  \bibfield  {author} {\bibinfo {author} {\bibfnamefont {Y.}~\bibnamefont
  {Tokura}}, \bibinfo {author} {\bibfnamefont {K.}~\bibnamefont {Yasuda}},\
  and\ \bibinfo {author} {\bibfnamefont {A.}~\bibnamefont {Tsukazaki}},\
  }\bibfield  {title} {\bibinfo {title} {Magnetic topological insulators},\
  }\href {https://doi.org/10.1038/s42254-018-0011-5} {\bibfield  {journal}
  {\bibinfo  {journal} {Nat. Rev. Phys.}\ }\textbf {\bibinfo {volume} {1}},\
  \bibinfo {pages} {126} (\bibinfo {year} {2019})}\BibitemShut {NoStop}%
\bibitem [{\citenamefont {Sekine}\ and\ \citenamefont
  {Nomura}(2021)}]{axion-review}%
  \BibitemOpen
  \bibfield  {author} {\bibinfo {author} {\bibfnamefont {A.}~\bibnamefont
  {Sekine}}\ and\ \bibinfo {author} {\bibfnamefont {K.}~\bibnamefont
  {Nomura}},\ }\bibfield  {title} {\bibinfo {title} {Axion electrodynamics in
  topological materials},\ }\href {https://doi.org/10.1063/5.0038804}
  {\bibfield  {journal} {\bibinfo  {journal} {J. Appl. Phys.}\ }\textbf
  {\bibinfo {volume} {129}},\ \bibinfo {pages} {141101} (\bibinfo {year}
  {2021})}\BibitemShut {NoStop}%
\bibitem [{\citenamefont {Chang}\ \emph {et~al.}(2023)\citenamefont {Chang},
  \citenamefont {Liu},\ and\ \citenamefont
  {MacDonald}}]{QAHE-RevModPhys.95.011002}%
  \BibitemOpen
  \bibfield  {author} {\bibinfo {author} {\bibfnamefont {C.-Z.}\ \bibnamefont
  {Chang}}, \bibinfo {author} {\bibfnamefont {C.-X.}\ \bibnamefont {Liu}},\
  and\ \bibinfo {author} {\bibfnamefont {A.~H.}\ \bibnamefont {MacDonald}},\
  }\bibfield  {title} {\bibinfo {title} {{Colloquium: Quantum anomalous Hall
  effect}},\ }\href {https://doi.org/10.1103/RevModPhys.95.011002} {\bibfield
  {journal} {\bibinfo  {journal} {Rev. Mod. Phys.}\ }\textbf {\bibinfo {volume}
  {95}},\ \bibinfo {pages} {011002} (\bibinfo {year} {2023})}\BibitemShut
  {NoStop}%
\bibitem [{\citenamefont {Haldane}(1988)}]{haldane-PhysRevLett.61.2015}%
  \BibitemOpen
  \bibfield  {author} {\bibinfo {author} {\bibfnamefont {F.~D.~M.}\
  \bibnamefont {Haldane}},\ }\bibfield  {title} {\bibinfo {title} {Model for a
  {Q}uantum {H}all {E}ffect without {L}andau {L}evels: Condensed-{M}atter
  realization of the "{P}arity {A}nomaly"},\ }\href
  {https://doi.org/10.1103/PhysRevLett.61.2015} {\bibfield  {journal} {\bibinfo
   {journal} {Phys. Rev. Lett.}\ }\textbf {\bibinfo {volume} {61}},\ \bibinfo
  {pages} {2015} (\bibinfo {year} {1988})}\BibitemShut {NoStop}%
\bibitem [{\citenamefont {Yu}\ \emph {et~al.}(2010)\citenamefont {Yu},
  \citenamefont {Zhang}, \citenamefont {Zhang}, \citenamefont {Zhang},
  \citenamefont {Dai},\ and\ \citenamefont
  {Fang}}]{yurui-doi:10.1126/science.1187485}%
  \BibitemOpen
  \bibfield  {author} {\bibinfo {author} {\bibfnamefont {R.}~\bibnamefont
  {Yu}}, \bibinfo {author} {\bibfnamefont {W.}~\bibnamefont {Zhang}}, \bibinfo
  {author} {\bibfnamefont {H.-J.}\ \bibnamefont {Zhang}}, \bibinfo {author}
  {\bibfnamefont {S.-C.}\ \bibnamefont {Zhang}}, \bibinfo {author}
  {\bibfnamefont {X.}~\bibnamefont {Dai}},\ and\ \bibinfo {author}
  {\bibfnamefont {Z.}~\bibnamefont {Fang}},\ }\bibfield  {title} {\bibinfo
  {title} {Quantized {A}nomalous {H}all {E}ffect in {M}agnetic {T}opological
  {I}nsulators},\ }\href {https://doi.org/10.1126/science.1187485} {\bibfield
  {journal} {\bibinfo  {journal} {Science}\ }\textbf {\bibinfo {volume}
  {329}},\ \bibinfo {pages} {61} (\bibinfo {year} {2010})}\BibitemShut
  {NoStop}%
\bibitem [{\citenamefont {Chang}\ \emph {et~al.}(2013)\citenamefont {Chang},
  \citenamefont {Zhang}, \citenamefont {Feng}, \citenamefont {Shen},
  \citenamefont {Zhang}, \citenamefont {Guo}, \citenamefont {Li}, \citenamefont
  {Ou}, \citenamefont {Wei}, \citenamefont {Wang}, \citenamefont {Ji},
  \citenamefont {Feng}, \citenamefont {Ji}, \citenamefont {Chen}, \citenamefont
  {Jia}, \citenamefont {Dai}, \citenamefont {Fang}, \citenamefont {Zhang},
  \citenamefont {He}, \citenamefont {Wang}, \citenamefont {Lu}, \citenamefont
  {Ma},\ and\ \citenamefont {Xue}}]{qahe-doi:10.1126/science.1234414}%
  \BibitemOpen
  \bibfield  {author} {\bibinfo {author} {\bibfnamefont {C.-Z.}\ \bibnamefont
  {Chang}}, \bibinfo {author} {\bibfnamefont {J.}~\bibnamefont {Zhang}},
  \bibinfo {author} {\bibfnamefont {X.}~\bibnamefont {Feng}}, \bibinfo {author}
  {\bibfnamefont {J.}~\bibnamefont {Shen}}, \bibinfo {author} {\bibfnamefont
  {Z.}~\bibnamefont {Zhang}}, \bibinfo {author} {\bibfnamefont
  {M.}~\bibnamefont {Guo}}, \bibinfo {author} {\bibfnamefont {K.}~\bibnamefont
  {Li}}, \bibinfo {author} {\bibfnamefont {Y.}~\bibnamefont {Ou}}, \bibinfo
  {author} {\bibfnamefont {P.}~\bibnamefont {Wei}}, \bibinfo {author}
  {\bibfnamefont {L.-L.}\ \bibnamefont {Wang}}, \bibinfo {author}
  {\bibfnamefont {Z.-Q.}\ \bibnamefont {Ji}}, \bibinfo {author} {\bibfnamefont
  {Y.}~\bibnamefont {Feng}}, \bibinfo {author} {\bibfnamefont {S.}~\bibnamefont
  {Ji}}, \bibinfo {author} {\bibfnamefont {X.}~\bibnamefont {Chen}}, \bibinfo
  {author} {\bibfnamefont {J.}~\bibnamefont {Jia}}, \bibinfo {author}
  {\bibfnamefont {X.}~\bibnamefont {Dai}}, \bibinfo {author} {\bibfnamefont
  {Z.}~\bibnamefont {Fang}}, \bibinfo {author} {\bibfnamefont {S.-C.}\
  \bibnamefont {Zhang}}, \bibinfo {author} {\bibfnamefont {K.}~\bibnamefont
  {He}}, \bibinfo {author} {\bibfnamefont {Y.}~\bibnamefont {Wang}}, \bibinfo
  {author} {\bibfnamefont {L.}~\bibnamefont {Lu}}, \bibinfo {author}
  {\bibfnamefont {X.-C.}\ \bibnamefont {Ma}},\ and\ \bibinfo {author}
  {\bibfnamefont {Q.-K.}\ \bibnamefont {Xue}},\ }\bibfield  {title} {\bibinfo
  {title} {Experimental {O}bservation of the {Q}uantum {A}nomalous {H}all
  {E}ffect in a {M}agnetic {T}opological {I}nsulator},\ }\href
  {https://doi.org/10.1126/science.1234414} {\bibfield  {journal} {\bibinfo
  {journal} {Science}\ }\textbf {\bibinfo {volume} {340}},\ \bibinfo {pages}
  {167} (\bibinfo {year} {2013})}\BibitemShut {NoStop}%
\bibitem [{\citenamefont {Mogi}\ \emph {et~al.}(2015)\citenamefont {Mogi},
  \citenamefont {Yoshimi}, \citenamefont {Tsukazaki}, \citenamefont {Yasuda},
  \citenamefont {Kozuka}, \citenamefont {Takahashi}, \citenamefont {Kawasaki},\
  and\ \citenamefont {Tokura}}]{apl-doi:10.1063/1.4935075}%
  \BibitemOpen
  \bibfield  {author} {\bibinfo {author} {\bibfnamefont {M.}~\bibnamefont
  {Mogi}}, \bibinfo {author} {\bibfnamefont {R.}~\bibnamefont {Yoshimi}},
  \bibinfo {author} {\bibfnamefont {A.}~\bibnamefont {Tsukazaki}}, \bibinfo
  {author} {\bibfnamefont {K.}~\bibnamefont {Yasuda}}, \bibinfo {author}
  {\bibfnamefont {Y.}~\bibnamefont {Kozuka}}, \bibinfo {author} {\bibfnamefont
  {K.~S.}\ \bibnamefont {Takahashi}}, \bibinfo {author} {\bibfnamefont
  {M.}~\bibnamefont {Kawasaki}},\ and\ \bibinfo {author} {\bibfnamefont
  {Y.}~\bibnamefont {Tokura}},\ }\bibfield  {title} {\bibinfo {title} {Magnetic
  modulation doping in topological insulators toward higher-temperature quantum
  anomalous hall effect},\ }\href {https://doi.org/10.1063/1.4935075}
  {\bibfield  {journal} {\bibinfo  {journal} {Appl. Phys. Lett.}\ }\textbf
  {\bibinfo {volume} {107}},\ \bibinfo {pages} {182401} (\bibinfo {year}
  {2015})}\BibitemShut {NoStop}%
\bibitem [{\citenamefont {Chang}\ \emph {et~al.}(2015)\citenamefont {Chang},
  \citenamefont {Zhao}, \citenamefont {Kim}, \citenamefont {Zhang},
  \citenamefont {Assaf}, \citenamefont {Heiman}, \citenamefont {Zhang},
  \citenamefont {Liu}, \citenamefont {Chan},\ and\ \citenamefont
  {Moodera}}]{Chang2015}%
  \BibitemOpen
  \bibfield  {author} {\bibinfo {author} {\bibfnamefont {C.-Z.}\ \bibnamefont
  {Chang}}, \bibinfo {author} {\bibfnamefont {W.}~\bibnamefont {Zhao}},
  \bibinfo {author} {\bibfnamefont {D.~Y.}\ \bibnamefont {Kim}}, \bibinfo
  {author} {\bibfnamefont {H.}~\bibnamefont {Zhang}}, \bibinfo {author}
  {\bibfnamefont {B.~A.}\ \bibnamefont {Assaf}}, \bibinfo {author}
  {\bibfnamefont {D.}~\bibnamefont {Heiman}}, \bibinfo {author} {\bibfnamefont
  {S.-C.}\ \bibnamefont {Zhang}}, \bibinfo {author} {\bibfnamefont
  {C.}~\bibnamefont {Liu}}, \bibinfo {author} {\bibfnamefont {M.~H.~W.}\
  \bibnamefont {Chan}},\ and\ \bibinfo {author} {\bibfnamefont {J.~S.}\
  \bibnamefont {Moodera}},\ }\bibfield  {title} {\bibinfo {title}
  {High-precision realization of robust quantum anomalous {H}all state in a
  hard ferromagnetic topological insulator},\ }\href
  {https://doi.org/10.1038/nmat4204} {\bibfield  {journal} {\bibinfo  {journal}
  {Nat. Mater.}\ }\textbf {\bibinfo {volume} {14}},\ \bibinfo {pages} {473}
  (\bibinfo {year} {2015})}\BibitemShut {NoStop}%
\bibitem [{\citenamefont {Qi}\ \emph {et~al.}(2008)\citenamefont {Qi},
  \citenamefont {Hughes},\ and\ \citenamefont
  {Zhang}}]{tft-PhysRevB.78.195424}%
  \BibitemOpen
  \bibfield  {author} {\bibinfo {author} {\bibfnamefont {X.-L.}\ \bibnamefont
  {Qi}}, \bibinfo {author} {\bibfnamefont {T.~L.}\ \bibnamefont {Hughes}},\
  and\ \bibinfo {author} {\bibfnamefont {S.-C.}\ \bibnamefont {Zhang}},\
  }\bibfield  {title} {\bibinfo {title} {Topological field theory of
  time-reversal invariant insulators},\ }\href
  {https://doi.org/10.1103/PhysRevB.78.195424} {\bibfield  {journal} {\bibinfo
  {journal} {Phys. Rev. B}\ }\textbf {\bibinfo {volume} {78}},\ \bibinfo
  {pages} {195424} (\bibinfo {year} {2008})}\BibitemShut {NoStop}%
\bibitem [{\citenamefont {Li}\ \emph {et~al.}(2010)\citenamefont {Li},
  \citenamefont {Wang}, \citenamefont {Qi},\ and\ \citenamefont
  {Zhang}}]{daf-Li2010}%
  \BibitemOpen
  \bibfield  {author} {\bibinfo {author} {\bibfnamefont {R.}~\bibnamefont
  {Li}}, \bibinfo {author} {\bibfnamefont {J.}~\bibnamefont {Wang}}, \bibinfo
  {author} {\bibfnamefont {X.-L.}\ \bibnamefont {Qi}},\ and\ \bibinfo {author}
  {\bibfnamefont {S.-C.}\ \bibnamefont {Zhang}},\ }\bibfield  {title} {\bibinfo
  {title} {Dynamical axion field in topological magnetic insulators},\ }\href
  {https://doi.org/10.1038/nphys1534} {\bibfield  {journal} {\bibinfo
  {journal} {Nat. Phys.}\ }\textbf {\bibinfo {volume} {6}},\ \bibinfo {pages}
  {284} (\bibinfo {year} {2010})}\BibitemShut {NoStop}%
\bibitem [{\citenamefont {Wang}\ \emph {et~al.}(2015)\citenamefont {Wang},
  \citenamefont {Lian}, \citenamefont {Qi},\ and\ \citenamefont
  {Zhang}}]{wj-axi-PhysRevB.92.081107}%
  \BibitemOpen
  \bibfield  {author} {\bibinfo {author} {\bibfnamefont {J.}~\bibnamefont
  {Wang}}, \bibinfo {author} {\bibfnamefont {B.}~\bibnamefont {Lian}}, \bibinfo
  {author} {\bibfnamefont {X.-L.}\ \bibnamefont {Qi}},\ and\ \bibinfo {author}
  {\bibfnamefont {S.-C.}\ \bibnamefont {Zhang}},\ }\bibfield  {title} {\bibinfo
  {title} {Quantized topological magnetoelectric effect of the zero-plateau
  quantum anomalous {H}all state},\ }\href
  {https://doi.org/10.1103/PhysRevB.92.081107} {\bibfield  {journal} {\bibinfo
  {journal} {Phys. Rev. B}\ }\textbf {\bibinfo {volume} {92}},\ \bibinfo
  {pages} {081107(R)} (\bibinfo {year} {2015})}\BibitemShut {NoStop}%
\bibitem [{\citenamefont {Morimoto}\ \emph {et~al.}(2015)\citenamefont
  {Morimoto}, \citenamefont {Furusaki},\ and\ \citenamefont
  {Nagaosa}}]{nagaosa-axi-PhysRevB.92.085113}%
  \BibitemOpen
  \bibfield  {author} {\bibinfo {author} {\bibfnamefont {T.}~\bibnamefont
  {Morimoto}}, \bibinfo {author} {\bibfnamefont {A.}~\bibnamefont {Furusaki}},\
  and\ \bibinfo {author} {\bibfnamefont {N.}~\bibnamefont {Nagaosa}},\
  }\bibfield  {title} {\bibinfo {title} {Topological magnetoelectric effects in
  thin films of topological insulators},\ }\href
  {https://doi.org/10.1103/PhysRevB.92.085113} {\bibfield  {journal} {\bibinfo
  {journal} {Phys. Rev. B}\ }\textbf {\bibinfo {volume} {92}},\ \bibinfo
  {pages} {085113} (\bibinfo {year} {2015})}\BibitemShut {NoStop}%
\bibitem [{\citenamefont {Mogi}\ \emph {et~al.}(2017)\citenamefont {Mogi},
  \citenamefont {Kawamura}, \citenamefont {Yoshimi}, \citenamefont {Tsukazaki},
  \citenamefont {Kozuka}, \citenamefont {Shirakawa}, \citenamefont {Takahashi},
  \citenamefont {Kawasaki},\ and\ \citenamefont
  {Tokura}}]{tokura-axi-Mogi2017}%
  \BibitemOpen
  \bibfield  {author} {\bibinfo {author} {\bibfnamefont {M.}~\bibnamefont
  {Mogi}}, \bibinfo {author} {\bibfnamefont {M.}~\bibnamefont {Kawamura}},
  \bibinfo {author} {\bibfnamefont {R.}~\bibnamefont {Yoshimi}}, \bibinfo
  {author} {\bibfnamefont {A.}~\bibnamefont {Tsukazaki}}, \bibinfo {author}
  {\bibfnamefont {Y.}~\bibnamefont {Kozuka}}, \bibinfo {author} {\bibfnamefont
  {N.}~\bibnamefont {Shirakawa}}, \bibinfo {author} {\bibfnamefont {K.~S.}\
  \bibnamefont {Takahashi}}, \bibinfo {author} {\bibfnamefont {M.}~\bibnamefont
  {Kawasaki}},\ and\ \bibinfo {author} {\bibfnamefont {Y.}~\bibnamefont
  {Tokura}},\ }\bibfield  {title} {\bibinfo {title} {A magnetic heterostructure
  of topological insulators as a candidate for an axion insulator},\ }\href
  {https://doi.org/10.1038/nmat4855} {\bibfield  {journal} {\bibinfo  {journal}
  {Nat. Mater.}\ }\textbf {\bibinfo {volume} {16}},\ \bibinfo {pages} {516}
  (\bibinfo {year} {2017})}\BibitemShut {NoStop}%
\bibitem [{\citenamefont {Xiao}\ \emph {et~al.}(2018)\citenamefont {Xiao},
  \citenamefont {Jiang}, \citenamefont {Shin}, \citenamefont {Wang},
  \citenamefont {Wang}, \citenamefont {Zhao}, \citenamefont {Liu},
  \citenamefont {Wu}, \citenamefont {Chan}, \citenamefont {Samarth},\ and\
  \citenamefont {Chang}}]{chang-axi-PhysRevLett.120.056801}%
  \BibitemOpen
  \bibfield  {author} {\bibinfo {author} {\bibfnamefont {D.}~\bibnamefont
  {Xiao}}, \bibinfo {author} {\bibfnamefont {J.}~\bibnamefont {Jiang}},
  \bibinfo {author} {\bibfnamefont {J.-H.}\ \bibnamefont {Shin}}, \bibinfo
  {author} {\bibfnamefont {W.}~\bibnamefont {Wang}}, \bibinfo {author}
  {\bibfnamefont {F.}~\bibnamefont {Wang}}, \bibinfo {author} {\bibfnamefont
  {Y.-F.}\ \bibnamefont {Zhao}}, \bibinfo {author} {\bibfnamefont
  {C.}~\bibnamefont {Liu}}, \bibinfo {author} {\bibfnamefont {W.}~\bibnamefont
  {Wu}}, \bibinfo {author} {\bibfnamefont {M.~H.~W.}\ \bibnamefont {Chan}},
  \bibinfo {author} {\bibfnamefont {N.}~\bibnamefont {Samarth}},\ and\ \bibinfo
  {author} {\bibfnamefont {C.-Z.}\ \bibnamefont {Chang}},\ }\bibfield  {title}
  {\bibinfo {title} {Realization of the {A}xion {I}nsulator {S}tate in
  {Q}uantum {A}nomalous {H}all {S}andwich {H}eterostructures},\ }\href
  {https://doi.org/10.1103/PhysRevLett.120.056801} {\bibfield  {journal}
  {\bibinfo  {journal} {Phys. Rev. Lett.}\ }\textbf {\bibinfo {volume} {120}},\
  \bibinfo {pages} {056801} (\bibinfo {year} {2018})}\BibitemShut {NoStop}%
\bibitem [{\citenamefont {Wan}\ \emph {et~al.}(2011)\citenamefont {Wan},
  \citenamefont {Turner}, \citenamefont {Vishwanath},\ and\ \citenamefont
  {Savrasov}}]{wan-wsm-PhysRevB.83.205101}%
  \BibitemOpen
  \bibfield  {author} {\bibinfo {author} {\bibfnamefont {X.}~\bibnamefont
  {Wan}}, \bibinfo {author} {\bibfnamefont {A.~M.}\ \bibnamefont {Turner}},
  \bibinfo {author} {\bibfnamefont {A.}~\bibnamefont {Vishwanath}},\ and\
  \bibinfo {author} {\bibfnamefont {S.~Y.}\ \bibnamefont {Savrasov}},\
  }\bibfield  {title} {\bibinfo {title} {Topological semimetal and fermi-arc
  surface states in the electronic structure of pyrochlore iridates},\ }\href
  {https://doi.org/10.1103/PhysRevB.83.205101} {\bibfield  {journal} {\bibinfo
  {journal} {Phys. Rev. B}\ }\textbf {\bibinfo {volume} {83}},\ \bibinfo
  {pages} {205101} (\bibinfo {year} {2011})}\BibitemShut {NoStop}%
\bibitem [{\citenamefont {Liu}\ \emph {et~al.}(2019)\citenamefont {Liu},
  \citenamefont {Liang}, \citenamefont {Liu}, \citenamefont {Xu}, \citenamefont
  {Li}, \citenamefont {Chen}, \citenamefont {Pei}, \citenamefont {Shi},
  \citenamefont {Mo}, \citenamefont {Dudin}, \citenamefont {Kim}, \citenamefont
  {Cacho}, \citenamefont {Li}, \citenamefont {Sun}, \citenamefont {Yang},
  \citenamefont {Liu}, \citenamefont {Parkin}, \citenamefont {Felser},\ and\
  \citenamefont {Chen}}]{Co3Sn2S2-doi:10.1126/science.aav2873}%
  \BibitemOpen
  \bibfield  {author} {\bibinfo {author} {\bibfnamefont {D.~F.}\ \bibnamefont
  {Liu}}, \bibinfo {author} {\bibfnamefont {A.~J.}\ \bibnamefont {Liang}},
  \bibinfo {author} {\bibfnamefont {E.~K.}\ \bibnamefont {Liu}}, \bibinfo
  {author} {\bibfnamefont {Q.~N.}\ \bibnamefont {Xu}}, \bibinfo {author}
  {\bibfnamefont {Y.~W.}\ \bibnamefont {Li}}, \bibinfo {author} {\bibfnamefont
  {C.}~\bibnamefont {Chen}}, \bibinfo {author} {\bibfnamefont {D.}~\bibnamefont
  {Pei}}, \bibinfo {author} {\bibfnamefont {W.~J.}\ \bibnamefont {Shi}},
  \bibinfo {author} {\bibfnamefont {S.~K.}\ \bibnamefont {Mo}}, \bibinfo
  {author} {\bibfnamefont {P.}~\bibnamefont {Dudin}}, \bibinfo {author}
  {\bibfnamefont {T.}~\bibnamefont {Kim}}, \bibinfo {author} {\bibfnamefont
  {C.}~\bibnamefont {Cacho}}, \bibinfo {author} {\bibfnamefont
  {G.}~\bibnamefont {Li}}, \bibinfo {author} {\bibfnamefont {Y.}~\bibnamefont
  {Sun}}, \bibinfo {author} {\bibfnamefont {L.~X.}\ \bibnamefont {Yang}},
  \bibinfo {author} {\bibfnamefont {Z.~K.}\ \bibnamefont {Liu}}, \bibinfo
  {author} {\bibfnamefont {S.~S.~P.}\ \bibnamefont {Parkin}}, \bibinfo {author}
  {\bibfnamefont {C.}~\bibnamefont {Felser}},\ and\ \bibinfo {author}
  {\bibfnamefont {Y.~L.}\ \bibnamefont {Chen}},\ }\bibfield  {title} {\bibinfo
  {title} {Magnetic {W}eyl semimetal phase in a {K}agomé crystal},\ }\href
  {https://doi.org/10.1126/science.aav2873} {\bibfield  {journal} {\bibinfo
  {journal} {Science}\ }\textbf {\bibinfo {volume} {365}},\ \bibinfo {pages}
  {1282} (\bibinfo {year} {2019})}\BibitemShut {NoStop}%
\bibitem [{\citenamefont {Liu}\ \emph {et~al.}(2018)\citenamefont {Liu},
  \citenamefont {Sun}, \citenamefont {Kumar}, \citenamefont {Muechler},
  \citenamefont {Sun}, \citenamefont {Jiao}, \citenamefont {Yang},
  \citenamefont {Liu}, \citenamefont {Liang}, \citenamefont {Xu}, \citenamefont
  {Kroder}, \citenamefont {S{\"u}{\ss}}, \citenamefont {Borrmann},
  \citenamefont {Shekhar}, \citenamefont {Wang}, \citenamefont {Xi},
  \citenamefont {Wang}, \citenamefont {Schnelle}, \citenamefont {Wirth},
  \citenamefont {Chen}, \citenamefont {Goennenwein},\ and\ \citenamefont
  {Felser}}]{Co3Sn2S2-Liu2018}%
  \BibitemOpen
  \bibfield  {author} {\bibinfo {author} {\bibfnamefont {E.}~\bibnamefont
  {Liu}}, \bibinfo {author} {\bibfnamefont {Y.}~\bibnamefont {Sun}}, \bibinfo
  {author} {\bibfnamefont {N.}~\bibnamefont {Kumar}}, \bibinfo {author}
  {\bibfnamefont {L.}~\bibnamefont {Muechler}}, \bibinfo {author}
  {\bibfnamefont {A.}~\bibnamefont {Sun}}, \bibinfo {author} {\bibfnamefont
  {L.}~\bibnamefont {Jiao}}, \bibinfo {author} {\bibfnamefont {S.-Y.}\
  \bibnamefont {Yang}}, \bibinfo {author} {\bibfnamefont {D.}~\bibnamefont
  {Liu}}, \bibinfo {author} {\bibfnamefont {A.}~\bibnamefont {Liang}}, \bibinfo
  {author} {\bibfnamefont {Q.}~\bibnamefont {Xu}}, \bibinfo {author}
  {\bibfnamefont {J.}~\bibnamefont {Kroder}}, \bibinfo {author} {\bibfnamefont
  {V.}~\bibnamefont {S{\"u}{\ss}}}, \bibinfo {author} {\bibfnamefont
  {H.}~\bibnamefont {Borrmann}}, \bibinfo {author} {\bibfnamefont
  {C.}~\bibnamefont {Shekhar}}, \bibinfo {author} {\bibfnamefont
  {Z.}~\bibnamefont {Wang}}, \bibinfo {author} {\bibfnamefont {C.}~\bibnamefont
  {Xi}}, \bibinfo {author} {\bibfnamefont {W.}~\bibnamefont {Wang}}, \bibinfo
  {author} {\bibfnamefont {W.}~\bibnamefont {Schnelle}}, \bibinfo {author}
  {\bibfnamefont {S.}~\bibnamefont {Wirth}}, \bibinfo {author} {\bibfnamefont
  {Y.}~\bibnamefont {Chen}}, \bibinfo {author} {\bibfnamefont {S.~T.~B.}\
  \bibnamefont {Goennenwein}},\ and\ \bibinfo {author} {\bibfnamefont
  {C.}~\bibnamefont {Felser}},\ }\bibfield  {title} {\bibinfo {title} {Giant
  anomalous {H}all effect in a ferromagnetic kagome-lattice semimetal},\ }\href
  {https://doi.org/10.1038/s41567-018-0234-5} {\bibfield  {journal} {\bibinfo
  {journal} {Nat. Phys.}\ }\textbf {\bibinfo {volume} {14}},\ \bibinfo {pages}
  {1125} (\bibinfo {year} {2018})}\BibitemShut {NoStop}%
\bibitem [{\citenamefont {Xu}\ \emph {et~al.}(2019)\citenamefont {Xu},
  \citenamefont {Song}, \citenamefont {Wang}, \citenamefont {Weng},\ and\
  \citenamefont {Dai}}]{euin2as2-PhysRevLett.122.256402}%
  \BibitemOpen
  \bibfield  {author} {\bibinfo {author} {\bibfnamefont {Y.}~\bibnamefont
  {Xu}}, \bibinfo {author} {\bibfnamefont {Z.}~\bibnamefont {Song}}, \bibinfo
  {author} {\bibfnamefont {Z.}~\bibnamefont {Wang}}, \bibinfo {author}
  {\bibfnamefont {H.}~\bibnamefont {Weng}},\ and\ \bibinfo {author}
  {\bibfnamefont {X.}~\bibnamefont {Dai}},\ }\bibfield  {title} {\bibinfo
  {title} {Higher-{O}rder {T}opology of the {A}xion {I}nsulator
  {E}u{I}n${}_2${A}s${}_2$},\ }\href
  {https://doi.org/10.1103/PhysRevLett.122.256402} {\bibfield  {journal}
  {\bibinfo  {journal} {Phys. Rev. Lett.}\ }\textbf {\bibinfo {volume} {122}},\
  \bibinfo {pages} {256402} (\bibinfo {year} {2019})}\BibitemShut {NoStop}%
\bibitem [{\citenamefont {Li}\ \emph {et~al.}(2020{\natexlab{a}})\citenamefont
  {Li}, \citenamefont {Li}, \citenamefont {Li}, \citenamefont {Ye},
  \citenamefont {Zheng}, \citenamefont {Zhang}, \citenamefont {Fu},
  \citenamefont {Duan},\ and\ \citenamefont
  {Xu}}]{lifese-PhysRevLett.125.086401}%
  \BibitemOpen
  \bibfield  {author} {\bibinfo {author} {\bibfnamefont {Y.}~\bibnamefont
  {Li}}, \bibinfo {author} {\bibfnamefont {J.}~\bibnamefont {Li}}, \bibinfo
  {author} {\bibfnamefont {Y.}~\bibnamefont {Li}}, \bibinfo {author}
  {\bibfnamefont {M.}~\bibnamefont {Ye}}, \bibinfo {author} {\bibfnamefont
  {F.}~\bibnamefont {Zheng}}, \bibinfo {author} {\bibfnamefont
  {Z.}~\bibnamefont {Zhang}}, \bibinfo {author} {\bibfnamefont
  {J.}~\bibnamefont {Fu}}, \bibinfo {author} {\bibfnamefont {W.}~\bibnamefont
  {Duan}},\ and\ \bibinfo {author} {\bibfnamefont {Y.}~\bibnamefont {Xu}},\
  }\bibfield  {title} {\bibinfo {title} {{High-Temperature Quantum Anomalous
  Hall Insulators in Lithium-Decorated Iron-Based Superconductor Materials}},\
  }\href {https://doi.org/10.1103/PhysRevLett.125.086401} {\bibfield  {journal}
  {\bibinfo  {journal} {Phys. Rev. Lett.}\ }\textbf {\bibinfo {volume} {125}},\
  \bibinfo {pages} {086401} (\bibinfo {year} {2020}{\natexlab{a}})}\BibitemShut
  {NoStop}%
\bibitem [{\citenamefont {You}\ \emph {et~al.}(2019)\citenamefont {You},
  \citenamefont {Zhang}, \citenamefont {Gu},\ and\ \citenamefont
  {Su}}]{pdbr3-PhysRevApplied.12.024063}%
  \BibitemOpen
  \bibfield  {author} {\bibinfo {author} {\bibfnamefont {J.-Y.}\ \bibnamefont
  {You}}, \bibinfo {author} {\bibfnamefont {Z.}~\bibnamefont {Zhang}}, \bibinfo
  {author} {\bibfnamefont {B.}~\bibnamefont {Gu}},\ and\ \bibinfo {author}
  {\bibfnamefont {G.}~\bibnamefont {Su}},\ }\bibfield  {title} {\bibinfo
  {title} {{Two-Dimensional Room-Temperature Ferromagnetic Semiconductors with
  Quantum Anomalous Hall Effect}},\ }\href
  {https://doi.org/10.1103/PhysRevApplied.12.024063} {\bibfield  {journal}
  {\bibinfo  {journal} {Phys. Rev. Applied}\ }\textbf {\bibinfo {volume}
  {12}},\ \bibinfo {pages} {024063} (\bibinfo {year} {2019})}\BibitemShut
  {NoStop}%
\bibitem [{\citenamefont {Li}\ \emph {et~al.}(2022)\citenamefont {Li},
  \citenamefont {Han},\ and\ \citenamefont
  {Qiao}}]{qiao-PhysRevLett.129.036801}%
  \BibitemOpen
  \bibfield  {author} {\bibinfo {author} {\bibfnamefont {Z.}~\bibnamefont
  {Li}}, \bibinfo {author} {\bibfnamefont {Y.}~\bibnamefont {Han}},\ and\
  \bibinfo {author} {\bibfnamefont {Z.}~\bibnamefont {Qiao}},\ }\bibfield
  {title} {\bibinfo {title} {{Chern Number Tunable Quantum Anomalous Hall
  Effect in Monolayer Transitional Metal Oxides via Manipulating Magnetization
  Orientation}},\ }\href {https://doi.org/10.1103/PhysRevLett.129.036801}
  {\bibfield  {journal} {\bibinfo  {journal} {Phys. Rev. Lett.}\ }\textbf
  {\bibinfo {volume} {129}},\ \bibinfo {pages} {036801} (\bibinfo {year}
  {2022})}\BibitemShut {NoStop}%
\bibitem [{\citenamefont {Gong}\ \emph {et~al.}(2019)\citenamefont {Gong},
  \citenamefont {Guo}, \citenamefont {Li}, \citenamefont {Zhu}, \citenamefont
  {Liao}, \citenamefont {Liu}, \citenamefont {Zhang}, \citenamefont {Gu},
  \citenamefont {Tang}, \citenamefont {Feng}, \citenamefont {Zhang},
  \citenamefont {Li}, \citenamefont {Song}, \citenamefont {Wang}, \citenamefont
  {Yu}, \citenamefont {Chen}, \citenamefont {Wang}, \citenamefont {Yao},
  \citenamefont {Duan}, \citenamefont {Xu}, \citenamefont {Zhang},
  \citenamefont {Ma}, \citenamefont {Xue},\ and\ \citenamefont
  {He}}]{mbt-exp-Yan}%
  \BibitemOpen
  \bibfield  {author} {\bibinfo {author} {\bibfnamefont {Y.}~\bibnamefont
  {Gong}}, \bibinfo {author} {\bibfnamefont {J.}~\bibnamefont {Guo}}, \bibinfo
  {author} {\bibfnamefont {J.}~\bibnamefont {Li}}, \bibinfo {author}
  {\bibfnamefont {K.}~\bibnamefont {Zhu}}, \bibinfo {author} {\bibfnamefont
  {M.}~\bibnamefont {Liao}}, \bibinfo {author} {\bibfnamefont {X.}~\bibnamefont
  {Liu}}, \bibinfo {author} {\bibfnamefont {Q.}~\bibnamefont {Zhang}}, \bibinfo
  {author} {\bibfnamefont {L.}~\bibnamefont {Gu}}, \bibinfo {author}
  {\bibfnamefont {L.}~\bibnamefont {Tang}}, \bibinfo {author} {\bibfnamefont
  {X.}~\bibnamefont {Feng}}, \bibinfo {author} {\bibfnamefont {D.}~\bibnamefont
  {Zhang}}, \bibinfo {author} {\bibfnamefont {W.}~\bibnamefont {Li}}, \bibinfo
  {author} {\bibfnamefont {C.}~\bibnamefont {Song}}, \bibinfo {author}
  {\bibfnamefont {L.}~\bibnamefont {Wang}}, \bibinfo {author} {\bibfnamefont
  {P.}~\bibnamefont {Yu}}, \bibinfo {author} {\bibfnamefont {X.}~\bibnamefont
  {Chen}}, \bibinfo {author} {\bibfnamefont {Y.}~\bibnamefont {Wang}}, \bibinfo
  {author} {\bibfnamefont {H.}~\bibnamefont {Yao}}, \bibinfo {author}
  {\bibfnamefont {W.}~\bibnamefont {Duan}}, \bibinfo {author} {\bibfnamefont
  {Y.}~\bibnamefont {Xu}}, \bibinfo {author} {\bibfnamefont {S.-C.}\
  \bibnamefont {Zhang}}, \bibinfo {author} {\bibfnamefont {X.}~\bibnamefont
  {Ma}}, \bibinfo {author} {\bibfnamefont {Q.-K.}\ \bibnamefont {Xue}},\ and\
  \bibinfo {author} {\bibfnamefont {K.}~\bibnamefont {He}},\ }\bibfield
  {title} {\bibinfo {title} {Experimental {R}ealization of an {I}ntrinsic
  {M}agnetic {T}opological {I}nsulator},\ }\href
  {https://doi.org/10.1088/0256-307X/36/7/076801} {\bibfield  {journal}
  {\bibinfo  {journal} {Chin. Phys. Lett.}\ }\textbf {\bibinfo {volume} {36}},\
  \bibinfo {eid} {076801} (\bibinfo {year} {2019})}\BibitemShut {NoStop}%
\bibitem [{\citenamefont {Otrokov}\ \emph
  {et~al.}(2019{\natexlab{a}})\citenamefont {Otrokov}, \citenamefont
  {Klimovskikh}, \citenamefont {Bentmann}, \citenamefont {Estyunin},
  \citenamefont {Zeugner}, \citenamefont {Aliev}, \citenamefont {Ga{\ss}},
  \citenamefont {Wolter}, \citenamefont {Koroleva}, \citenamefont {Shikin},
  \citenamefont {Blanco-Rey}, \citenamefont {Hoffmann}, \citenamefont
  {Rusinov}, \citenamefont {Vyazovskaya}, \citenamefont {Eremeev},
  \citenamefont {Koroteev}, \citenamefont {Kuznetsov}, \citenamefont {Freyse},
  \citenamefont {S{\'a}nchez-Barriga}, \citenamefont {Amiraslanov},
  \citenamefont {Babanly}, \citenamefont {Mamedov}, \citenamefont {Abdullayev},
  \citenamefont {Zverev}, \citenamefont {Alfonsov}, \citenamefont {Kataev},
  \citenamefont {B{\"u}chner}, \citenamefont {Schwier}, \citenamefont {Kumar},
  \citenamefont {Kimura}, \citenamefont {Petaccia}, \citenamefont {Di~Santo},
  \citenamefont {Vidal}, \citenamefont {Schatz}, \citenamefont {Ki{\ss}ner},
  \citenamefont {{\"U}nzelmann}, \citenamefont {Min}, \citenamefont {Moser},
  \citenamefont {Peixoto}, \citenamefont {Reinert}, \citenamefont {Ernst},
  \citenamefont {Echenique}, \citenamefont {Isaeva},\ and\ \citenamefont
  {Chulkov}}]{mbt-exp-Otrokov2019}%
  \BibitemOpen
  \bibfield  {author} {\bibinfo {author} {\bibfnamefont {M.~M.}\ \bibnamefont
  {Otrokov}}, \bibinfo {author} {\bibfnamefont {I.~I.}\ \bibnamefont
  {Klimovskikh}}, \bibinfo {author} {\bibfnamefont {H.}~\bibnamefont
  {Bentmann}}, \bibinfo {author} {\bibfnamefont {D.}~\bibnamefont {Estyunin}},
  \bibinfo {author} {\bibfnamefont {A.}~\bibnamefont {Zeugner}}, \bibinfo
  {author} {\bibfnamefont {Z.~S.}\ \bibnamefont {Aliev}}, \bibinfo {author}
  {\bibfnamefont {S.}~\bibnamefont {Ga{\ss}}}, \bibinfo {author} {\bibfnamefont
  {A.~U.~B.}\ \bibnamefont {Wolter}}, \bibinfo {author} {\bibfnamefont {A.~V.}\
  \bibnamefont {Koroleva}}, \bibinfo {author} {\bibfnamefont {A.~M.}\
  \bibnamefont {Shikin}}, \bibinfo {author} {\bibfnamefont {M.}~\bibnamefont
  {Blanco-Rey}}, \bibinfo {author} {\bibfnamefont {M.}~\bibnamefont
  {Hoffmann}}, \bibinfo {author} {\bibfnamefont {I.~P.}\ \bibnamefont
  {Rusinov}}, \bibinfo {author} {\bibfnamefont {A.~Y.}\ \bibnamefont
  {Vyazovskaya}}, \bibinfo {author} {\bibfnamefont {S.~V.}\ \bibnamefont
  {Eremeev}}, \bibinfo {author} {\bibfnamefont {Y.~M.}\ \bibnamefont
  {Koroteev}}, \bibinfo {author} {\bibfnamefont {V.~M.}\ \bibnamefont
  {Kuznetsov}}, \bibinfo {author} {\bibfnamefont {F.}~\bibnamefont {Freyse}},
  \bibinfo {author} {\bibfnamefont {J.}~\bibnamefont {S{\'a}nchez-Barriga}},
  \bibinfo {author} {\bibfnamefont {I.~R.}\ \bibnamefont {Amiraslanov}},
  \bibinfo {author} {\bibfnamefont {M.~B.}\ \bibnamefont {Babanly}}, \bibinfo
  {author} {\bibfnamefont {N.~T.}\ \bibnamefont {Mamedov}}, \bibinfo {author}
  {\bibfnamefont {N.~A.}\ \bibnamefont {Abdullayev}}, \bibinfo {author}
  {\bibfnamefont {V.~N.}\ \bibnamefont {Zverev}}, \bibinfo {author}
  {\bibfnamefont {A.}~\bibnamefont {Alfonsov}}, \bibinfo {author}
  {\bibfnamefont {V.}~\bibnamefont {Kataev}}, \bibinfo {author} {\bibfnamefont
  {B.}~\bibnamefont {B{\"u}chner}}, \bibinfo {author} {\bibfnamefont {E.~F.}\
  \bibnamefont {Schwier}}, \bibinfo {author} {\bibfnamefont {S.}~\bibnamefont
  {Kumar}}, \bibinfo {author} {\bibfnamefont {A.}~\bibnamefont {Kimura}},
  \bibinfo {author} {\bibfnamefont {L.}~\bibnamefont {Petaccia}}, \bibinfo
  {author} {\bibfnamefont {G.}~\bibnamefont {Di~Santo}}, \bibinfo {author}
  {\bibfnamefont {R.~C.}\ \bibnamefont {Vidal}}, \bibinfo {author}
  {\bibfnamefont {S.}~\bibnamefont {Schatz}}, \bibinfo {author} {\bibfnamefont
  {K.}~\bibnamefont {Ki{\ss}ner}}, \bibinfo {author} {\bibfnamefont
  {M.}~\bibnamefont {{\"U}nzelmann}}, \bibinfo {author} {\bibfnamefont {C.~H.}\
  \bibnamefont {Min}}, \bibinfo {author} {\bibfnamefont {S.}~\bibnamefont
  {Moser}}, \bibinfo {author} {\bibfnamefont {T.~R.~F.}\ \bibnamefont
  {Peixoto}}, \bibinfo {author} {\bibfnamefont {F.}~\bibnamefont {Reinert}},
  \bibinfo {author} {\bibfnamefont {A.}~\bibnamefont {Ernst}}, \bibinfo
  {author} {\bibfnamefont {P.~M.}\ \bibnamefont {Echenique}}, \bibinfo {author}
  {\bibfnamefont {A.}~\bibnamefont {Isaeva}},\ and\ \bibinfo {author}
  {\bibfnamefont {E.~V.}\ \bibnamefont {Chulkov}},\ }\bibfield  {title}
  {\bibinfo {title} {Prediction and observation of an antiferromagnetic
  topological insulator},\ }\href {https://doi.org/10.1038/s41586-019-1840-9}
  {\bibfield  {journal} {\bibinfo  {journal} {Nature}\ }\textbf {\bibinfo
  {volume} {576}},\ \bibinfo {pages} {416} (\bibinfo {year}
  {2019}{\natexlab{a}})}\BibitemShut {NoStop}%
\bibitem [{\citenamefont {Li}\ \emph {et~al.}(2019{\natexlab{a}})\citenamefont
  {Li}, \citenamefont {Li}, \citenamefont {Du}, \citenamefont {Wang},
  \citenamefont {Gu}, \citenamefont {Zhang}, \citenamefont {He}, \citenamefont
  {Duan},\ and\ \citenamefont {Xu}}]{mbt-sciadv-doi:10.1126/sciadv.aaw5685}%
  \BibitemOpen
  \bibfield  {author} {\bibinfo {author} {\bibfnamefont {J.}~\bibnamefont
  {Li}}, \bibinfo {author} {\bibfnamefont {Y.}~\bibnamefont {Li}}, \bibinfo
  {author} {\bibfnamefont {S.}~\bibnamefont {Du}}, \bibinfo {author}
  {\bibfnamefont {Z.}~\bibnamefont {Wang}}, \bibinfo {author} {\bibfnamefont
  {B.-L.}\ \bibnamefont {Gu}}, \bibinfo {author} {\bibfnamefont {S.-C.}\
  \bibnamefont {Zhang}}, \bibinfo {author} {\bibfnamefont {K.}~\bibnamefont
  {He}}, \bibinfo {author} {\bibfnamefont {W.}~\bibnamefont {Duan}},\ and\
  \bibinfo {author} {\bibfnamefont {Y.}~\bibnamefont {Xu}},\ }\bibfield
  {title} {\bibinfo {title} {Intrinsic magnetic topological insulators in van
  der {W}aals layered {M}n{B}i${}_2${T}e${}_4$-family materials},\ }\href
  {https://doi.org/10.1126/sciadv.aaw5685} {\bibfield  {journal} {\bibinfo
  {journal} {Sci. Adv.}\ }\textbf {\bibinfo {volume} {5}},\ \bibinfo {pages}
  {eaaw5685} (\bibinfo {year} {2019}{\natexlab{a}})}\BibitemShut {NoStop}%
\bibitem [{\citenamefont {Li}\ \emph {et~al.}(2019{\natexlab{b}})\citenamefont
  {Li}, \citenamefont {Wang}, \citenamefont {Zhang}, \citenamefont {Gu},
  \citenamefont {Duan},\ and\ \citenamefont
  {Xu}}]{ljh-maggap-PhysRevB.100.121103}%
  \BibitemOpen
  \bibfield  {author} {\bibinfo {author} {\bibfnamefont {J.}~\bibnamefont
  {Li}}, \bibinfo {author} {\bibfnamefont {C.}~\bibnamefont {Wang}}, \bibinfo
  {author} {\bibfnamefont {Z.}~\bibnamefont {Zhang}}, \bibinfo {author}
  {\bibfnamefont {B.-L.}\ \bibnamefont {Gu}}, \bibinfo {author} {\bibfnamefont
  {W.}~\bibnamefont {Duan}},\ and\ \bibinfo {author} {\bibfnamefont
  {Y.}~\bibnamefont {Xu}},\ }\bibfield  {title} {\bibinfo {title}
  {{Magnetically controllable topological quantum phase transitions in the
  antiferromagnetic topological insulator
  ${\mathrm{MnBi}}_{2}{\mathrm{Te}}_{4}$}},\ }\href
  {https://doi.org/10.1103/PhysRevB.100.121103} {\bibfield  {journal} {\bibinfo
   {journal} {Phys. Rev. B}\ }\textbf {\bibinfo {volume} {100}},\ \bibinfo
  {pages} {121103(R)} (\bibinfo {year} {2019}{\natexlab{b}})}\BibitemShut
  {NoStop}%
\bibitem [{\citenamefont {Deng}\ \emph {et~al.}(2020)\citenamefont {Deng},
  \citenamefont {Yu}, \citenamefont {Shi}, \citenamefont {Guo}, \citenamefont
  {Xu}, \citenamefont {Wang}, \citenamefont {Chen},\ and\ \citenamefont
  {Zhang}}]{mbt-exp-zyb-doi:10.1126/science.aax8156}%
  \BibitemOpen
  \bibfield  {author} {\bibinfo {author} {\bibfnamefont {Y.}~\bibnamefont
  {Deng}}, \bibinfo {author} {\bibfnamefont {Y.}~\bibnamefont {Yu}}, \bibinfo
  {author} {\bibfnamefont {M.~Z.}\ \bibnamefont {Shi}}, \bibinfo {author}
  {\bibfnamefont {Z.}~\bibnamefont {Guo}}, \bibinfo {author} {\bibfnamefont
  {Z.}~\bibnamefont {Xu}}, \bibinfo {author} {\bibfnamefont {J.}~\bibnamefont
  {Wang}}, \bibinfo {author} {\bibfnamefont {X.~H.}\ \bibnamefont {Chen}},\
  and\ \bibinfo {author} {\bibfnamefont {Y.}~\bibnamefont {Zhang}},\ }\bibfield
   {title} {\bibinfo {title} {Quantum anomalous {H}all effect in intrinsic
  magnetic topological insulator {M}n{B}i${}_2${T}e${}_4$},\ }\href
  {https://doi.org/10.1126/science.aax8156} {\bibfield  {journal} {\bibinfo
  {journal} {Science}\ }\textbf {\bibinfo {volume} {367}},\ \bibinfo {pages}
  {895} (\bibinfo {year} {2020})}\BibitemShut {NoStop}%
\bibitem [{\citenamefont {Ge}\ \emph {et~al.}(2020)\citenamefont {Ge},
  \citenamefont {Liu}, \citenamefont {Li}, \citenamefont {Li}, \citenamefont
  {Luo}, \citenamefont {Wu}, \citenamefont {Xu},\ and\ \citenamefont
  {Wang}}]{nsr-10.1093/nsr/nwaa089}%
  \BibitemOpen
  \bibfield  {author} {\bibinfo {author} {\bibfnamefont {J.}~\bibnamefont
  {Ge}}, \bibinfo {author} {\bibfnamefont {Y.}~\bibnamefont {Liu}}, \bibinfo
  {author} {\bibfnamefont {J.}~\bibnamefont {Li}}, \bibinfo {author}
  {\bibfnamefont {H.}~\bibnamefont {Li}}, \bibinfo {author} {\bibfnamefont
  {T.}~\bibnamefont {Luo}}, \bibinfo {author} {\bibfnamefont {Y.}~\bibnamefont
  {Wu}}, \bibinfo {author} {\bibfnamefont {Y.}~\bibnamefont {Xu}},\ and\
  \bibinfo {author} {\bibfnamefont {J.}~\bibnamefont {Wang}},\ }\bibfield
  {title} {\bibinfo {title} {{High-Chern-number and high-temperature quantum
  Hall effect without Landau levels}},\ }\href
  {https://doi.org/10.1093/nsr/nwaa089} {\bibfield  {journal} {\bibinfo
  {journal} {Natl. Sci. Rev.}\ }\textbf {\bibinfo {volume} {7}},\ \bibinfo
  {pages} {1280} (\bibinfo {year} {2020})}\BibitemShut {NoStop}%
\bibitem [{\citenamefont {Otrokov}\ \emph
  {et~al.}(2019{\natexlab{b}})\citenamefont {Otrokov}, \citenamefont {Rusinov},
  \citenamefont {Blanco-Rey}, \citenamefont {Hoffmann}, \citenamefont
  {Vyazovskaya}, \citenamefont {Eremeev}, \citenamefont {Ernst}, \citenamefont
  {Echenique}, \citenamefont {Arnau},\ and\ \citenamefont
  {Chulkov}}]{mbt-otrokov-PhysRevLett.122.107202}%
  \BibitemOpen
  \bibfield  {author} {\bibinfo {author} {\bibfnamefont {M.~M.}\ \bibnamefont
  {Otrokov}}, \bibinfo {author} {\bibfnamefont {I.~P.}\ \bibnamefont
  {Rusinov}}, \bibinfo {author} {\bibfnamefont {M.}~\bibnamefont {Blanco-Rey}},
  \bibinfo {author} {\bibfnamefont {M.}~\bibnamefont {Hoffmann}}, \bibinfo
  {author} {\bibfnamefont {A.~Y.}\ \bibnamefont {Vyazovskaya}}, \bibinfo
  {author} {\bibfnamefont {S.~V.}\ \bibnamefont {Eremeev}}, \bibinfo {author}
  {\bibfnamefont {A.}~\bibnamefont {Ernst}}, \bibinfo {author} {\bibfnamefont
  {P.~M.}\ \bibnamefont {Echenique}}, \bibinfo {author} {\bibfnamefont
  {A.}~\bibnamefont {Arnau}},\ and\ \bibinfo {author} {\bibfnamefont {E.~V.}\
  \bibnamefont {Chulkov}},\ }\bibfield  {title} {\bibinfo {title} {Unique
  {T}hickness-{D}ependent {P}roperties of the van der {W}aals {I}nterlayer
  {A}ntiferromagnet {M}n{B}i${}_2${T}e${}_4$ {F}ilms},\ }\href
  {https://doi.org/10.1103/PhysRevLett.122.107202} {\bibfield  {journal}
  {\bibinfo  {journal} {Phys. Rev. Lett.}\ }\textbf {\bibinfo {volume} {122}},\
  \bibinfo {pages} {107202} (\bibinfo {year} {2019}{\natexlab{b}})}\BibitemShut
  {NoStop}%
\bibitem [{\citenamefont {Zhang}\ \emph {et~al.}(2019)\citenamefont {Zhang},
  \citenamefont {Shi}, \citenamefont {Zhu}, \citenamefont {Xing}, \citenamefont
  {Zhang},\ and\ \citenamefont {Wang}}]{mbt-wj-PhysRevLett.122.206401}%
  \BibitemOpen
  \bibfield  {author} {\bibinfo {author} {\bibfnamefont {D.}~\bibnamefont
  {Zhang}}, \bibinfo {author} {\bibfnamefont {M.}~\bibnamefont {Shi}}, \bibinfo
  {author} {\bibfnamefont {T.}~\bibnamefont {Zhu}}, \bibinfo {author}
  {\bibfnamefont {D.}~\bibnamefont {Xing}}, \bibinfo {author} {\bibfnamefont
  {H.}~\bibnamefont {Zhang}},\ and\ \bibinfo {author} {\bibfnamefont
  {J.}~\bibnamefont {Wang}},\ }\bibfield  {title} {\bibinfo {title}
  {Topological {A}xion {S}tates in the {M}agnetic {I}nsulator
  {M}n{B}i${}_2${T}e${}_4$ with the {Q}uantized {M}agnetoelectric {E}ffect},\
  }\href {https://doi.org/10.1103/PhysRevLett.122.206401} {\bibfield  {journal}
  {\bibinfo  {journal} {Phys. Rev. Lett.}\ }\textbf {\bibinfo {volume} {122}},\
  \bibinfo {pages} {206401} (\bibinfo {year} {2019})}\BibitemShut {NoStop}%
\bibitem [{\citenamefont {Liu}\ \emph {et~al.}(2020)\citenamefont {Liu},
  \citenamefont {Wang}, \citenamefont {Li}, \citenamefont {Wu}, \citenamefont
  {Li}, \citenamefont {Li}, \citenamefont {He}, \citenamefont {Xu},
  \citenamefont {Zhang},\ and\ \citenamefont {Wang}}]{mbt-exp-Liu2020}%
  \BibitemOpen
  \bibfield  {author} {\bibinfo {author} {\bibfnamefont {C.}~\bibnamefont
  {Liu}}, \bibinfo {author} {\bibfnamefont {Y.}~\bibnamefont {Wang}}, \bibinfo
  {author} {\bibfnamefont {H.}~\bibnamefont {Li}}, \bibinfo {author}
  {\bibfnamefont {Y.}~\bibnamefont {Wu}}, \bibinfo {author} {\bibfnamefont
  {Y.}~\bibnamefont {Li}}, \bibinfo {author} {\bibfnamefont {J.}~\bibnamefont
  {Li}}, \bibinfo {author} {\bibfnamefont {K.}~\bibnamefont {He}}, \bibinfo
  {author} {\bibfnamefont {Y.}~\bibnamefont {Xu}}, \bibinfo {author}
  {\bibfnamefont {J.}~\bibnamefont {Zhang}},\ and\ \bibinfo {author}
  {\bibfnamefont {Y.}~\bibnamefont {Wang}},\ }\bibfield  {title} {\bibinfo
  {title} {Robust axion insulator and {C}hern insulator phases in a
  two-dimensional antiferromagnetic topological insulator},\ }\href
  {https://doi.org/10.1038/s41563-019-0573-3} {\bibfield  {journal} {\bibinfo
  {journal} {Nat. Mater.}\ }\textbf {\bibinfo {volume} {19}},\ \bibinfo {pages}
  {522} (\bibinfo {year} {2020})}\BibitemShut {NoStop}%
\bibitem [{\citenamefont {Mermin}\ and\ \citenamefont
  {Wagner}(1966)}]{mermin-PhysRevLett.17.1133}%
  \BibitemOpen
  \bibfield  {author} {\bibinfo {author} {\bibfnamefont {N.~D.}\ \bibnamefont
  {Mermin}}\ and\ \bibinfo {author} {\bibfnamefont {H.}~\bibnamefont
  {Wagner}},\ }\bibfield  {title} {\bibinfo {title} {{Absence of Ferromagnetism
  or Antiferromagnetism in One- or Two-Dimensional Isotropic Heisenberg
  Models}},\ }\href {https://doi.org/10.1103/PhysRevLett.17.1133} {\bibfield
  {journal} {\bibinfo  {journal} {Phys. Rev. Lett.}\ }\textbf {\bibinfo
  {volume} {17}},\ \bibinfo {pages} {1133} (\bibinfo {year}
  {1966})}\BibitemShut {NoStop}%
\bibitem [{\citenamefont {Huang}\ \emph {et~al.}(1994)\citenamefont {Huang},
  \citenamefont {Kief}, \citenamefont {Mankey},\ and\ \citenamefont
  {Willis}}]{fewlayers-PhysRevB.49.3962}%
  \BibitemOpen
  \bibfield  {author} {\bibinfo {author} {\bibfnamefont {F.}~\bibnamefont
  {Huang}}, \bibinfo {author} {\bibfnamefont {M.~T.}\ \bibnamefont {Kief}},
  \bibinfo {author} {\bibfnamefont {G.~J.}\ \bibnamefont {Mankey}},\ and\
  \bibinfo {author} {\bibfnamefont {R.~F.}\ \bibnamefont {Willis}},\ }\bibfield
   {title} {\bibinfo {title} {{Magnetism in the few-monolayers limit: A surface
  magneto-optic Kerr-effect study of the magnetic behavior of ultrathin films
  of Co, Ni, and Co-Ni alloys on Cu(100) and Cu(111)}},\ }\href
  {https://doi.org/10.1103/PhysRevB.49.3962} {\bibfield  {journal} {\bibinfo
  {journal} {Phys. Rev. B}\ }\textbf {\bibinfo {volume} {49}},\ \bibinfo
  {pages} {3962} (\bibinfo {year} {1994})}\BibitemShut {NoStop}%
\bibitem [{\citenamefont {Zhang}\ \emph {et~al.}(2020)\citenamefont {Zhang},
  \citenamefont {Wang}, \citenamefont {Shi}, \citenamefont {Zhu}, \citenamefont
  {Zhang},\ and\ \citenamefont {Wang}}]{225cpl-Zhang_2020}%
  \BibitemOpen
  \bibfield  {author} {\bibinfo {author} {\bibfnamefont {J.}~\bibnamefont
  {Zhang}}, \bibinfo {author} {\bibfnamefont {D.}~\bibnamefont {Wang}},
  \bibinfo {author} {\bibfnamefont {M.}~\bibnamefont {Shi}}, \bibinfo {author}
  {\bibfnamefont {T.}~\bibnamefont {Zhu}}, \bibinfo {author} {\bibfnamefont
  {H.}~\bibnamefont {Zhang}},\ and\ \bibinfo {author} {\bibfnamefont
  {J.}~\bibnamefont {Wang}},\ }\bibfield  {title} {\bibinfo {title} {{Large
  Dynamical Axion Field in Topological Antiferromagnetic Insulator
  {M}n${}_2${B}i${}_2${T}e${}_5$}},\ }\href
  {https://doi.org/10.1088/0256-307x/37/7/077304} {\bibfield  {journal}
  {\bibinfo  {journal} {Chin. Phys. Lett.}\ }\textbf {\bibinfo {volume} {37}},\
  \bibinfo {pages} {077304} (\bibinfo {year} {2020})}\BibitemShut {NoStop}%
\bibitem [{\citenamefont {Li}\ \emph {et~al.}(2020{\natexlab{b}})\citenamefont
  {Li}, \citenamefont {Jiang}, \citenamefont {Zhang}, \citenamefont {Liu},
  \citenamefont {Yang},\ and\ \citenamefont
  {Wang}}]{225prb-PhysRevB.102.121107}%
  \BibitemOpen
  \bibfield  {author} {\bibinfo {author} {\bibfnamefont {Y.}~\bibnamefont
  {Li}}, \bibinfo {author} {\bibfnamefont {Y.}~\bibnamefont {Jiang}}, \bibinfo
  {author} {\bibfnamefont {J.}~\bibnamefont {Zhang}}, \bibinfo {author}
  {\bibfnamefont {Z.}~\bibnamefont {Liu}}, \bibinfo {author} {\bibfnamefont
  {Z.}~\bibnamefont {Yang}},\ and\ \bibinfo {author} {\bibfnamefont
  {J.}~\bibnamefont {Wang}},\ }\bibfield  {title} {\bibinfo {title} {{Intrinsic
  topological phases in ${\mathrm{Mn}}_{2}{\mathrm{Bi}}_{2}{\mathrm{Te}}_{5}$
  tuned by the layer magnetization}},\ }\href
  {https://doi.org/10.1103/PhysRevB.102.121107} {\bibfield  {journal} {\bibinfo
   {journal} {Phys. Rev. B}\ }\textbf {\bibinfo {volume} {102}},\ \bibinfo
  {pages} {121107(R)} (\bibinfo {year} {2020}{\natexlab{b}})}\BibitemShut
  {NoStop}%
\bibitem [{\citenamefont {Eremeev}\ \emph {et~al.}(2022)\citenamefont
  {Eremeev}, \citenamefont {Otrokov}, \citenamefont {Ernst},\ and\
  \citenamefont {Chulkov}}]{otrokov-prb-PhysRevB.105.195105}%
  \BibitemOpen
  \bibfield  {author} {\bibinfo {author} {\bibfnamefont {S.~V.}\ \bibnamefont
  {Eremeev}}, \bibinfo {author} {\bibfnamefont {M.~M.}\ \bibnamefont
  {Otrokov}}, \bibinfo {author} {\bibfnamefont {A.}~\bibnamefont {Ernst}},\
  and\ \bibinfo {author} {\bibfnamefont {E.~V.}\ \bibnamefont {Chulkov}},\
  }\bibfield  {title} {\bibinfo {title} {{Magnetic ordering and topology in
  ${\mathrm{Mn}}_{2}{\mathrm{Bi}}_{2}{\mathrm{Te}}_{5}$ and
  ${\mathrm{Mn}}_{2}{\mathrm{Sb}}_{2}{\mathrm{Te}}_{5}$ van der Waals
  materials}},\ }\href {https://doi.org/10.1103/PhysRevB.105.195105} {\bibfield
   {journal} {\bibinfo  {journal} {Phys. Rev. B}\ }\textbf {\bibinfo {volume}
  {105}},\ \bibinfo {pages} {195105} (\bibinfo {year} {2022})}\BibitemShut
  {NoStop}%
\bibitem [{\citenamefont {Cao}\ \emph {et~al.}(2021)\citenamefont {Cao},
  \citenamefont {Han}, \citenamefont {Lv}, \citenamefont {Wang}, \citenamefont
  {Luo}, \citenamefont {Zhang}, \citenamefont {Yao}, \citenamefont {Zhou},
  \citenamefont {Chen}, \citenamefont {Zhang},\ and\ \citenamefont
  {Chen}}]{225exp-PhysRevB.104.054421}%
  \BibitemOpen
  \bibfield  {author} {\bibinfo {author} {\bibfnamefont {L.}~\bibnamefont
  {Cao}}, \bibinfo {author} {\bibfnamefont {S.}~\bibnamefont {Han}}, \bibinfo
  {author} {\bibfnamefont {Y.-Y.}\ \bibnamefont {Lv}}, \bibinfo {author}
  {\bibfnamefont {D.}~\bibnamefont {Wang}}, \bibinfo {author} {\bibfnamefont
  {Y.-C.}\ \bibnamefont {Luo}}, \bibinfo {author} {\bibfnamefont {Y.-Y.}\
  \bibnamefont {Zhang}}, \bibinfo {author} {\bibfnamefont {S.-H.}\ \bibnamefont
  {Yao}}, \bibinfo {author} {\bibfnamefont {J.}~\bibnamefont {Zhou}}, \bibinfo
  {author} {\bibfnamefont {Y.~B.}\ \bibnamefont {Chen}}, \bibinfo {author}
  {\bibfnamefont {H.}~\bibnamefont {Zhang}},\ and\ \bibinfo {author}
  {\bibfnamefont {Y.-F.}\ \bibnamefont {Chen}},\ }\bibfield  {title} {\bibinfo
  {title} {{Growth and characterization of the dynamical axion insulator
  candidate ${\mathrm{Mn}}_{2}{\mathrm{Bi}}_{2}{\mathrm{Te}}_{5}$ with
  intrinsic antiferromagnetism}},\ }\href
  {https://doi.org/10.1103/PhysRevB.104.054421} {\bibfield  {journal} {\bibinfo
   {journal} {Phys. Rev. B}\ }\textbf {\bibinfo {volume} {104}},\ \bibinfo
  {pages} {054421} (\bibinfo {year} {2021})}\BibitemShut {NoStop}%
\bibitem [{\citenamefont {Zhang}\ \emph {et~al.}(2022)\citenamefont {Zhang},
  \citenamefont {Zhang}, \citenamefont {Zhang}, \citenamefont {Yang},
  \citenamefont {Wang}, \citenamefont {Xu},\ and\ \citenamefont
  {Liu}}]{nanoscale}%
  \BibitemOpen
  \bibfield  {author} {\bibinfo {author} {\bibfnamefont {H.}~\bibnamefont
  {Zhang}}, \bibinfo {author} {\bibfnamefont {J.}~\bibnamefont {Zhang}},
  \bibinfo {author} {\bibfnamefont {Y.}~\bibnamefont {Zhang}}, \bibinfo
  {author} {\bibfnamefont {W.}~\bibnamefont {Yang}}, \bibinfo {author}
  {\bibfnamefont {Y.}~\bibnamefont {Wang}}, \bibinfo {author} {\bibfnamefont
  {X.}~\bibnamefont {Xu}},\ and\ \bibinfo {author} {\bibfnamefont
  {F.}~\bibnamefont {Liu}},\ }\bibfield  {title} {\bibinfo {title} {{A generic
  dual d-band model for interlayer ferromagnetic coupling in a transition-metal
  doped MnBi${}_2$Te${}_4$ family of materials}},\ }\href
  {https://doi.org/10.1039/D2NR03283J} {\bibfield  {journal} {\bibinfo
  {journal} {Nanoscale}\ }\textbf {\bibinfo {volume} {14}},\ \bibinfo {pages}
  {13689} (\bibinfo {year} {2022})}\BibitemShut {NoStop}%
\bibitem [{\citenamefont {Kresse}\ and\ \citenamefont
  {Furthm\"uller}(1996)}]{vasp-PhysRevB.54.11169}%
  \BibitemOpen
  \bibfield  {author} {\bibinfo {author} {\bibfnamefont {G.}~\bibnamefont
  {Kresse}}\ and\ \bibinfo {author} {\bibfnamefont {J.}~\bibnamefont
  {Furthm\"uller}},\ }\bibfield  {title} {\bibinfo {title} {Efficient iterative
  schemes for ab initio total-energy calculations using a plane-wave basis
  set},\ }\href {https://doi.org/10.1103/PhysRevB.54.11169} {\bibfield
  {journal} {\bibinfo  {journal} {Phys. Rev. B}\ }\textbf {\bibinfo {volume}
  {54}},\ \bibinfo {pages} {11169} (\bibinfo {year} {1996})}\BibitemShut
  {NoStop}%
\bibitem [{\citenamefont {Perdew}\ \emph {et~al.}(1996)\citenamefont {Perdew},
  \citenamefont {Burke},\ and\ \citenamefont
  {Ernzerhof}}]{gga-PhysRevLett.77.3865}%
  \BibitemOpen
  \bibfield  {author} {\bibinfo {author} {\bibfnamefont {J.~P.}\ \bibnamefont
  {Perdew}}, \bibinfo {author} {\bibfnamefont {K.}~\bibnamefont {Burke}},\ and\
  \bibinfo {author} {\bibfnamefont {M.}~\bibnamefont {Ernzerhof}},\ }\bibfield
  {title} {\bibinfo {title} {{Generalized Gradient Approximation Made
  Simple}},\ }\href {https://doi.org/10.1103/PhysRevLett.77.3865} {\bibfield
  {journal} {\bibinfo  {journal} {Phys. Rev. Lett.}\ }\textbf {\bibinfo
  {volume} {77}},\ \bibinfo {pages} {3865} (\bibinfo {year}
  {1996})}\BibitemShut {NoStop}%
\bibitem [{\citenamefont {Becke}\ and\ \citenamefont
  {Johnson}(2006)}]{mbj-doi:10.1063/1.2213970}%
  \BibitemOpen
  \bibfield  {author} {\bibinfo {author} {\bibfnamefont {A.~D.}\ \bibnamefont
  {Becke}}\ and\ \bibinfo {author} {\bibfnamefont {E.~R.}\ \bibnamefont
  {Johnson}},\ }\bibfield  {title} {\bibinfo {title} {A simple effective
  potential for exchange},\ }\href {https://doi.org/10.1063/1.2213970}
  {\bibfield  {journal} {\bibinfo  {journal} {J. Chem. Phys.}\ }\textbf
  {\bibinfo {volume} {124}},\ \bibinfo {pages} {221101} (\bibinfo {year}
  {2006})}\BibitemShut {NoStop}%
\bibitem [{\citenamefont {Grimme}\ \emph {et~al.}(2010)\citenamefont {Grimme},
  \citenamefont {Antony}, \citenamefont {Ehrlich},\ and\ \citenamefont
  {Krieg}}]{dft-3d-doi:10.1063/1.3382344}%
  \BibitemOpen
  \bibfield  {author} {\bibinfo {author} {\bibfnamefont {S.}~\bibnamefont
  {Grimme}}, \bibinfo {author} {\bibfnamefont {J.}~\bibnamefont {Antony}},
  \bibinfo {author} {\bibfnamefont {S.}~\bibnamefont {Ehrlich}},\ and\ \bibinfo
  {author} {\bibfnamefont {H.}~\bibnamefont {Krieg}},\ }\bibfield  {title}
  {\bibinfo {title} {{A consistent and accurate ab initio parametrization of
  density functional dispersion correction (DFT-D) for the 94 elements H-Pu}},\
  }\href {https://doi.org/10.1063/1.3382344} {\bibfield  {journal} {\bibinfo
  {journal} {J. Chem. Phys.}\ }\textbf {\bibinfo {volume} {132}},\ \bibinfo
  {pages} {154104} (\bibinfo {year} {2010})}\BibitemShut {NoStop}%
\bibitem [{\citenamefont {Wang}\ \emph {et~al.}(2021)\citenamefont {Wang},
  \citenamefont {Xu}, \citenamefont {Liu}, \citenamefont {Tang},\ and\
  \citenamefont {Geng}}]{vaspkit}%
  \BibitemOpen
  \bibfield  {author} {\bibinfo {author} {\bibfnamefont {V.}~\bibnamefont
  {Wang}}, \bibinfo {author} {\bibfnamefont {N.}~\bibnamefont {Xu}}, \bibinfo
  {author} {\bibfnamefont {J.-C.}\ \bibnamefont {Liu}}, \bibinfo {author}
  {\bibfnamefont {G.}~\bibnamefont {Tang}},\ and\ \bibinfo {author}
  {\bibfnamefont {W.-T.}\ \bibnamefont {Geng}},\ }\bibfield  {title} {\bibinfo
  {title} {{VASPKIT: A user-friendly interface facilitating high-throughput
  computing and analysis using VASP code}},\ }\href
  {https://doi.org/https://doi.org/10.1016/j.cpc.2021.108033} {\bibfield
  {journal} {\bibinfo  {journal} {Comput. Phys. Commun.}\ }\textbf {\bibinfo
  {volume} {267}},\ \bibinfo {pages} {108033} (\bibinfo {year}
  {2021})}\BibitemShut {NoStop}%
\bibitem [{\citenamefont {Togo}\ and\ \citenamefont {Tanaka}(2015)}]{phonopy}%
  \BibitemOpen
  \bibfield  {author} {\bibinfo {author} {\bibfnamefont {A.}~\bibnamefont
  {Togo}}\ and\ \bibinfo {author} {\bibfnamefont {I.}~\bibnamefont {Tanaka}},\
  }\bibfield  {title} {\bibinfo {title} {First principles phonon calculations
  in materials science},\ }\href
  {https://doi.org/https://doi.org/10.1016/j.scriptamat.2015.07.021} {\bibfield
   {journal} {\bibinfo  {journal} {Scr. Mater.}\ }\textbf {\bibinfo {volume}
  {108}},\ \bibinfo {pages} {1} (\bibinfo {year} {2015})}\BibitemShut {NoStop}%
\bibitem [{\citenamefont {Momma}\ and\ \citenamefont {Izumi}(2011)}]{VESTA}%
  \BibitemOpen
  \bibfield  {author} {\bibinfo {author} {\bibfnamefont {K.}~\bibnamefont
  {Momma}}\ and\ \bibinfo {author} {\bibfnamefont {F.}~\bibnamefont {Izumi}},\
  }\bibfield  {title} {\bibinfo {title} {{{\it VESTA3} for three-dimensional
  visualization of crystal, volumetric and morphology data}},\ }\href
  {https://doi.org/10.1107/S0021889811038970} {\bibfield  {journal} {\bibinfo
  {journal} {J. Appl. Crystallogr.}\ }\textbf {\bibinfo {volume} {44}},\
  \bibinfo {pages} {1272} (\bibinfo {year} {2011})}\BibitemShut {NoStop}%
\bibitem [{\citenamefont {Mostofi}\ \emph {et~al.}(2014)\citenamefont
  {Mostofi}, \citenamefont {Yates}, \citenamefont {Pizzi}, \citenamefont {Lee},
  \citenamefont {Souza}, \citenamefont {Vanderbilt},\ and\ \citenamefont
  {Marzari}}]{wannier90-MOSTOFI20142309}%
  \BibitemOpen
  \bibfield  {author} {\bibinfo {author} {\bibfnamefont {A.~A.}\ \bibnamefont
  {Mostofi}}, \bibinfo {author} {\bibfnamefont {J.~R.}\ \bibnamefont {Yates}},
  \bibinfo {author} {\bibfnamefont {G.}~\bibnamefont {Pizzi}}, \bibinfo
  {author} {\bibfnamefont {Y.-S.}\ \bibnamefont {Lee}}, \bibinfo {author}
  {\bibfnamefont {I.}~\bibnamefont {Souza}}, \bibinfo {author} {\bibfnamefont
  {D.}~\bibnamefont {Vanderbilt}},\ and\ \bibinfo {author} {\bibfnamefont
  {N.}~\bibnamefont {Marzari}},\ }\bibfield  {title} {\bibinfo {title} {{An
  updated version of wannier90: A tool for obtaining maximally-localised
  Wannier functions}},\ }\href
  {https://doi.org/https://doi.org/10.1016/j.cpc.2014.05.003} {\bibfield
  {journal} {\bibinfo  {journal} {Comput. Phys. Commun.}\ }\textbf {\bibinfo
  {volume} {185}},\ \bibinfo {pages} {2309} (\bibinfo {year}
  {2014})}\BibitemShut {NoStop}%
\bibitem [{\citenamefont {Marzari}\ and\ \citenamefont
  {Vanderbilt}(1997)}]{mlwf1-PhysRevB.56.12847}%
  \BibitemOpen
  \bibfield  {author} {\bibinfo {author} {\bibfnamefont {N.}~\bibnamefont
  {Marzari}}\ and\ \bibinfo {author} {\bibfnamefont {D.}~\bibnamefont
  {Vanderbilt}},\ }\bibfield  {title} {\bibinfo {title} {{Maximally localized
  generalized Wannier functions for composite energy bands}},\ }\href
  {https://doi.org/10.1103/PhysRevB.56.12847} {\bibfield  {journal} {\bibinfo
  {journal} {Phys. Rev. B}\ }\textbf {\bibinfo {volume} {56}},\ \bibinfo
  {pages} {12847} (\bibinfo {year} {1997})}\BibitemShut {NoStop}%
\bibitem [{\citenamefont {Souza}\ \emph {et~al.}(2001)\citenamefont {Souza},
  \citenamefont {Marzari},\ and\ \citenamefont
  {Vanderbilt}}]{mlwf2-PhysRevB.65.035109}%
  \BibitemOpen
  \bibfield  {author} {\bibinfo {author} {\bibfnamefont {I.}~\bibnamefont
  {Souza}}, \bibinfo {author} {\bibfnamefont {N.}~\bibnamefont {Marzari}},\
  and\ \bibinfo {author} {\bibfnamefont {D.}~\bibnamefont {Vanderbilt}},\
  }\bibfield  {title} {\bibinfo {title} {{Maximally localized Wannier functions
  for entangled energy bands}},\ }\href
  {https://doi.org/10.1103/PhysRevB.65.035109} {\bibfield  {journal} {\bibinfo
  {journal} {Phys. Rev. B}\ }\textbf {\bibinfo {volume} {65}},\ \bibinfo
  {pages} {035109} (\bibinfo {year} {2001})}\BibitemShut {NoStop}%
\bibitem [{\citenamefont {Wu}\ \emph {et~al.}(2018)\citenamefont {Wu},
  \citenamefont {Zhang}, \citenamefont {Song}, \citenamefont {Troyer},\ and\
  \citenamefont {Soluyanov}}]{wanniertools-WU2018405}%
  \BibitemOpen
  \bibfield  {author} {\bibinfo {author} {\bibfnamefont {Q.}~\bibnamefont
  {Wu}}, \bibinfo {author} {\bibfnamefont {S.}~\bibnamefont {Zhang}}, \bibinfo
  {author} {\bibfnamefont {H.-F.}\ \bibnamefont {Song}}, \bibinfo {author}
  {\bibfnamefont {M.}~\bibnamefont {Troyer}},\ and\ \bibinfo {author}
  {\bibfnamefont {A.~A.}\ \bibnamefont {Soluyanov}},\ }\bibfield  {title}
  {\bibinfo {title} {Wanniertools: An open-source software package for novel
  topological materials},\ }\href
  {https://doi.org/https://doi.org/10.1016/j.cpc.2017.09.033} {\bibfield
  {journal} {\bibinfo  {journal} {Comput. Phys. Commun.}\ }\textbf {\bibinfo
  {volume} {224}},\ \bibinfo {pages} {405} (\bibinfo {year}
  {2018})}\BibitemShut {NoStop}%
\bibitem [{\citenamefont {Eremeev}\ \emph {et~al.}(2018)\citenamefont
  {Eremeev}, \citenamefont {Otrokov},\ and\ \citenamefont
  {Chulkov}}]{otrokov-nanolett-doi:10.1021/acs.nanolett.8b03057}%
  \BibitemOpen
  \bibfield  {author} {\bibinfo {author} {\bibfnamefont {S.~V.}\ \bibnamefont
  {Eremeev}}, \bibinfo {author} {\bibfnamefont {M.~M.}\ \bibnamefont
  {Otrokov}},\ and\ \bibinfo {author} {\bibfnamefont {E.~V.}\ \bibnamefont
  {Chulkov}},\ }\bibfield  {title} {\bibinfo {title} {{New Universal Type of
  Interface in the Magnetic Insulator/Topological Insulator
  Heterostructures}},\ }\href {https://doi.org/10.1021/acs.nanolett.8b03057}
  {\bibfield  {journal} {\bibinfo  {journal} {Nano Lett.}\ }\textbf {\bibinfo
  {volume} {18}},\ \bibinfo {pages} {6521} (\bibinfo {year}
  {2018})}\BibitemShut {NoStop}%
\bibitem [{\citenamefont {Hirahara}\ \emph {et~al.}(2020)\citenamefont
  {Hirahara}, \citenamefont {Otrokov}, \citenamefont {Sasaki}, \citenamefont
  {Sumida}, \citenamefont {Tomohiro}, \citenamefont {Kusaka}, \citenamefont
  {Okuyama}, \citenamefont {Ichinokura}, \citenamefont {Kobayashi},
  \citenamefont {Takeda}, \citenamefont {Amemiya}, \citenamefont {Shirasawa},
  \citenamefont {Ideta}, \citenamefont {Miyamoto}, \citenamefont {Tanaka},
  \citenamefont {Kuroda}, \citenamefont {Okuda}, \citenamefont {Hono},
  \citenamefont {Eremeev},\ and\ \citenamefont
  {Chulkov}}]{otrokov-nc-Hirahara2020}%
  \BibitemOpen
  \bibfield  {author} {\bibinfo {author} {\bibfnamefont {T.}~\bibnamefont
  {Hirahara}}, \bibinfo {author} {\bibfnamefont {M.~M.}\ \bibnamefont
  {Otrokov}}, \bibinfo {author} {\bibfnamefont {T.~T.}\ \bibnamefont {Sasaki}},
  \bibinfo {author} {\bibfnamefont {K.}~\bibnamefont {Sumida}}, \bibinfo
  {author} {\bibfnamefont {Y.}~\bibnamefont {Tomohiro}}, \bibinfo {author}
  {\bibfnamefont {S.}~\bibnamefont {Kusaka}}, \bibinfo {author} {\bibfnamefont
  {Y.}~\bibnamefont {Okuyama}}, \bibinfo {author} {\bibfnamefont
  {S.}~\bibnamefont {Ichinokura}}, \bibinfo {author} {\bibfnamefont
  {M.}~\bibnamefont {Kobayashi}}, \bibinfo {author} {\bibfnamefont
  {Y.}~\bibnamefont {Takeda}}, \bibinfo {author} {\bibfnamefont
  {K.}~\bibnamefont {Amemiya}}, \bibinfo {author} {\bibfnamefont
  {T.}~\bibnamefont {Shirasawa}}, \bibinfo {author} {\bibfnamefont
  {S.}~\bibnamefont {Ideta}}, \bibinfo {author} {\bibfnamefont
  {K.}~\bibnamefont {Miyamoto}}, \bibinfo {author} {\bibfnamefont
  {K.}~\bibnamefont {Tanaka}}, \bibinfo {author} {\bibfnamefont
  {S.}~\bibnamefont {Kuroda}}, \bibinfo {author} {\bibfnamefont
  {T.}~\bibnamefont {Okuda}}, \bibinfo {author} {\bibfnamefont
  {K.}~\bibnamefont {Hono}}, \bibinfo {author} {\bibfnamefont {S.~V.}\
  \bibnamefont {Eremeev}},\ and\ \bibinfo {author} {\bibfnamefont {E.~V.}\
  \bibnamefont {Chulkov}},\ }\bibfield  {title} {\bibinfo {title} {{Fabrication
  of a novel magnetic topological heterostructure and temperature evolution of
  its massive Dirac cone}},\ }\href
  {https://doi.org/10.1038/s41467-020-18645-9} {\bibfield  {journal} {\bibinfo
  {journal} {Nat. Commun.}\ }\textbf {\bibinfo {volume} {11}},\ \bibinfo
  {pages} {4821} (\bibinfo {year} {2020})}\BibitemShut {NoStop}%
\bibitem [{\citenamefont {Jain}\ \emph {et~al.}(2013)\citenamefont {Jain},
  \citenamefont {Ong}, \citenamefont {Hautier}, \citenamefont {Chen},
  \citenamefont {Richards}, \citenamefont {Dacek}, \citenamefont {Cholia},
  \citenamefont {Gunter}, \citenamefont {Skinner}, \citenamefont {Ceder},\ and\
  \citenamefont {Persson}}]{MaterialsProject}%
  \BibitemOpen
  \bibfield  {author} {\bibinfo {author} {\bibfnamefont {A.}~\bibnamefont
  {Jain}}, \bibinfo {author} {\bibfnamefont {S.~P.}\ \bibnamefont {Ong}},
  \bibinfo {author} {\bibfnamefont {G.}~\bibnamefont {Hautier}}, \bibinfo
  {author} {\bibfnamefont {W.}~\bibnamefont {Chen}}, \bibinfo {author}
  {\bibfnamefont {W.~D.}\ \bibnamefont {Richards}}, \bibinfo {author}
  {\bibfnamefont {S.}~\bibnamefont {Dacek}}, \bibinfo {author} {\bibfnamefont
  {S.}~\bibnamefont {Cholia}}, \bibinfo {author} {\bibfnamefont
  {D.}~\bibnamefont {Gunter}}, \bibinfo {author} {\bibfnamefont
  {D.}~\bibnamefont {Skinner}}, \bibinfo {author} {\bibfnamefont
  {G.}~\bibnamefont {Ceder}},\ and\ \bibinfo {author} {\bibfnamefont {K.~A.}\
  \bibnamefont {Persson}},\ }\bibfield  {title} {\bibinfo {title} {{Commentary:
  The Materials Project: A materials genome approach to accelerating materials
  innovation}},\ }\href {https://doi.org/10.1063/1.4812323} {\bibfield
  {journal} {\bibinfo  {journal} {APL Mater.}\ }\textbf {\bibinfo {volume}
  {1}},\ \bibinfo {pages} {011002} (\bibinfo {year} {2013})}\BibitemShut
  {NoStop}%
\bibitem [{\citenamefont {Han}\ \emph {et~al.}(2021)\citenamefont {Han},
  \citenamefont {Sun}, \citenamefont {Qi}, \citenamefont {Xu},\ and\
  \citenamefont {Qiao}}]{Qiao-PhysRevB.103.245403}%
  \BibitemOpen
  \bibfield  {author} {\bibinfo {author} {\bibfnamefont {Y.}~\bibnamefont
  {Han}}, \bibinfo {author} {\bibfnamefont {S.}~\bibnamefont {Sun}}, \bibinfo
  {author} {\bibfnamefont {S.}~\bibnamefont {Qi}}, \bibinfo {author}
  {\bibfnamefont {X.}~\bibnamefont {Xu}},\ and\ \bibinfo {author}
  {\bibfnamefont {Z.}~\bibnamefont {Qiao}},\ }\bibfield  {title} {\bibinfo
  {title} {{Interlayer ferromagnetism and high-temperature quantum anomalous
  Hall effect in $p$-doped $\mathrm{Mn}{\mathrm{Bi}}_{2}{\mathrm{Te}}_{4}$
  multilayers}},\ }\href {https://doi.org/10.1103/PhysRevB.103.245403}
  {\bibfield  {journal} {\bibinfo  {journal} {Phys. Rev. B}\ }\textbf {\bibinfo
  {volume} {103}},\ \bibinfo {pages} {245403} (\bibinfo {year}
  {2021})}\BibitemShut {NoStop}%
\bibitem [{\citenamefont {Tan}\ \emph {et~al.}(2021)\citenamefont {Tan},
  \citenamefont {Liu}, \citenamefont {Wang},\ and\ \citenamefont
  {Yan}}]{Yan-PhysRevLett.127.046401}%
  \BibitemOpen
  \bibfield  {author} {\bibinfo {author} {\bibfnamefont {H.}~\bibnamefont
  {Tan}}, \bibinfo {author} {\bibfnamefont {Y.}~\bibnamefont {Liu}}, \bibinfo
  {author} {\bibfnamefont {Z.}~\bibnamefont {Wang}},\ and\ \bibinfo {author}
  {\bibfnamefont {B.}~\bibnamefont {Yan}},\ }\bibfield  {title} {\bibinfo
  {title} {Charge density waves and electronic properties of superconducting
  kagome metals},\ }\href {https://doi.org/10.1103/PhysRevLett.127.046401}
  {\bibfield  {journal} {\bibinfo  {journal} {Phys. Rev. Lett.}\ }\textbf
  {\bibinfo {volume} {127}},\ \bibinfo {pages} {046401} (\bibinfo {year}
  {2021})}\BibitemShut {NoStop}%
\bibitem [{\citenamefont {Pavarini}\ \emph {et~al.}(2012)\citenamefont
  {Pavarini}, \citenamefont {Koch}, \citenamefont {Anders},\ and\ \citenamefont
  {Jarrell}}]{book}%
  \BibitemOpen
  \bibfield  {author} {\bibinfo {author} {\bibfnamefont {E.}~\bibnamefont
  {Pavarini}}, \bibinfo {author} {\bibfnamefont {E.}~\bibnamefont {Koch}},
  \bibinfo {author} {\bibfnamefont {F.}~\bibnamefont {Anders}},\ and\ \bibinfo
  {author} {\bibfnamefont {M.}~\bibnamefont {Jarrell}},\ }\href@noop {} {\emph
  {\bibinfo {title} {Correlated Electrons: From Models to Materials}}}\
  (\bibinfo {year} {2012})\BibitemShut {NoStop}%
\bibitem [{\citenamefont {Li}\ \emph {et~al.}(2020{\natexlab{c}})\citenamefont
  {Li}, \citenamefont {Li}, \citenamefont {He}, \citenamefont {Wan},
  \citenamefont {Duan},\ and\ \citenamefont {Xu}}]{lizhe-PhysRevB.102.081107}%
  \BibitemOpen
  \bibfield  {author} {\bibinfo {author} {\bibfnamefont {Z.}~\bibnamefont
  {Li}}, \bibinfo {author} {\bibfnamefont {J.}~\bibnamefont {Li}}, \bibinfo
  {author} {\bibfnamefont {K.}~\bibnamefont {He}}, \bibinfo {author}
  {\bibfnamefont {X.}~\bibnamefont {Wan}}, \bibinfo {author} {\bibfnamefont
  {W.}~\bibnamefont {Duan}},\ and\ \bibinfo {author} {\bibfnamefont
  {Y.}~\bibnamefont {Xu}},\ }\bibfield  {title} {\bibinfo {title} {{Tunable
  interlayer magnetism and band topology in van der Waals heterostructures of
  {M}n{B}i${}_2${T}e${}_4$-family materials}},\ }\href
  {https://doi.org/10.1103/PhysRevB.102.081107} {\bibfield  {journal} {\bibinfo
   {journal} {Phys. Rev. B}\ }\textbf {\bibinfo {volume} {102}},\ \bibinfo
  {pages} {081107(R)} (\bibinfo {year} {2020}{\natexlab{c}})}\BibitemShut
  {NoStop}%
\bibitem [{\citenamefont {Zhang}\ \emph {et~al.}(2009)\citenamefont {Zhang},
  \citenamefont {Liu}, \citenamefont {Qi}, \citenamefont {Dai}, \citenamefont
  {Fang},\ and\ \citenamefont {Zhang}}]{Bi2Te3Zhang2009}%
  \BibitemOpen
  \bibfield  {author} {\bibinfo {author} {\bibfnamefont {H.}~\bibnamefont
  {Zhang}}, \bibinfo {author} {\bibfnamefont {C.-X.}\ \bibnamefont {Liu}},
  \bibinfo {author} {\bibfnamefont {X.-L.}\ \bibnamefont {Qi}}, \bibinfo
  {author} {\bibfnamefont {X.}~\bibnamefont {Dai}}, \bibinfo {author}
  {\bibfnamefont {Z.}~\bibnamefont {Fang}},\ and\ \bibinfo {author}
  {\bibfnamefont {S.-C.}\ \bibnamefont {Zhang}},\ }\bibfield  {title} {\bibinfo
  {title} {{Topological insulators in Bi$ {}_2 $Se$ {}_3 $, Bi$ {}_2 $Te$ {}_3
  $ and Sb$ {}_2 $Te$ {}_3 $ with a single Dirac cone on the surface}},\ }\href
  {https://doi.org/10.1038/nphys1270} {\bibfield  {journal} {\bibinfo
  {journal} {Nat. Phys.}\ }\textbf {\bibinfo {volume} {5}},\ \bibinfo {pages}
  {438} (\bibinfo {year} {2009})}\BibitemShut {NoStop}%
\bibitem [{\citenamefont {Liu}\ \emph {et~al.}(2010)\citenamefont {Liu},
  \citenamefont {Qi}, \citenamefont {Zhang}, \citenamefont {Dai}, \citenamefont
  {Fang},\ and\ \citenamefont {Zhang}}]{modelPhysRevB.82.045122}%
  \BibitemOpen
  \bibfield  {author} {\bibinfo {author} {\bibfnamefont {C.-X.}\ \bibnamefont
  {Liu}}, \bibinfo {author} {\bibfnamefont {X.-L.}\ \bibnamefont {Qi}},
  \bibinfo {author} {\bibfnamefont {H.-J.}\ \bibnamefont {Zhang}}, \bibinfo
  {author} {\bibfnamefont {X.}~\bibnamefont {Dai}}, \bibinfo {author}
  {\bibfnamefont {Z.}~\bibnamefont {Fang}},\ and\ \bibinfo {author}
  {\bibfnamefont {S.-C.}\ \bibnamefont {Zhang}},\ }\bibfield  {title} {\bibinfo
  {title} {Model hamiltonian for topological insulators},\ }\href
  {https://doi.org/10.1103/PhysRevB.82.045122} {\bibfield  {journal} {\bibinfo
  {journal} {Phys. Rev. B}\ }\textbf {\bibinfo {volume} {82}},\ \bibinfo
  {pages} {045122} (\bibinfo {year} {2010})}\BibitemShut {NoStop}%
\bibitem [{\citenamefont {Wang}\ \emph {et~al.}(2020)\citenamefont {Wang},
  \citenamefont {Wang}, \citenamefont {Yang}, \citenamefont {Shi},
  \citenamefont {Ruan}, \citenamefont {Xing}, \citenamefont {Wang},\ and\
  \citenamefont {Zhang}}]{pt-PhysRevB.101.081109}%
  \BibitemOpen
  \bibfield  {author} {\bibinfo {author} {\bibfnamefont {H.}~\bibnamefont
  {Wang}}, \bibinfo {author} {\bibfnamefont {D.}~\bibnamefont {Wang}}, \bibinfo
  {author} {\bibfnamefont {Z.}~\bibnamefont {Yang}}, \bibinfo {author}
  {\bibfnamefont {M.}~\bibnamefont {Shi}}, \bibinfo {author} {\bibfnamefont
  {J.}~\bibnamefont {Ruan}}, \bibinfo {author} {\bibfnamefont {D.}~\bibnamefont
  {Xing}}, \bibinfo {author} {\bibfnamefont {J.}~\bibnamefont {Wang}},\ and\
  \bibinfo {author} {\bibfnamefont {H.}~\bibnamefont {Zhang}},\ }\bibfield
  {title} {\bibinfo {title} {{Dynamical axion state with hidden pseudospin
  Chern numbers in ${\mathrm{MnBi}}_{2}{\mathrm{Te}}_{4}$-based
  heterostructures}},\ }\href {https://doi.org/10.1103/PhysRevB.101.081109}
  {\bibfield  {journal} {\bibinfo  {journal} {Phys. Rev. B}\ }\textbf {\bibinfo
  {volume} {101}},\ \bibinfo {pages} {081109(R)} (\bibinfo {year}
  {2020})}\BibitemShut {NoStop}%
\bibitem [{\citenamefont {Xu}\ \emph {et~al.}(2011)\citenamefont {Xu},
  \citenamefont {Weng}, \citenamefont {Wang}, \citenamefont {Dai},\ and\
  \citenamefont {Fang}}]{HgCr2Se4-PhysRevLett.107.186806}%
  \BibitemOpen
  \bibfield  {author} {\bibinfo {author} {\bibfnamefont {G.}~\bibnamefont
  {Xu}}, \bibinfo {author} {\bibfnamefont {H.}~\bibnamefont {Weng}}, \bibinfo
  {author} {\bibfnamefont {Z.}~\bibnamefont {Wang}}, \bibinfo {author}
  {\bibfnamefont {X.}~\bibnamefont {Dai}},\ and\ \bibinfo {author}
  {\bibfnamefont {Z.}~\bibnamefont {Fang}},\ }\bibfield  {title} {\bibinfo
  {title} {{Chern Semimetal and the Quantized Anomalous Hall Effect in
  ${\mathrm{HgCr}}_{2}{\mathrm{Se}}_{4}$}},\ }\href
  {https://doi.org/10.1103/PhysRevLett.107.186806} {\bibfield  {journal}
  {\bibinfo  {journal} {Phys. Rev. Lett.}\ }\textbf {\bibinfo {volume} {107}},\
  \bibinfo {pages} {186806} (\bibinfo {year} {2011})}\BibitemShut {NoStop}%
\bibitem [{\citenamefont {Burkov}\ and\ \citenamefont
  {Balents}(2011)}]{burkov-prl-PhysRevLett.107.127205}%
  \BibitemOpen
  \bibfield  {author} {\bibinfo {author} {\bibfnamefont {A.~A.}\ \bibnamefont
  {Burkov}}\ and\ \bibinfo {author} {\bibfnamefont {L.}~\bibnamefont
  {Balents}},\ }\bibfield  {title} {\bibinfo {title} {{Weyl Semimetal in a
  Topological Insulator Multilayer}},\ }\href
  {https://doi.org/10.1103/PhysRevLett.107.127205} {\bibfield  {journal}
  {\bibinfo  {journal} {Phys. Rev. Lett.}\ }\textbf {\bibinfo {volume} {107}},\
  \bibinfo {pages} {127205} (\bibinfo {year} {2011})}\BibitemShut {NoStop}%
\bibitem [{\citenamefont {Zyuzin}\ \emph {et~al.}(2012)\citenamefont {Zyuzin},
  \citenamefont {Wu},\ and\ \citenamefont
  {Burkov}}]{burkov-pt-PhysRevB.85.165110}%
  \BibitemOpen
  \bibfield  {author} {\bibinfo {author} {\bibfnamefont {A.~A.}\ \bibnamefont
  {Zyuzin}}, \bibinfo {author} {\bibfnamefont {S.}~\bibnamefont {Wu}},\ and\
  \bibinfo {author} {\bibfnamefont {A.~A.}\ \bibnamefont {Burkov}},\ }\bibfield
   {title} {\bibinfo {title} {Weyl semimetal with broken time reversal and
  inversion symmetries},\ }\href {https://doi.org/10.1103/PhysRevB.85.165110}
  {\bibfield  {journal} {\bibinfo  {journal} {Phys. Rev. B}\ }\textbf {\bibinfo
  {volume} {85}},\ \bibinfo {pages} {165110} (\bibinfo {year}
  {2012})}\BibitemShut {NoStop}%
\end{thebibliography}%

\end{document}